\documentclass[final]{ws-ijmpe}
\usepackage{bbm}
\usepackage{amsmath}
\usepackage{amsfonts,amsbsy}
\usepackage{amssymb}
\usepackage{feynman}

\def\empile#1\over#2{\mathrel{\mathop{\kern 0pt#1}\limits_{#2}}}
\def\bs{\boldsymbol}
\def\wt#1{\widetilde{#1}}

\newcommand{\slv}{\raise.15ex\hbox{$/$}\kern-.53em\hbox{$v$}}
\newcommand{\slF}{\raise.15ex\hbox{$/$}\kern-.53em\hbox{$F$}}
\newcommand{\slL}{\raise.15ex\hbox{$/$}\kern-.53em\hbox{$L$}}
\newcommand{\slP}{\raise.15ex\hbox{$/$}\kern-.53em\hbox{$P$}}
\newcommand{\slp}{\raise.15ex\hbox{$/$}\kern-.53em\hbox{$p$}}
\newcommand{\slq}{\raise.15ex\hbox{$/$}\kern-.53em\hbox{$q$}}
\newcommand{\slR}{\raise.15ex\hbox{$/$}\kern-.53em\hbox{$R$}}
\newcommand{\slQ}{\raise.15ex\hbox{$/$}\kern-.53em\hbox{$Q$}}
\newcommand{\slK}{\raise.15ex\hbox{$/$}\kern-.53em\hbox{$K$}}
\newcommand{\slk}{\raise.15ex\hbox{$/$}\kern-.53em\hbox{$k$}}
\newcommand{\slD}{\raise.15ex\hbox{$/$}\kern-.53em\hbox{$D$}}
\newcommand{\slC}{\raise.15ex\hbox{$/$}\kern-.53em\hbox{$C$}}
\newcommand{\slA}{\raise.15ex\hbox{$/$}\kern-.53em\hbox{$A$}}
\newcommand{\slSigma}{\raise.15ex\hbox{$/$}\kern-.53em\hbox{$\Sigma$}}
\newcommand{\slpartial}{\raise.15ex\hbox{$/$}\kern-.53em\hbox{$\partial$}}
\newcommand{\slcalP}{\raise.15ex\hbox{$/$}\kern-.63em\hbox{$\cal P$}}
\newcommand{\slcalA}{\raise.15ex\hbox{$/$}\kern-.63em\hbox{$\cal A$}}

\def\p{{\boldsymbol p}}
\def\q{{\boldsymbol q}}
\def\P{{\boldsymbol P}}

\def\k{{\boldsymbol k}}

\def\x{{\boldsymbol x}}
\def\y{{\boldsymbol y}}

\def\z{{\boldsymbol z}}
\def\u{{\boldsymbol u}}
\def\v{{\boldsymbol v}}

\def\colora{}
\def\colorb{}
\def\colorc{}
\def\colord{}
\def\colore{}

\def\colorh{}
\def\vec{}

\catcode`\@=11

\newcount\@tempcntc
\def\@citex[#1]#2{\if@filesw\immediate\write\@auxout{\string\citation{#2}}\fi
  \@tempcnta\z@\@tempcntb\m@ne\def\@citea{}\@cite{%
        \@for\@citeb:=#2\do%
    {\@ifundefined{b@\@citeb}%
        {\@citeo\@tempcntb\m@ne\@citea%
                \def\@citea{,\penalty\@m\ }{\bf ?}\@warning%
                {Citation `\@citeb' on page \thepage \space undefined}}%
        {\setbox\z@\hbox{\global\@tempcntc0\csname b@\@citeb\endcsname\relax}
     \ifnum\@tempcntc=\z@ \@citeo\@tempcntb\m@ne%
       \@citea\def\@citea{,\penalty\@m}%
       \hbox{\csname b@\@citeb\endcsname}%
     \else%
      \advance\@tempcntb\@ne%
      \ifnum\@tempcntb=\@tempcntc%
      \else\advance\@tempcntb\m@ne\@citeo%
      \@tempcnta\@tempcntc\@tempcntb\@tempcntc\fi\fi}}\@citeo}{#1}}%

\def\@citeo{\ifnum\@tempcnta>\@tempcntb\else\@citea
  \def\@citea{,\penalty\@m}%
  \ifnum\@tempcnta=\@tempcntb\the\@tempcnta\else
   {\advance\@tempcnta\@ne\ifnum\@tempcnta=\@tempcntb \else
\def\@citea{--}\fi
    \advance\@tempcnta\m@ne\the\@tempcnta\@citea\the\@tempcntb}\fi\fi}

\catcode`\@=12

\begin{document}


\catchline{}{}{}{}{}

\title{HIGH ENERGY SCATTERING\\ IN QUANTUM
CHROMODYNAMICS\footnote{Lectures given at the Xth
Hadron Physics Workshop, March 2007, Florianopolis, Brazil.}}

\author{\footnotesize FRANCOIS GELIS}

\address{Theory Division,
  PH-TH, Case C01600, CERN,\\
  CH-1211 Geneva 23, Switzerland\\
francois.gelis@cern.ch}

\author{TUOMAS LAPPI, RAJU VENUGOPALAN}

\address{Brookhaven National Laboratory,
  Physics Department\\
  Upton, NY-11973, USA\\
 tvv@quark.phy.bnl.gov, raju@bnl.gov}

\maketitle


\begin{abstract}
In this series of three lectures, we discuss several aspects of high
energy scattering among hadrons in Quantum Chromodynamics. The first
lecture is devoted to a description of the parton model, Bjorken
scaling and the scaling violations due to the evolution of parton
distributions with the transverse resolution scale. The second lecture
describes parton evolution at small momentum fraction $x$, the
phenomenon of gluon saturation and the Color Glass Condensate
(CGC). In the third lecture, we present the application of the CGC to
the study of high energy hadronic collisions, with emphasis on
nucleus-nucleus collisions. In particular, we provide the outline of a
proof of high energy factorization for inclusive gluon production.
\end{abstract}
\vskip 5mm
\begin{flushright}
Preprint CERN-PH-TH/2007-131
\end{flushright}

\section{Introduction}
Quantum Chromodynamics (QCD) is very successful at describing hadronic
scatterings involving very large momentum transfers. A crucial element
in these successes is the asymptotic freedom of QCD
\cite{running_alpha}, that renders the coupling weaker as the momentum
transfer scale increases, thereby making perturbation theory more and
more accurate. The other important property of QCD when comparing key
theoretical predictions to experimental measurements is the
factorization of the short distance physics which can be computed
reliably in perturbation theory from the long distance strong coupling
physics related to confinement. The latter are organized into
non-perturbative parton distributions, that depend on the scales of
time and transverse space at which the hadron is resolved in the
process under consideration. In fact, QCD not only enables one to
compute the perturbative hard cross-section, but also predicts the
scale dependence of the parton distributions.
\begin{figure}[htbp]
\centerline{
\hskip 15mm
\resizebox*{4cm}{!}{\includegraphics{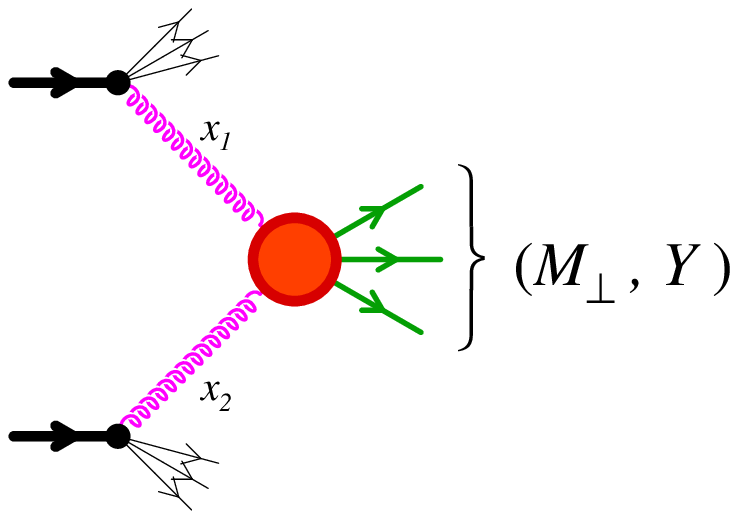}}
\hfill
\resizebox*{4cm}{!}{\includegraphics{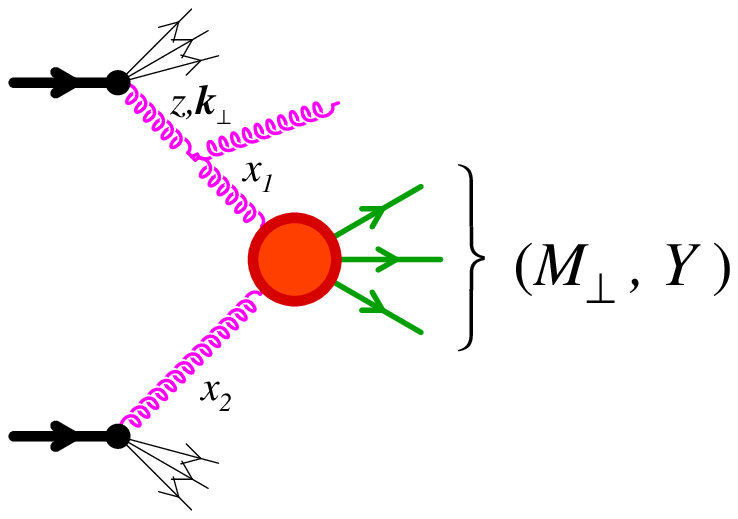}}
\hskip 15mm
}
\caption{\label{fig:logs}Generic hard process in the scattering of two
hadrons. Left: Leading Order. Right: Next-to-Leading Order correction
involving gluon radiation in the initial state.}
\end{figure}
A generic issue in the application of perturbative QCD to the study of
hadronic scatterings is the occurrence of logarithmic corrections in
higher orders of the perturbative expansion. These logarithms can be
large enough to compensate the extra coupling constant $\alpha_s$ they
come accompanied with, thus voiding the naive, fixed order,
application of perturbation theory. Consider for instance a generic
gluon-gluon fusion process, as illustrated on the left of figure
\ref{fig:logs}, producing a final state of momentum $P^\mu$. The two
gluons have longitudinal momentum fractions $x_{1,2}$ given by
\begin{equation}
x_{1,2}=\frac{M_\perp}{\sqrt{s}}\,e^{\pm Y}\; ,
\end{equation}
where $M_\perp\equiv\sqrt{\P_\perp^2+P^2}$ ($P^2\equiv P_\mu P^\mu$ is
the invariant mass of the final state) and $Y\equiv\ln(P^+/P^-)/2$.
On the right of figure \ref{fig:logs} is represented a radiative
correction to this process, where a gluon is emitted from one of the
incoming lines. Roughly speaking, such a correction is accompanied by a 
factor 
\begin{equation}
\alpha_s\;\int_{x_1}\frac{dz}{z}\;\int^{M_\perp}\frac{d^2\k_\perp}{k_\perp^2}
\; ,
\end{equation}
where $z$ is the momentum fraction of the gluon before the splitting,
and $\k_\perp$ its transverse momentum. Such corrections produce
logarithms, $\log(1/x_1)$ and $\log(M_\perp)$, that respectively
become large when $x_1$ is small or when $M_\perp$ is large compared
to typical hadronic mass scales.  These logarithms tell us that parton
distributions must depend on the momentum fraction $x$ and on a
transverse resolution scale $M_\perp$, that are set by the process
under consideration.  In the linear regime\footnote{We use the
denomination ``linear'' here to distinguish it from the saturation
regime discussed later that is characterized by non-linear evolution
equations.}, there are ``factorization theorems'' -- {\sl $k_{\rm
t}$-factorization} \cite{ktfact} in the first case and {\sl collinear
factorization} \cite{collfact} in the second case -- that tell us that
the logarithms are universal and can be systematically absorbed in the
definition of parton distributions~\footnote{The latter is currently
more rigorously established than the former.}.  The $x$ dependence
that results from resumming the logarithms of $1/x$ is taken into
account by the BFKL equation \cite{bfkl}. Similarly, the dependence on
the transverse resolution scale $M_\perp$ is accounted for by the
DGLAP equation \cite{dglap}.

The application of QCD is a lot less straightforward for scattering at
very large center of mass energy, and moderate momentum
transfers. This kinematics in fact dominates the bulk of the
cross-section at collider energies. A striking example of this
kinematics is encountered in Heavy Ion Collisions (HIC), when one
attempts to calculate the multiplicity of produced particles.  There,
despite the very large center of mass energy\footnote{At RHIC, center
of mass energies range up to $\sqrt{s}=200$~GeV/nucleon; the LHC will
collide nuclei at $\sqrt{s}=5.5$~TeV/nucleon.}, typical momentum
transfers are small\footnote{For instance, in a collision at
$\sqrt{s}=200$~GeV between gold nuclei at RHIC, 99\% of the
multiplicity comes from hadrons whose $p_\perp$ is below 2~GeV.}, of
the order of a few GeVs at most. In this kinematics, two phenomena
that become dominant are
\begin{itemize}
  \item {\bf Gluon saturation}~: the linear evolution equations (DGLAP
  or BFKL) for the parton distributions implicitly assume that the
  parton densities in the hadron are small and that the only important
  processes are splittings. However, at low values of $x$, the gluon
  density may become so large that gluon recombinations are an
  important effect.
  \item {\bf Multiple scatterings}~: processes involving more than one
  parton from a given projectile become sizeable.
\end{itemize}
It is highly non trivial that this dominant regime of hadronic
interactions is amenable to a controlled perturbative treatment within
QCD, and the realization of this possibility is a major theoretical
advance in the last decade.  The goal of these three lectures is to
present the framework in which such calculations can be carried out.

In the first lecture, we will review key aspects of the parton
model. Our recurring example will be the Deep Inelastic Scattering
(DIS) process of scattering a high energy electron at high momentum
transfers off a proton. Beginning with the inclusive DIS
cross-section, we will arrive at the parton model (firstly in its most
naive incarnation, and then within QCD), and subsequently at the DGLAP
evolution equations that control the scaling violations measured
experimentally.

In the second lecture, we will address the evolution of the parton
model to small values of the momentum fraction $x$ and the saturation
of the gluon distribution. After illustrating the tremendous
simplification of high energy scattering in the eikonal limit, we will
derive the BFKL equation and its non-linear extension, the BK
equation. We then discuss how these evolution equations arise in the
Color Glass Condensate effective theory. We conclude the lecture
with a discussion of the close analogy between the energy dependence
of scattering amplitudes in QCD and the temporal evolution of
reaction-diffusion processes in statistical mechanics.

The third lecture is devoted to the study of nucleus-nucleus
collisions at high energy. Our main focus is the study of bulk
particle production in these reactions within the CGC framework.
After an exposition of the power counting rules in the saturated
regime, we explain how to keep track of the infinite sets of diagrams
that contribute to the inclusive gluon spectrum. Specifically, we
demonstrate how these can be resummed at leading and next-to-leading
order by solving classical equations of motion for the gauge fields
The inclusive quark spectrum is discussed as well. We conclude the
lecture with a discussion of the inclusive gluon spectrum at
next-to-leading order and outline a proof of high energy factorization
in this context. Understanding this factorization may hold the key to
understanding early thermalization in heavy ion collisions. Some
recent progress in this direction is briefly discussed.

\section{Lecture I~: Parton model, Bjorken scaling, scaling violations}
In this lecture, we will begin with the simple parton model and develop the 
conventional Operator Product Expansion (OPE) approach and the associated
DGLAP evolution equations. To keep things as simple as possible, we will
use Deep Inelastic Scattering to illustrate the ideas in this lecture.
\subsection{Kinematics of DIS}
\begin{figure}[htbp]
\begin{center}
\resizebox*{4.5cm}{!}{\includegraphics{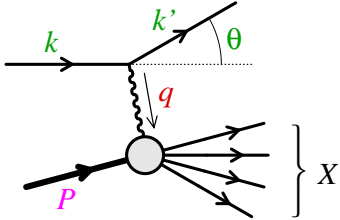}}
\end{center}
\caption{\label{fig:DIS_kin}Kinematical variables in the Deep
Inelastic Scattering process. $k$ and $P$ are known from the
experimental setup, and $k^\prime$ is obtained by measuring the
deflected lepton.}
\end{figure}
The basic idea of Deep Inelastic Scattering (DIS) is to use a well understood lepton 
probe (that does not involve strong interactions) to
study a hadron. The interaction is via the exchange of a virtual
photon\footnote{If the virtuality of the photon is small (in {\sl
photo-production} reactions for instance), the assertion that the
photon is a ``well known probe that does not involve strong
interactions'' is not valid anymore. Indeed, the photon may fluctuate, for instance, 
into a $\rho$ meson.}. Variants of this reaction involve
the exchange of a $W^\pm$ or $Z^0$ boson which become increasingly important at large momentum transfers. The kinematics of DIS is
characterized by a few Lorentz invariants (see figure
\ref{fig:DIS_kin} for the notations), traditionally defined as
\begin{eqnarray}
  {\colorb\nu}&\equiv& P\cdot q\nonumber\\
  {\colorb s}&\equiv& (P+k)^2\nonumber\\
  {\colorb M_{_X}^2}&\equiv& (P+q)^2=m_{_N}^2+2\nu+q^2\; ,
\end{eqnarray}
where $m_{_N}$ is the nucleon mass (assuming that the target is a
proton) and $M_{_X}$ the invariant mass of the hadronic final
state. Because the exchanged photon is space-like, one usually
introduces $Q^2\equiv -q^2 >0$, and also ${\colorb x\equiv Q^2/2\nu}$.
Note that since $M_{_X}^2 \ge m_{_N}^2$, we must have ${\colord 0\le
x\le 1}$ -- the value $x=1$ being reached only in the case where the
proton is scattered elastically.

The simplest cross-section one can measure in a DIS experiment is the
total inclusive electron+proton cross-section, where one sums over all possible hadronic final
states~:
\begin{equation}
{\colora E^\prime}\frac{d{\colord\sigma}_{\colorc e^-N}}{d^3{\colora\vec{\k}^\prime}}
=\sum_{{\colorb{\rm states\ }X}}
{\colora E^\prime}
\frac{d{\colord\sigma}_{{\colorc e^-N}\to {\colorc e^-}{\colorb X}}}
{d^3{\colora\vec{\k}^\prime}}\; .
\end{equation}
The partial cross-section associated to a given final state $X$ can be
written as
\begin{equation}
{\colora E^\prime}
\frac{d{\colord\sigma}_{{\colorc e^-N}\to {\colorc e^-}{\colorb X}}}
{d^3{\colora\vec{\k}^\prime}} 
= \!\int\!\frac{{\colorb\left[d\Phi_{_X}\right]}}
{32\pi^3({\colorb s}\!-\!{\colorc m_{_N}^2})}
(2\pi)^4\delta({\colorc P}\!+\!{\colora k}\!-\!{\colora k^\prime}\!-\!{\colorb P_{_X}})
\left<
\left|
{\colord{\cal M}_{_{\colorb X}}}
\right|^2
\right>_{\rm spin}\; ,
\end{equation}
where $[d\Phi_{_X}]$ denotes the invariant phase-space element for the
final state $X$ and ${\cal M}_{_X}$ is the corresponding transition
amplitude. The ``spin" symbol denotes an average over all spin polarizations of the initial state and a sum over 
those in the final state. The transition amplitude is decomposed into an electromagnetic part
and a hadronic matrix element as 
\begin{equation}
{\colorc{\cal M}_{_{\colorb X}}} = 
\frac{i{\colord e}}{{\colorb q^2}}\, \left[
{\overline u}({\colora\vec\k^\prime})\gamma^\mu u({\colora\vec\k})\right]\,
\big<{\colorb X}\big|J_\mu(0)\big|{\colorc N(P)}\big>\; .
\end{equation}
In this equation $J_\mu$ is the hadron electromagnetic current that couples
to the photon, and $\big|{\colorc N(P)}\big>$ denotes a state
containing a nucleon of momentum $P$. 

Squaring this amplitude and collecting all the factors, the inclusive DIS cross-section can be expressed as 
\begin{equation}
{\colora E^\prime}
\frac{d{\colord\sigma}_{{\colorc e^-N}}}{d^3{\colora\vec{\k}^\prime}}
=\frac{1}{32\pi^3({\colorb s}-{\colorc m_{_N}^2})} 
\frac{{\colord e^2}}{{\colorb q^4}} 4\pi {\colord L^{\mu\nu}} 
{\colorb W_{\mu\nu}}\; ,
\end{equation}
where the {\sl leptonic tensor} (neglecting the electron mass) is
\begin{eqnarray}
{\colord L^{\mu\nu}}&\equiv& 
\left<
{\overline u}({\colora\vec\k^\prime})\gamma^\mu u({\colora\vec\k})
{\overline u}({\colora\vec\k})\gamma^\nu u({\colora\vec\k^\prime})
\right>_{\rm spin}
\nonumber\\
&=& 2({\colora k^\mu k^{\prime\nu}}
+{\colora k^\nu k^{\prime\mu}}-g^{\mu\nu}\,{\colora k}\cdot{\colora k^\prime})
\; .
\end{eqnarray}
and $W_{\mu\nu}$ -- the {\sl hadronic tensor} -- is defined as
\begin{eqnarray}
4\pi {\colorb W_{\mu\nu}}&\equiv& 
\sum_{{\colorb{\rm states\ }X}} \int {\colorb\left[d\Phi_{_X}\right]}
{\colorh{(2\pi)^4\delta(P+q-P_{_X})}}
\nonumber\\
&&\qquad\times
\left<
\big<{\colorc N(P)}\big|J_\nu^\dagger(0)\big|{\colorb X}\big>
\big<{\colorb X}\big|J_\mu(0)\big|{\colorc N(P)}\big>
\right>_{\rm spin}
\nonumber\\
&=&\int d^4y\; 
e^{iq\cdot y}
\;\left<
\big<{\colorc N(P)}\big|
J_\nu^\dagger(y)J_\mu(0)\big|{\colorc N(P)}\big>
\right>_{\rm spin}
\; .
\label{eq:W1}
\end{eqnarray}
The second equality is obtained using the complete basis of hadronic states $X$. Thus, the hadronic tensor is the Fourier
transform of the expectation value of the product of two currents in
the nucleon state. An important point is that this object cannot be
calculated by perturbative methods. This rank-2 tensor can be expressed simply in terms of 
two independent {\sl structure functions} as a consequence of 
\begin{itemize}
  \item Conservation of the electromagnetic current~: 
    $q_\mu W^{\mu\nu}=q_\nu W^{\mu\nu}=0$
  \item Parity and time-reversal symmetry~: $W^{\mu\nu}=W^{\nu\mu}$
  \item Electromagnetic currents conserve parity~: the Levi-Civita
  tensor $\epsilon^{\mu\nu\rho\sigma}$ cannot appear\footnote{This
  property is not true in DIS reactions involving the exchange of a
  weak current;  an additional structure function $F_3$ is needed
  in this case.} in the tensorial decomposition of $W^{\mu\nu}$
\end{itemize}
When one works out these constraints, the most general tensor one can
construct from $P^\mu, q^\mu$ and $g^{\mu\nu}$ reads~:
\begin{equation}
{\colorb W_{\mu\nu}}=
-{\colorb F_1}\left(g_{\mu\nu}-\frac{q_\mu q_\nu}{q^2}\right)
+\frac{\colorb F_2}{\colorc P\cdot q}
\left(P_\mu-q_\mu\frac{P\cdot q}{q^2}\right)
\left(P_\nu-q_\nu\frac{P\cdot q}{q^2}\right)\; ,
\end{equation}
where $F_{1,2}$ are the two structure functions\footnote{The structure
function $F_2$ differs slightly from the $W_2$ defined in
\cite{DrellW1}~: $F_2=\nu W_2 /m_{_N}^2$.}. As scalars,
they only depend on Lorentz invariants, namely, the variables $x$
and $Q^2$. The inclusive DIS
cross-section in the rest frame of the proton can be expressed in terms of $F_{1,2}$ as 
\begin{equation}
\frac{d{\colord\sigma}_{{\colorc e^-N}}}{d{\colora E^\prime} d{\colora\Omega}}
=
\frac{{\colord\alpha_{\rm em}^2}}{4{\colorc m_{_N}} {\colora E^2} 
\sin^4({\colora\theta}/2)}
\left[
2\,F_1\,\sin^2\frac{\colora\theta}{2}
+
\frac{m_{_N}^2}{\nu}\,F_2\,\cos^2\frac{\colora\theta}{2}
\right]\; ,
\label{eq:DIS1}
\end{equation}
where $\Omega$ represents the solid angle of the scattered electron
and $E^\prime$ its energy. 

\subsection{Experimental facts}
Two major experimental results from SLAC \cite{slac} in the late
1960's played a crucial role in the development of the parton model.
\begin{figure}[htbp]
\begin{center}
\hskip 5mm
\resizebox*{5.5cm}{!}{\includegraphics{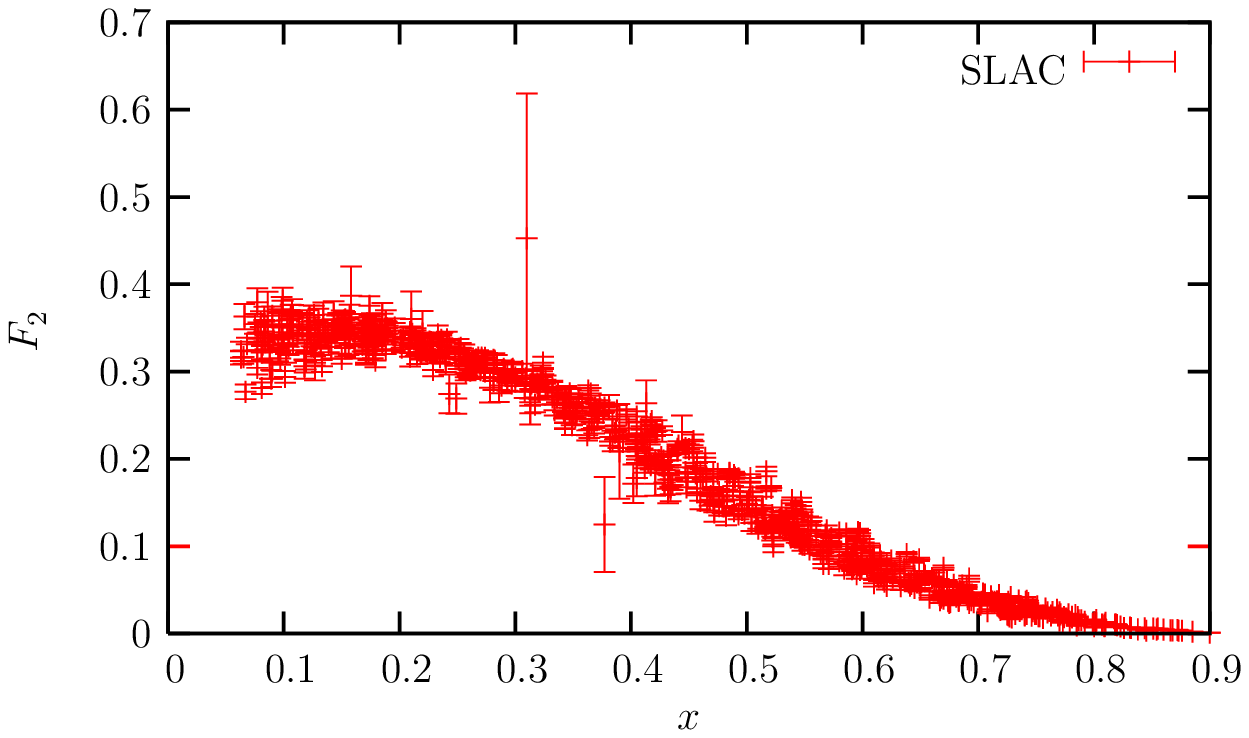}}
\hfill
\resizebox*{5.5cm}{!}{\includegraphics{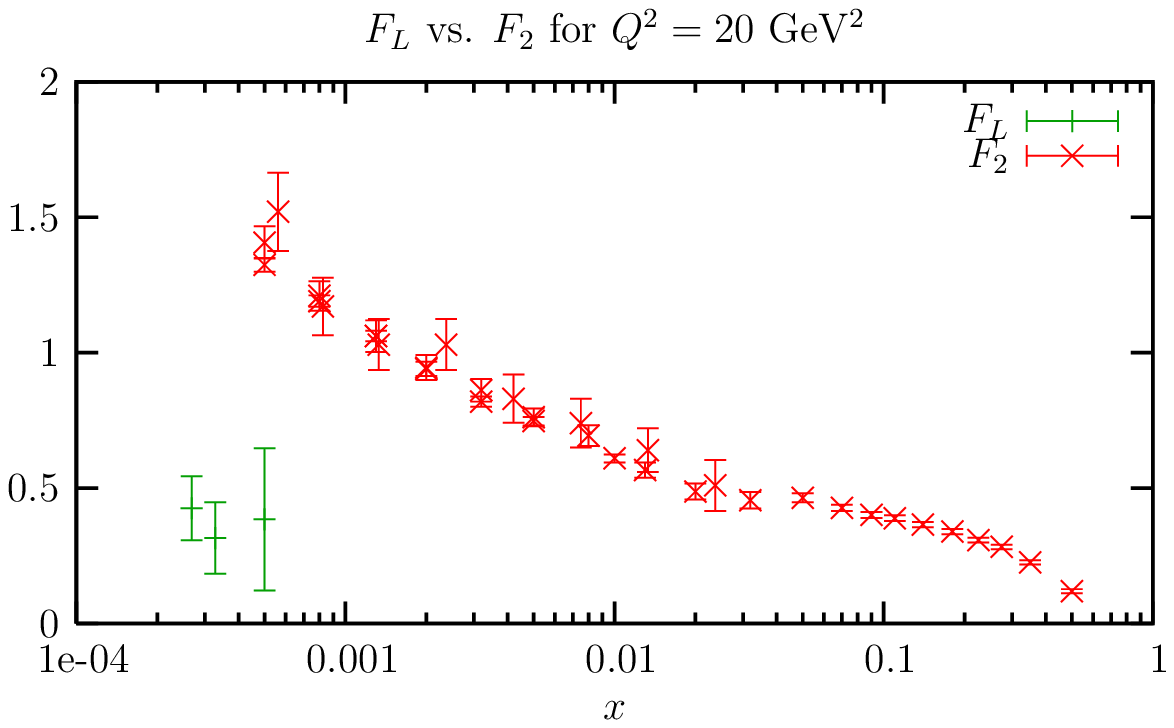}}
\hskip 5mm
\end{center}
\caption{\label{fig:slac}SLAC results on DIS.}
\end{figure}
The left plot of figure \ref{fig:slac} shows the measured values of
$F_2(x,Q^2)$ as a function of $x$. Even though the data covers a
significant range in $Q^2$, all the data points seem to line up on a
single curve, indicating that $F_2$ depends very little on $Q^2$ in
this regime. This property is now known as {\sl Bjorken scaling}
\cite{bjorken}. In the right plot of figure \ref{fig:slac}, one
sees a comparison of $F_2$ with the combination\footnote{$F_{_L}$, the
longitudinal structure function, describes the inclusive cross-section
between the proton and a longitudinally polarized proton.}
$F_{_L}\equiv F_2-2x\,F_1$. Although there are few data points for
$F_{_L}$, one can see that it is significantly lower than $F_2$ and
close to zero~\footnote{From current algebra, it was predicted that
$F_2=2xF_1$; this relation is known as the {\sl Callan-Gross relation}
\cite{CallaG1}.}. As we shall see shortly, these two experimental
facts already tell us a lot about the internal structure of the
proton.

\subsection{Naive parton model}
In order to get a first insight into the inner structure of the
proton, it is interesting to compare the DIS cross-section in
eq.~(\ref{eq:DIS1}) and the $e^-\mu^-$ cross-section (also expressed
in the rest frame of the muon),
\begin{equation}
\frac{d{\colord\sigma}_{{\colorc e^-\mu^-}}}{d{\colora E^\prime}d{\colora\Omega}}
=
\frac{{\colord\alpha_{\rm em}^2}{\colorb\delta(1-x)}}{4{\colorc m_\mu} {\colora E^2} \sin^4\frac{\colora\theta}{2}}
\left[
\sin^2\frac{\colora\theta}{2}
+
\frac{\colorc m_\mu^2}{\nu}\cos^2\frac{\colora\theta}{2}
\right]\; .
\label{eq:el-mu}
\end{equation}
Note that, since this reaction is elastic, the corresponding $x$
variable is equal to $1$, hence the delta function in the
prefactor. The comparison of this formula with eq.~(\ref{eq:DIS1}),
and in particular its angular dependence, is suggestive of the proton
being composed of {\sl point like fermions} -- named {\sl partons} by
Feynman -- off which the virtual photon scatters. If the constituent
struck by the photon carries the momentum $p_c$, this comparison
suggests that
\begin{equation}
2\,F_1\sim F_2\sim \delta(1-x_c)\qquad\mbox{with}\quad
x_c\equiv\frac{Q^2}{2q\cdot p_c}\; .
\end{equation}
Assuming that this parton carries the fraction $x_{_F}$ of the
momentum of the proton, i.e. $p_c=x_{_F}P$, the relation between the
variables $x$ and $x_c$ is $x_c=x/x_{_F}$. Therefore, we get~:
\begin{equation}
2\,F_1\sim F_2\sim x_{_F}\delta(x-x_{_F})\; .
\end{equation}
In other words, the kinematical variable $x$ measured from the
scattering angle of the electron would be equal to the fraction of
momentum carried by the struck constituent. Note that Bjorken
scaling appears quite naturally in this picture.

Having gained intuition into what may constitute a proton, we shall 
 now compute the hadronic tensor $W^{\mu\nu}$ for the DIS reaction
on a free fermion $i$ carrying the fraction $x_{_F}$ of the proton
momentum. Because we ignore interactions for the time
being, this calculation (in contrast to that for a proton target) can be done in closed form. We obtain, 
\begin{eqnarray}
4\pi {\colorb W_i^{\mu\nu}}&\equiv&
{\colorh{\int\frac{d^4p^\prime}{(2\pi)^4}}}
2\pi\delta({\colorh{p^{\prime2}}})
{\colorh{(2\pi)^4\delta(x_{_F}P+q-p^\prime)}}
\nonumber\\
&&\qquad\times
\left<\big<{\colorc x_{_F}P}\big|J^{\mu\dagger}(0)\big|
{\colorh{p^\prime}}\big> 
\big<{\colorh{p^\prime}}\big|
J^\nu(0)\big|{\colorc x_{_F}P}\big>\right>_{\rm spin}
\nonumber\\
&=&2\pi{\colorh{x_{_F}}}\delta(x-x_{_F})\,
e_i^2\;{\colorh{
\left[\!-\!\left(g^{\mu\nu}\!-\!\frac{q^\mu q^\nu}{q^2}\right)
\!\!+\!\frac{2x_{_F}}{P\cdot q}
\left(\!P^\mu\!-\!q^\mu\frac{P\cdot q}{q^2}\right)\!\!\!
\left(\!P^\nu\!-\!q^\nu\frac{P\cdot q}{q^2}\right)\!\!\right]}}\; ,
\nonumber
\end{eqnarray}
where $e_i$ is the electric charge of the parton under
consideration. Let us now assume that in a proton there are
$f_i(x_{_F})dx_{_F}$ partons of type $i$ with a momentum fraction
between $x_{_F}$ and $x_{_F}+dx_{_F}$, and that the photon scatters
incoherently off each of them. We would thus have
\begin{equation}
{\colorb W^{\mu\nu}}
=\sum_i\int_0^1\frac{\colord dx_{_F}}{\colord x_{_F}}\;
{\colorb f_i(x_{_F})}\;{\colorb W_i^{\mu\nu}}\; .
\end{equation}
(The factor $x_{_F}$ in the denominator is a ``flux factor''.) At this
point, we can simply read the values of $F_{1,2}$,
\begin{equation}
{\colorb F_1}=\frac{1}{2}\sum_i e_i^2 {\colorb f_i(x)}\quad,\quad 
{\colorb F_2}=2\,x{\colorb F_1}\; .
\end{equation}
We thus  see that the two experimental observations of i) Bjorken scaling and ii) the Callan-Gross relation 
are automatically realized in this naive picture of the proton\footnote{In
particular, $F_{_L}=0$ in this model is intimately
related to the spin $1/2$ structure of the scattered partons. Scalar partons, for instance, would give $F_1=0$, at variance with experimental results.}.

Despite its success, this model is quite puzzling, because it
assumes that  partons are free inside the proton -- while the
rather large mass of the proton suggests a strong binding of these
constituents inside the proton. Our task for the rest of this lecture
is to study DIS in a quantum field theory of strong interactions,
thereby turning the naive parton model into a systematic description
of hadronic reactions. Before we proceed further, let us describe in
qualitative terms (see \cite{partons} for instance) what a
proton constituted of fermionic constituents bound by interactions
involving the exchange of gauge bosons may look like.
\begin{figure}[htbp]
\begin{center}
\hskip 0mm
\resizebox*{!}{17.5mm}{\includegraphics{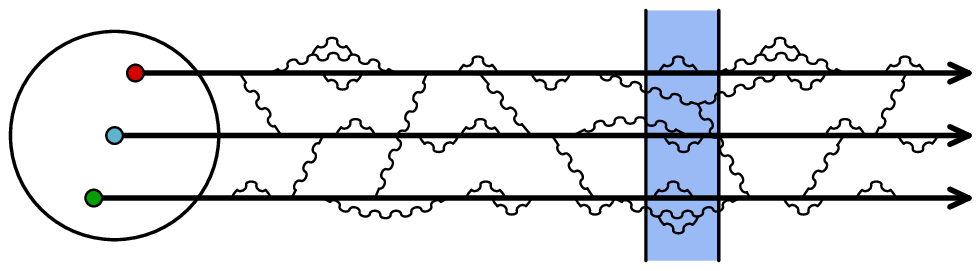}}
\hfill
\resizebox*{!}{17.5mm}{\includegraphics{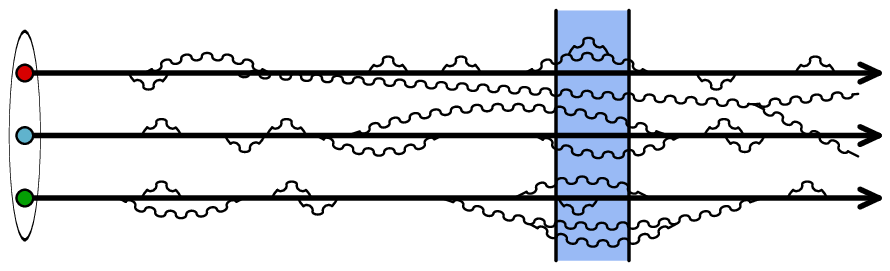}}
\hskip 0mm
\end{center}
\caption{\label{fig:proton}Cartoons of the valence partons of a
proton, and their interactions and fluctuations. Left: proton at low
energy. Right: proton at high energy.}
\end{figure}
In the left panel of figure \ref{fig:proton} are represented the three
valence partons (quarks) of the proton. These quarks interact by gluon
exchanges, and can also fluctuate into states that contain additional
gluons (and also quark-antiquark pairs). These fluctuations can exist
at any space-time scale smaller than the proton size ($\sim$~1~fermi).
(In this picture, one should think of the horizontal axis as the time
axis.) When one probes the proton in a scattering experiment, the
probe (e.g. the virtual photon in DIS) is characterized by certain
resolutions in time and in transverse coordinate. The shaded area in
the picture is meant to represent the time resolution of the probe~:
any fluctuation which is shorter lived than this resolution cannot be
seen by the probe, because it appears and dies out too quickly.

In the right panel of figure \ref{fig:proton}, the same proton is
represented after a boost, while the probe has not changed. The main
difference is that all the internal time scales are Lorentz
dilated. As a consequence, the interactions among
the quarks now take place over times much larger than the resolution
of the probe. The probe therefore sees only free constituents.
Moreover, this time dilation allows more fluctuations to be resolved by the
probe; thus, a high energy proton appears to contain more gluons than
a proton at low energy\footnote{Equivalently, if the energy of the
proton is fixed, there are more gluons at lower values of the momentum
fraction $x_{_F}$.}.

\subsection{Bjorken scaling from free field theory}
We will now derive Bjorken scaling and the Callan-Gross
relation from quantum field theory. We will consider a theory involving
fermions (quarks) and bosons (gluons), but shall at first consider the free field theory 
limit by neglecting all their interactions. We will consider a kinematical regime in DIS that involves a large value of the
momentum transfer $Q^2$ and of the center of mass energy $\sqrt{s}$
of the collision, while the value of $x$ is kept constant. This limit
is known as the Bjorken limit. 

\begin{figure}[htbp]
\begin{center}
\resizebox*{5cm}{!}{\includegraphics{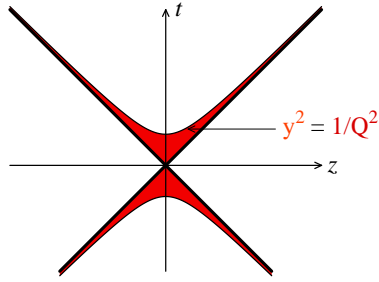}}
\end{center}
\caption{\label{fig:OPE}Region of $y^\mu$ that dominates in the Bjorken
limit.}
\end{figure}

To appreciate strong interaction physics in the Bjorken limit, consider a frame in which the 4-momentum of
the photon can be written as
\begin{equation}
q^\mu=\frac{1}{\colorc m_{_N}}({\colorb\nu}
,0,0,\sqrt{{\colorb\nu^2}+{\colorc m_{_N}^2}{\colorb Q^2}})\; .
\end{equation}
From the combinations of the components of $q^\mu$
\begin{eqnarray}
&& 
{\colord q^+}\equiv \frac{q^0+q^3}{\sqrt{2}}\sim 
\frac{\colorb\nu}{\colorc m_{_N}}\to +\infty
\nonumber\\
&&
{\colord q^-}\equiv \frac{q^0-q^3}{\sqrt{2}}\sim 
{\colorc m_{_N}}{\colorb x} \to \mbox{constant}\; ,
\end{eqnarray}
and because $q\cdot y = q^+y^-+q^-y^+-\vec\q_\perp\cdot\vec\y_\perp$, the
integration over $y^\mu$ in $W^{\mu\nu}$ is dominated by
\begin{eqnarray}
&&{\colord y^-}\sim \frac{\colorc m_{_N}}{\colorb\nu}\to 0
\quad,\quad {\colord y^+}\sim ({\colorc m_{_N}}{\colorb x})^{-1}\; .
\end{eqnarray}
Therefore, the invariant separation between the points at which the
two currents are evaluated is ${\colord y^2} \le 2{\colord y^+y^-}
\sim 1/{\colorb Q^2} \to 0$. Noting that in eq.~(\ref{eq:W1}) the
product of the two currents can be replaced by their commutator, and
recalling that expectation values of commutators vanish for space-like
separations, we also see that $y^2\ge 0$. Thus, the Bjorken limit
corresponds to a time-like separation between the two currents, with
the invariant separation $y^2$ going to zero, as illustrated in figure
\ref{fig:OPE}.
It is important to note that in this limit, although the invariant
$y^2$ goes to zero, the components of $y^\mu$ do not necessarily
become small. This will have important ramifications when we apply
the Operator Product Expansion to $W^{\mu\nu}$.

For our forthcoming discussion, consider the forward Compton amplitude
$T^{\mu\nu}$ 
\begin{equation}
4\pi {\colorb T_{\mu\nu}}\equiv i\int d^4y e^{iq\cdot y} 
\left<
\big<{\colorc N(P)}\big|
{\colorb T}(J_\mu^\dagger(y) J_\nu(0))
\big|{\colorc N(P)}\big>
\right>_{\rm spin}\; .
\end{equation}
\begin{figure}[htbp]
\begin{center}
\resizebox*{3cm}{!}{\includegraphics{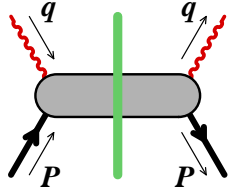}}
\end{center}
\caption{\label{fig:compton}Forward Compton amplitude. We have also
represented a cut contributing to $W^{\mu\nu}$.}
\end{figure}
It differs from $W^{\mu\nu}$ by the fact that the two
currents are time-ordered, and as
illustrated in figure \ref{fig:compton},  one can recover
$W^{\mu\nu}$ from its imaginary part, 
\begin{equation}
{\colorb W_{\mu\nu}}=2 \,{\rm
Im}\,{\colorb T_{\mu\nu}}\; .
\end{equation}
At fixed $Q^2$, $T^{\mu\nu}$ is analytic in the variable $\nu$, except
for two cuts on the real axis that start at $\nu=\pm Q^2/2$. The cut
at positive $\nu$ corresponds to the threshold $(P+q)^2\ge m_{_N}^2$
above which the DIS reaction becomes possible, and the cut at negative
$\nu$ can be inferred from the fact that $T^{\mu\nu}$ is unchanged
under the exchange $(\mu\leftrightarrow\nu,q\leftrightarrow-q)$. It is
also possible to decompose the tensor $T^{\mu\nu}$ in terms of two
structure functions $T_{1,2}$~:
\begin{equation}
{\colorb T_{\mu\nu}}=
-{\colorb T_1}\left(g_{\mu\nu}-\frac{q_\mu q_\nu}{q^2}\right)
+\frac{\colorb T_2}{P\cdot q}
\left(P_\mu-q_\mu\frac{P\cdot q}{q^2}\right)
\left(P_\nu-q_\nu\frac{P\cdot q}{q^2}\right)\; ,
\end{equation}
and the DIS structure functions $F_{1,2}$ can be expressed in terms of
the discontinuity of $T_{1,2}$ across the cuts.

We now remind the reader of some basic results about the {\sl
Operator Product Expansion} (OPE) \cite{Wilso1,PeskiS1}. Consider a
correlator $\big<{\colorb{\cal A}}(0){\colorb{\cal
B}}(y)\phi(x_1)\cdots\phi(x_n)\big>$, where ${\colorb{\cal A}}$ and
${\colorb{\cal B}}$ are two local operators (possibly composite) and
the $\phi$'s are unspecified field operators. In the limit $y^\mu\to
0$, this object is usually singular, because products of operators
evaluated at the same point are ill-defined. The OPE states that the
nature of these singularities is a property of the operators
${\colorb{\cal A}}$ and ${\colorb{\cal B}}$, and is not influenced by
the nature and localization of the $\phi(x_i)$'s. This singular
behavior can be expressed as
\begin{equation}
{\colorb{\cal A}}(0){\colorb{\cal B}}(y)\empile{=}\over{y^\mu\to 0}
\sum_i {\colorc C_i}(y)\;{\colord{\cal O}_i}(0)\; ,
\end{equation}
where the $C_i(y)$ are numbers (known as the {\sl Wilson
coefficients}) that contain the singular $y^\mu$ dependence and the
${\colord{\cal O}_i}(0)$ are local operators that have the same
quantum numbers as the product ${\cal A}{\cal B}$. This expansion --
known as the OPE -- can then be used to obtain the limit $y^\mu\to 0$
of any correlator containing the product ${\colorb{\cal
A}}(0){\colorb{\cal B}}(y)$. If $\colorb{\rm d}({\cal O}_i), {\rm
d}({\cal A})$,and ${\rm d}({\cal B})$ are the respective mass
dimensions of the operators ${\cal O}_i, {\cal A}$ and ${\cal B}$, a
simple dimensional argument tells us that
\begin{equation}
{\colorc C_i}(y)\empile{\sim}\over{y^\mu\to 0}
|y|^{\colorb{\rm d}({\cal O}_i)-{\rm d}({\cal A})-{\rm d}({\cal B})}\qquad\mbox{(up\ to\ logarithms)}\; .
\end{equation}
(Here $|y|=\sqrt{y_\mu y^\mu}$.) From this relation, we see that the
operators ${\cal O}_i$ having the lowest dimension lead to the most
singular behavior in the limit $y^\mu\to 0$. Thus, only a small number
of operators are relevant in the analysis of this limit and one can
ignore the higher dimensional operators.

Things are however a bit more complicated in the case of DIS, because
only the invariant $y^2$ goes to zero, while the components $y^\mu$ do
not go to zero. The local operators that may appear in the OPE of
${\colorb T(J_\mu^\dagger(y)J_\nu(0))}$ can be classified according to
the representation of the Lorentz group to which they belong. Let us
denote them ${\colord{\cal O}}^{{\colora\mu_1\cdots\mu_s}}_{{\colora
s},{\colorc i}}$, where ${\colora s}$ is the ``spin'' of the operator
(the number of Lorentz indices it carries), and the index ${\colorc
i}$ labels the various operators having the same Lorentz
structure. The OPE can be written as~:
\begin{equation}
\sum_{\colorc s,i}
C^{{\colora s},{\colorc i}}_{\colora\mu_1\cdots\mu_s}(y)\;
{\colord{\cal O}}^{{\colora\mu_1\cdots\mu_s}}_{{\colora s},{\colorc i}}(0)\; .
\end{equation}
Because they depend only on the 4-vector $y^\mu$, the Wilson
coefficients must be of the form\footnote{There could also be terms
where one or more pairs $y^{\mu_i}y^{\mu_j}$ are replaced by
$y^2\,g^{\mu_i\mu_j}$, but such terms are less singulars in the
Bjorken limit.}
\begin{equation}
C^{{\colora s},{\colorc i}}_{\colora\mu_1\cdots\mu_s}(y)
\equiv
y_{\colora\mu_1}\cdots y_{\colora\mu_s}\;C_{{\colora s},{\colorc i}}(y^2)\; ,
\end{equation}
where $C_{{\colora s},{\colorc i}}(y^2)$ depends only on the invariant
$y^2$. Similarly, the expectation value of the operators
${\colord{\cal O}}^{{\colora\mu_1\cdots\mu_s}}_{{\colora s}}$ in the
proton state can only depend on the proton momentum $P^\mu$, and the
leading part in the Bjorken limit is\footnote{Here also, there could
be terms where a pair $P^{\mu_i}P^{\mu_j}$ is replaced by
$m_{_N}^2\,g^{\mu_i\mu_j}$, but they too lead to subleading contributions
in the Bjorken limit.}
\begin{equation}
\Big<\big<{\colorc N(P)}\big|
{\colord{\cal O}}^{{\colora\mu_1\cdots\mu_s}}_{{\colora s},{\colorc i}}(0)
\big|{\colorc N(P)}\big>\Big>_{\rm spin}
\!\!={\colorc P}^{\colora\mu_1}\cdots {\colorc P}^{\colora\mu_s}\;
\big<{\colord{\cal O}}_{{\colora s},{\colorc i}}\big>\; ,
\end{equation}
where the $\big<{\colord{\cal O}}_{{\colora s},{\colorc i}}\big>$ are
some non-perturbative matrix elements.

Let us now denote by ${d}_{{\colora s},{\colorc i}}$ the mass
dimension of the operator ${\colord{\cal
O}}^{{\colora\mu_1\cdots\mu_s}}_{{\colora s},{\colorc i}}$. Then, the
dimension of $C_{{\colora s},{\colorc i}}(y^2)$ is $6+{\colora
s}-{d}_{{\colora s},{\colorc i}}$, which means that it scales like
\begin{equation}
C_{{\colora s},{\colorc i}}(y^2)
\empile{\sim}\over{y^2\to 0}
{\colorb(y^2)^{({\bs
d}_{{\colora s},{\colorc i}}-{\colora s}-6)/2}}\; .
\end{equation}
Because the individual components of $y^\mu$ do not go to zero, it is
this scaling alone that determines the behavior of the hadronic tensor
in the Bjorken limit. Contrary to the standard OPE, the scaling
depends on the difference between the dimension of the operator and
its spin, called its {\sl twist} $t_{{\colora s},{\colorc i}}\equiv
{\colorb {d}_{{\colora s},{\colorc i}}-s}$, rather than its dimension
alone. The Bjorken limit of DIS is dominated by the operators that
have the lowest possible twist. As we shall see, there is an infinity
of these lowest twist operators, because the dimension can be
compensated by the spin of the operator. If we go back to the
structure functions $T_{1,2}$, we can write
\begin{eqnarray}
{\colorb T_r}(x,Q^2)=\sum_{s}x^{a_r-s}\sum_i \big<{\colord {\cal O}}_{{\colora s},{\colorc i}}\big>\;
{D}_{r;{\colora s},{\colorc i}}(Q^2)\qquad (r=1,2)
\; ,
\label{eq:OPE2}
\end{eqnarray}
where $a_1=0$ and $a_2=1$.  The difference by one power of $x$ (at
fixed $Q^2$) between $T_1$ and $T_2$ comes from their respective
definitions from $T^{\mu\nu}$ that differ by one power of the proton
momentum $P$. Eq.~(\ref{eq:OPE2}) gives the structure functions
$T_{1,2}$ as a series of terms, each of which has factorized $x$ and
$Q^2$ dependences. (The functions ${D}_{r;{\colora s},{\colorc i}}$
($r=1,2$) are related to the Fourier transform of $C_{{\colora
s},{\colorc i}}(y^2)$, and thus can only depend on the invariant
$Q^2$). Moreover, for dimensional reasons, the functions
${D}_{r;{\colora s},{\colorc i}}$ must scale like ${\colorb
Q^{2+s-{\bs d}_{{\colora s},{\colorc i}}}}$. {\sl Therefore, it follows that 
Bjorken scaling arises from twist 2 operators}. It is important to
keep in mind that in eq.~(\ref{eq:OPE2}), the functions
${D}_{r;{\colora s},{\colorc i}}$ are in principle calculable in
perturbation theory and do not depend on the nature of the target,
while the $\big<{\colord {\cal O}}_{{\colora s},{\colorc i}}\big>$'s
are non perturbative matrix elements that depend on the target. Thus, the
OPE approach in our present implementation cannot provide quantitative
results beyond simple scaling laws.

\begin{figure}[htbp]
\begin{center}
\resizebox*{6cm}{!}{\includegraphics{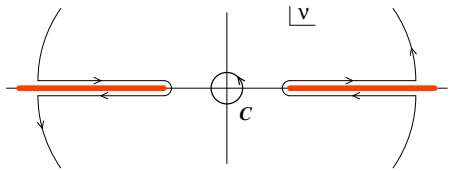}}
\end{center}
\caption{\label{fig:contour}Contour in the complex $\nu$ plane, and
its deformation to pick up the contribution of the cuts.}
\end{figure}

It is easy to check that $T_1$ is even in $x$ while $T_2$ is odd; this
means that only even values of the spin $s$ can appear in the sum in
eq.~(\ref{eq:OPE2}). We shall now rewrite this equation in a more
compact form to see what it tells us about the structure
functions $F_{1,2}$. Writing 
\begin{equation}
{\colorb T_r}=\sum_{{\rm even\ }{\colora s}} {\colorb t_r}({\colora s},Q^2)\;{\colord
x^{a_r-{\colora s}}}=\sum_{{\rm even\ }{\colora s}} {\colorb t_r}({\colora s},Q^2)\;
\left(\frac{2}{Q^2}\right)^{{\colora s}-a_r}\; \nu^{{\colora s}-a_r}\; ,
\end{equation}
we get (for ${\colora s}$ even)
\begin{equation}
{\colorb t_r}({\colora s},Q^2)
=
\frac{1}{2\pi i}\left(\frac{Q^2}{2}\right)^{{\colora s}-a_r}
\int_{\cal C} \frac{d\nu}{\nu}\;
 \nu^{a_r-{\colora s}} \; {\colora T_r}(\nu,Q^2)
\; ,
\end{equation}
where ${\cal C}$ is a small circle around the origin in the complex
$\nu$ plane (see figure \ref{fig:contour}).
This contour can then be deformed and wrapped around the cuts along
the real axis, as illustrated in the figure \ref{fig:contour}. Because
the structure function $F_r$ is the discontinuity of $T_r$ across the
cut, we can write
\begin{equation}
{\colorb t_r}({\colora s},Q^2)=\frac{2}{\pi}\int_0^1
\frac{dx}{x}\;x^{{\colora s}-a_r}\;
{\colorb F_r}(x,Q^2)\; .
\end{equation}
Therefore, we see that {\sl the OPE gives the $x$-moments of the DIS
structure functions}.

In order to go further and calculate the perturbative Wilson
coefficients ${D}_{r;{\colora s},{\colorc i}}$, we must now identify
the twist 2 operators that may contribute to DIS. In a theory of
fermions and gauge bosons, we can construct two kinds of twist 2
operators~:
\begin{eqnarray}
&&
{\colord{\colord {\cal O}}}_{{\colora s},f}^{\colora\mu_1\cdots\mu_s}\equiv
\overline{\psi}_f \gamma^{\{{\colora\mu_1}}
\partial^{\colora\mu_2}
\cdots
\partial^{{\colora\mu_s}\}}\psi_f
\nonumber\\
&&
{\colord{\colord {\cal O}}}_{{\colora s},g}^{\colora\mu_1\cdots\mu_s}\equiv
F_\alpha{}^{\{{\colora\mu_1}}
\partial^{\colora\mu_2}
\cdots
\partial^{\colora\mu_{s-1}}
F^{{\colora\mu_s}\}\alpha}
\; ,
\end{eqnarray}
where the brackets $\{\cdots\}$ denote a symmetrization of the indices
$\colora\mu_1\cdots\mu_s$ and a subtraction of the traced terms on
those indices. To compute the Wilson coefficients, the simplest method
is to exploit the fact that they are independent of the
target. Therefore, we can take as the ``target'' an elementary object,
like a quark or a gluon, for which everything can be computed in
closed form (including the $\big<{\colord {\cal O}}_{{\colora
s},{\colorc i}}\big>$). Consider first a quark state as the target, of
a given flavor $f$ and spin $\sigma$. At lowest order, one has
\begin{eqnarray}
&&
\big<f,\sigma\big|{\colord {\cal O}}_{{\colora s},f^\prime}^{{\colora\mu_1\cdots\mu_s}}\big|f,\sigma\big>
=
\delta_{ff^\prime}
\overline{u}_{\sigma}(P)\gamma^{\{\mu_1}u_{\sigma}(P)
P^{\mu_2}\cdots P^{\mu_s\}}
\nonumber\\
&&
\big<f,\sigma\big|{\colord {\cal O}}_{{\colora s},g}^{{\colora\mu_1\cdots\mu_s}}\big|f,\sigma\big>
=0
\; .
\end{eqnarray}
Averaging over the spin, and comparing with $P^{\mu_1}\cdots
P^{\mu_s}\big<{\colord {\cal O}}_{{\colora s},{\colorc i}}\big>$, we
get
\begin{equation}
\big<{\colord {\cal O}}_{{\colora s},f^\prime}\big>_f=\delta_{ff^\prime}
\quad,\qquad
\big<{\colord {\cal O}}_{{\colora s},g}\big>_f=0\; .
\end{equation}
On the other hand, we have already calculated directly the hadronic
tensor for a single quark. By computing the moments of the
corresponding $F_{1,2}$, we get the ${\colorb t_r}({\colora s},Q^2)$
for $s$ even~:
\begin{equation}
{\colorb t_1}({\colora s},Q^2)=\frac{1}{\pi}\,e_f^2\quad,\qquad
{\colorb t_2}({\colora s},Q^2)=\frac{2}{\pi}\,e_f^2\; .
\end{equation}
From this, the bare Wilson coefficients for the operators
involving quarks are
\begin{equation}
D_{1;{\colora s},f}(Q^2)=\frac{1}{\pi}\,e_f^2\quad,\qquad
D_{2;{\colora s},f}(Q^2)=\frac{2}{\pi}\,e_f^2\; .
\end{equation}
By repeating the same steps with a vector boson state, those
involving only gluons are 
\begin{equation}
D_{1;{\colora s},g}(Q^2)=D_{2;{\colora s},g}(Q^2)=0\; ,
\end{equation}
if the vector bosons are assumed to be electrically neutral.

Going back to a nucleon target, we cannot compute the $\big<{\colord
{\cal O}}_{{\colora s},{\colorc i}}\big>$. However, we can hide
momentarily our ignorance by defining functions $f_f(x)$ and
$f_{\bar{f}}(x)$ (respectively the quark and antiquark distributions)
such that\footnote{DIS with exchange of a photon cannot disentangle
the quarks from the antiquarks. In order to do that, one could scatter
a neutrino off the target, so that the interaction proceeds via a weak
charged current.}
\begin{equation}
\int_0^1\frac{dx}{x}\;x^{{\colora s}}\;\Big[{\colord f_f(x)}+{\colord f_{\bar{f}}(x)}\Big]
=\left<{\colord {\cal O}}_{{\colora s},{\colorc f}}\right>\; .
\end{equation}
(The sum ${\colord f_f(x)}+{\colord f_{\bar{f}}(x)}$ is known as the
{\sl singlet quark distribution of flavor $f$}.) Thus, the OPE formulas for $F_1$ and $F_2$ on a nucleon in terms of these quark
distributions are 
\begin{equation}
{\colorb F_1}(x)=\frac{1}{2}\sum_f e_f^2 \Big[{\colord f_f(x)}+{\colord f_{\bar{f}}(x)}\Big]\quad,\qquad
{\colorb F_2}(x)=2x{\colorb F_1}(x)\; .
\label{eq:F12}
\end{equation}
We see that these formulas have the required properties: (i)
Bjorken scaling and (ii) the Callan-Gross relation.

Despite the fact that the OPE in a free theory of quarks and gluons
leads to a result which is embarrassingly similar to the much simpler
calculation we performed in the naive parton model, this exercise has
taught us several important things~:
\begin{itemize}
\item We can derive an {\colora operator definition of the parton
distributions} ${\colord f_i(x)}$ (albeit it is not calculable
perturbatively)
\item Bjorken scaling can be derived from first principles in a field
theory of free quarks and gluons. This was a puzzle pre-QCD because clearly 
these partons are constituents of a strongly bound state.
\item The puzzle could be resolved if the field theory of strong
interactions became a free theory in the limit $Q^2\to +\infty$, a
property known as {\colorb\sl asymptotic freedom}.
\end{itemize}
As shown by Gross, Politzer and Wilczek in 1973, non-Abelian gauge
theories with a reasonable number of fermionic fields (e.g. QCD with 6
flavors of quarks) are asymptotically free\cite{running_alpha} and
were therefore a natural candidate for being the right theory of the
strong interactions.

\subsection{Scaling violations}
Although it was interesting to see that a free quantum field theory
reproduces the Bjorken scaling, this fact alone does not tell much
about the detailed nature of the strong interactions at the level of
quarks and gluons. Much more interesting are the {\sl violations} of
this scaling that arise from these interactions and it is the detailed
comparison of these to experiments that played a crucial role in
establishing QCD as the theory of the strong interactions.

The effect of interactions can be evaluated perturbatively in the
framework of the OPE, thanks to renormalization group equations. In
the previous discussion, we implicitly assumed that there is no scale
dependence in the moments $\big<{\colord {\cal O}}_{{\colora
s},{\colorc i}}\big>$ of the quark distribution functions. But this is
not entirely true; when interactions are taken into account, {\colora
they depend on a renormalization scale} ${\colorb\mu^2}$. The parton
distributions become scale dependent as well. However, since $F_{1,2}$
are observable quantities that can be extracted from a cross-section,
they cannot depend on any renormalization scale. Thus, there must also
be a $\mu^2$ dependence in the Wilson coefficients, that exactly
compensates the $\mu^2$ dependence originating from the $\big<{\colord
{\cal O}}_{{\colora s},{\colorc i}}\big>$. By dimensional analysis,
the Wilson coefficients have an overall power of $Q^2$ set by their
dimension (see the discussion following eq.~(\ref{eq:OPE2})),
multiplied by a dimensionless function that can only depend on the
ratio $Q^2/\mu^2$. By comparing the {\sl Callan-Symanzik
equations}\cite{PeskiS1} for $T^{\mu\nu}$ with those for the
expectation values $\big<{\colord {\cal O}}_{{\colora s},{\colorc
i}}\big>$, the renormalization group equation\cite{PeskiS1} obeyed by
the Wilson coefficients is
\footnote{We have used the fact that the electromagnetic currents are
conserved and therefore have a vanishing anomalous dimension. Note
also that we have exploited the fact that for twist 2 operators
$D_{r;{\colora s},{\colorc j}}$ depends only on $Q^2/\mu^2$, so that
we can replace $\mu\partial_\mu$ by $-Q\partial_{_Q}$.}
\begin{equation}
	\left[\left(-{\colorb Q\partial_{_Q}}+{\colorc\beta(g)}\partial_g\right)\delta_{ij}
	  -{\colorc\gamma_{{\colora s},ji}(g)}\right]D_{r;{\colora s},{\colorc j}}({\colorb Q/\mu},g)=0\; ,
\end{equation}
where $\beta(g)$ is the beta function, and ${\colorc\gamma_{{\colora
s},ji}(g)}$ is the matrix of anomalous dimensions for the operators of
spin $s$ (it is not diagonal because operators with identical quantum
numbers can mix through renormalization).

In order to solve these equations, let us first introduce the {\sl
running coupling} ${\colord\overline{g}}({\colorb Q},g)$ such that
\begin{equation}
\ln({\colorb Q}/Q_0)=
\int_g^{{\colord\overline{g}}({\colorb Q},g)}\frac{dg^\prime}
{\colorc\beta(g^\prime)}\; .
\end{equation}
Note that this is equivalent to ${\colorb
Q\partial_{_Q}}{\colord\overline{g}}({\colorb
Q},g)={\colorc\beta}({\colord\overline{g}}({\colorb Q},g))$ and
${\colord\overline{g}}(Q_0,g)=g$; in other words,
${\colord\overline{g}}({\colorb Q},g)$ is the value at the scale $Q$
of the coupling whose value at the scale $Q_0$ is $g$. The usefulness
of the running coupling stems from the fact that any function that
depends on $Q$ and $g$ only through the combination
${\colord\overline{g}}({\colorb Q},g)$ obeys the equation
\begin{equation}
\left[-{\colorb
Q\partial_{_Q}}+{\colorc\beta(g)}\partial_g\right]F({\colord\overline{g}}({\colorb Q},g))=0\; .
\end{equation}
It is convenient to express the Wilson coefficients at the
scale $Q$ from those at the scale $Q_0$ as 
\begin{equation}
D_{r;{\colora s},{\colorc i}}({\colorb Q/\mu},g) = D_{r;{\colora
s},{\colorc j}}(Q_0/{\colorb\mu},{\colord\overline{g}}({\colorb Q},g))
\left[e^{-\int_{Q_0}^{\colorb Q}\frac{dM}{M}{\colorc \gamma_{\colora s}}({\colord\overline{g}}(M,g))}\right]_{\colorc ji}\; .
\end{equation}
In QCD, which is asymptotically free, we can approximate
the anomalous dimensions and running coupling at one loop by
\begin{equation}
{\colorc\gamma}_{{\colora s},{\colorc ij}}({\colord\overline{g}})={\colord\overline{g}^2} A_{ij}(s)\quad,\qquad 
{\colord\overline{g}^2}({\colorb Q},g)=\frac{8\pi^2}{{\colorc\beta_0}\ln({\colorb Q/\Lambda_{_{QCD}}})}\; .
\end{equation}
(The $A_{ij}(s)$ are obtained from a 1-loop perturbative calculation.)
In this case, the scale dependence of the Wilson coefficients can be
expressed in closed form as 
\begin{equation}
D_{r;{\colora s},{\colorc i}}({\colorb Q/\mu},g)
=
D_{r;{\colora s},{\colorc j}}(Q_0/{\colorb \mu},{\colord\overline{g}}({\colorb Q},g))
\left[\left(
\frac{\ln({\colorb Q/\Lambda_{_{QCD}}})}{\ln(Q_0/{\colorb\Lambda_{_{QCD}}})}
\right)^{-\frac{8\pi^2}{{\colorc\beta_0}}A(s)}\right]_{\colorc ji}\; .
\end{equation}
From this formula, we can write the moments of the structure functions,
\begin{equation}
\int_0^1\frac{dx}{x}\,x^{\colora s}\;{\colorb F_1}(x,{\colorb Q^2})
=\sum_{{\colorc i},f}\frac{e_f^2}{2} \left[\left(
\frac{\ln({\colorb Q/\Lambda_{_{QCD}}})}{\ln(Q_0/{\colorb\Lambda_{_{QCD}}})}
\right)^{-\frac{8\pi^2}{{\colorc\beta_0}}A(s)}\right]_{f{\colorc i}} \left<{\colord {\cal O}}_{{\colora s},{\colorc i}}\right>_{_{Q_0}}\; ,
\end{equation}
(and a similar formula for $F_2$). We see that we can preserve the
relationship between $F_1$ and the quark distributions,
eq.~(\ref{eq:F12}), provided that we let the quark distributions
become scale dependent in such a way that their moments read
\begin{equation}
\int_0^1
\frac{dx}{x}\,x^{\colora s}\,
\Big[
{\colord f_f}(x,{\colorb
Q^2})
+
{\colord f_{\bar{f}}}(x,{\colorb
Q^2})
\Big]
\equiv \sum_{\colorc i} \left[\left(
\frac{\ln({\colorb Q/\Lambda_{_{QCD}}})}{\ln(Q_0/{\colorb\Lambda_{_{QCD}}})}
\right)^{-\frac{8\pi^2}{{\colorc\beta_0}}A(s)}\right]_{f{\colorc i}} \!\!\!
\left<{\colord {\cal
O}}_{{\colora s},{\colorc i}}\right>_{_{Q_0}}\; .
\end{equation}
By also calculating the scale dependence of $F_2$, one could verify
that the Callan-Gross relation ${\colorb F_2}(x,{\colorb
Q^2})=2x{\colorb F_1}(x,{\colorb Q^2})$ is preserved at the 1-loop
order. It is crucial to note that, although we do not know how to
compute the expectation values $\left<{\colord {\cal O}}_{{\colora
s},{\colorc i}}\right>_{_{Q_0}}$ at the starting scale $Q_0$, QCD
predicts how the quark distribution varies when one changes the scale
$Q$. We also see that, in addition to a dependence on $Q^2$, the
singlet quark distribution now depends on the expectation value of
operators that involve only gluons (when the index $i=g$ in the
previous formula).

The scale dependence of the parton distributions can also be
reformulated in the more familiar form of the {\sl DGLAP
equations}. In order to do this, one should also introduce a {\sl
gluon distribution} $f_g$, also defined by its moments,
\begin{equation}
\int_0^1\frac{dx}{x}\,x^{\colora s}\,{\colord f_g}(x,{\colorb Q^2})\equiv
\sum_{\colorc i}
\left[\left(
\frac{\ln({\colorb Q/\Lambda_{_{QCD}}})}{\ln(Q_0/{\colorb\Lambda_{_{QCD}}})}
\right)^{-\frac{8\pi^2}{{\colorc\beta_0}}A(s)}\right]_{g{\colorc i}} \left<{\colord {\cal O}}_{{\colora s},{\colorc i}}\right>_{_{Q_0}}\; .
\end{equation}
Then one can check that the derivatives of the moments of the parton
distributions with respect to the scale $Q^2$ are given by
\begin{equation}
{\colorb Q^2}\frac{{\colorb\partial} {\colord {\bs f}_i}({\colora s},{\colorb Q^2})}{{\colorb\partial Q^2}}
=
-\frac{{\colord\overline{g}^2}({\colorb Q},g)}{2}
A_{ji}(s) {\colord {\bs f}_j}({\colora s},{\colorb Q^2})\; ,
\label{eq:moments}
\end{equation}
where we have used the shorthands ${\bs f}_f\equiv
f_f+f_{\bar{f}}\;,\; {\bs f}_g\equiv f_g$. In order to turn this
equation into an equation for the parton distributions themselves, one
can use
\begin{equation}
A({\colora s}){\bs f}({\colora s})=\int_0^1\frac{dx}{x}\,x^{\colora s}
\int_x^1 \frac{dy}{y}\,A({\colorc x/y}){\bs f}({\colorc y})\; ,
\end{equation}
that relates the product of the moments of two functions to the moment
of a particular convolution of these functions. Using this result, and
defining {\sl splitting function} $P_{ij}$ from their moments,
\begin{equation}
\int_0^1\frac{dx}{x}\,x^{\colora s}\,{\colorc P_{ij}(x)}
\equiv -4\pi^2 A_{\colorc ij}({\colora s})\; ,
\end{equation}
it is easy to derive the DGLAP
equation\cite{dglap},
\begin{equation}
{\colorb Q^2}\frac{{\colorb\partial} {\colord {\bs f}_i}(x,{\colorb Q^2})}{{\colorb\partial Q^2}}
=
\frac{{\colord\overline{g}^2}({\colorb Q},g)}{8\pi^2}\int_x^1 \frac{dy}{y}\, 
{\colorc P_{ji}(x/y)} {\colord {\bs f}_j}(y,{\colorb Q^2})\; ,
\label{eq:dglap}
\end{equation}
that resums powers of $\alpha_s\log(Q^2/Q_0^2)$.  This equation for
the parton distributions has a probabilistic interpretation~: the
splitting function $\overline{g}^2 P_{ji}(z)\ln(Q^2)$ can be seen as
the probability that a parton $j$ splits into two partons separated by
at least $Q^{-1}$ (so that a process with a transverse scale $Q$ will
see two partons), one of them being a parton $i$ that carries the
fraction $z$ of the momentum of the original parton.

At 1-loop, the coefficients $A_{ij}({\colora s})$ in the anomalous
dimensions are
\begin{eqnarray}
&&
A_{gg}({\colora s})\!=\!\frac{1}{2\pi^2}\left\{
3\left[\frac{1}{12}\!-\!\frac{1}{{\colora s}({\colora s}\!-\!1)}
\!-\!\frac{1}{({\colora s}\!+\!1)({\colora s}\!+\!2)}\!+\!\sum_{j=2}^{{\colora s}}\frac{1}{j}\right]
\!\!+\!\frac{N_f}{6}\right\}
\nonumber\\
&&
A_{gf}({\colora s})=-\frac{1}{4\pi^2}\left\{\frac{1}{{\colora s}+2}+\frac{2}{{\colora s}({\colora s}+1)({\colora s}+2)}\right\}
\nonumber\\
&&
A_{fg}({\colora s})=-\frac{1}{3\pi^2}\left\{
\frac{1}{{\colora s}+1}+\frac{2}{{\colora s}({\colora s}-1)}
\right\}
\nonumber\\
&&
A_{ff^\prime}({\colora s})=\frac{1}{6\pi^2}\left\{
1-\frac{2}{{\colora s}({\colora s}+1)}+4\sum_{j=2}^{{\colora s}}\frac{1}{j}
\right\}\delta_{ff^\prime}\; ,
\label{eq:anom}
\end{eqnarray}
where $N_f$ is the number of flavors of quarks. On can note that,
since $A_{gf}(s)$ is flavor independent, the {\colora
non-singlet\footnote{Here, the word ``singlet'' refers to the flavor
of the quarks.} linear combinations} ($\sum_f a_f {\colord{\cal
O}}_{{\colora s},f}$ with $\sum_f a_f=0$) are eigenvectors of the
matrix of anomalous dimensions, with an eigenvalue $A_{ff}({\colora
s})$. These linear combinations {\colora do not mix with the remaining
two operators}, ~~~$\sum_f{\colord{\cal O}}_{{\colora s},f}$ and
${\colord {\cal O}}_{{\colora s},g}$, through renormalization. By
examining these anomalous dimensions for $s=1$, we can see that the
eigenvalue for the non-singlet quark operators is vanishing~:
$A_{ff}({\colora s}=1)=0$. Going back to the eq.~(\ref{eq:moments}),
this implies that
\begin{equation}
{\colorb\frac{\partial}{\partial Q^2}} \left\{
\int_0^1dx \sum_f a_f \Big[
{\colord {f}_f}(x,{\colorb Q^2})
+
{\colord {f}_{\bar{f}}}(x,{\colorb Q^2})
\Big]
\right\}=0
\end{equation}
for any linear combination such that $\sum_f a_f=0$. This relation
implies for instance that the number of $u+\overline{u}$ quarks minus
the number of $d+\overline{d}$ quarks does not depend on the scale
$Q$, which is due to the fact that the splittings $g\to q\overline{q}$
produce quarks of all flavors in equal numbers (if one neglects the
quark masses). An interesting relation can also be obtained for
$s=2$. For this moment, the matrix of anomalous dimensions in the
singlet sector,
\begin{equation}
\begin{pmatrix}
A_{ff}(2) & A_{fg}(2)\cr
N_f A_{gf}(2) & A_{gg}(2)\cr
\end{pmatrix}
=
\frac{1}{\pi^2}
\begin{pmatrix}
\frac{4}{9} & -\frac{4}{9}\cr
-\frac{N_f}{12} & \frac{N_f}{12}\cr
\end{pmatrix}\; ,
\end{equation}
has a vanishing eigenvalue, which means that {\colora a linear
combination of the flavor singlet operators is not renormalized}~:
${\colord {\cal O}}_{2,g}^{\mu\nu}+\sum_f {\colord {\cal
O}}_{2,f}^{\mu\nu}$. This leads also to a sum rule
\begin{equation}
{\colorb\frac{\partial}{\partial Q^2}} \left\{
\int_0^1dx\, x\left[\sum_f \Big[{\colord f_f}(x,{\colorb Q^2})+
{\colord f_{\bar{f}}}(x,{\colorb Q^2})
\Big]
+{\colord f_g}(x,{\colorb Q^2})\right]\right\}=0\; ,
\end{equation}
whose physical interpretation is the conservation of the total
momentum of the proton -- which therefore cannot depend on the
resolution scale $Q$. (Collinear splittings, that are responsible for
the $Q$ dependence of the number of partons, do not alter their total
momentum.)

\begin{figure}[htbp]
\begin{center}
\resizebox*{6cm}{!}{\includegraphics{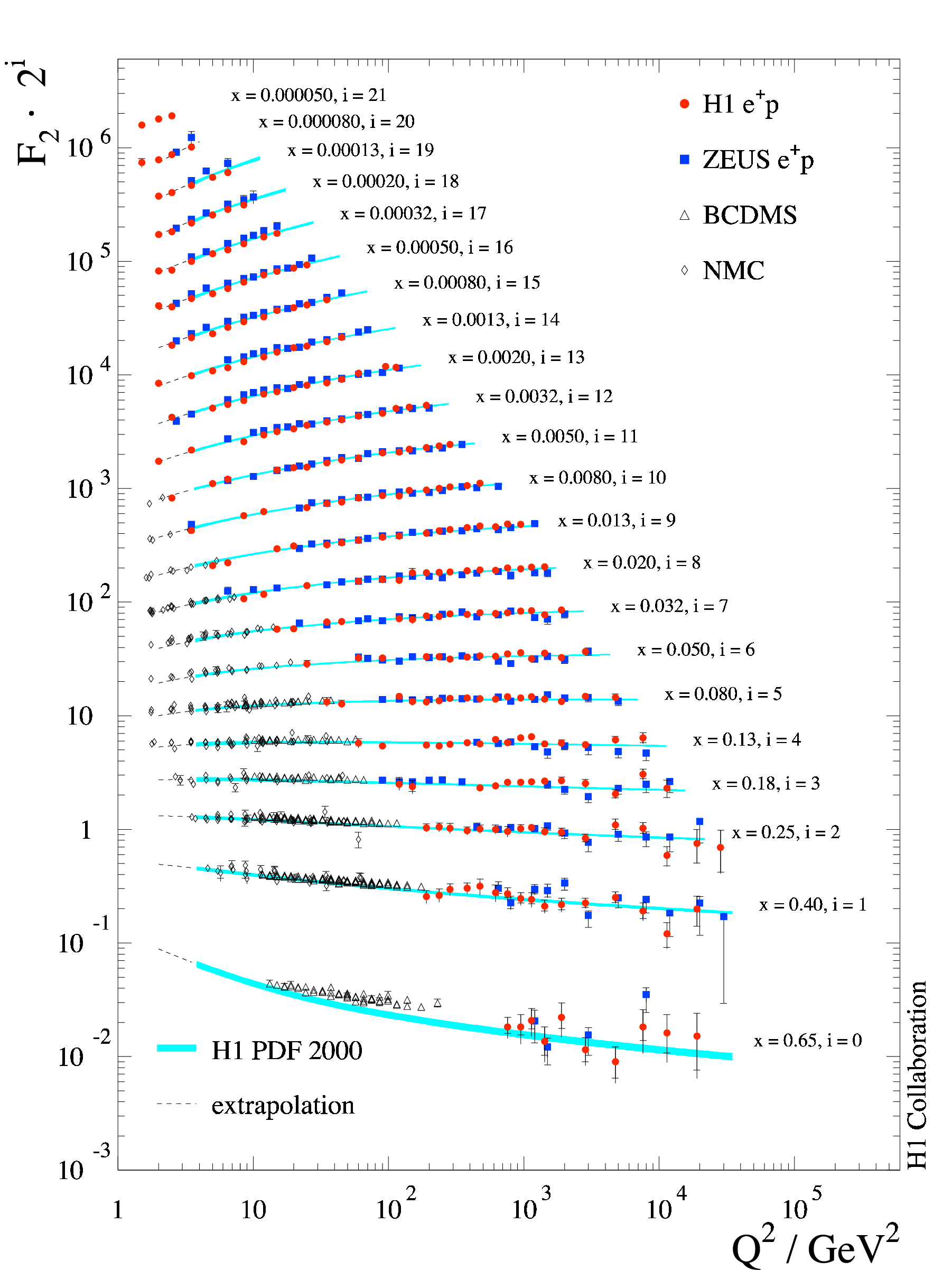}}
\end{center}
\caption{\label{fig:hera}Comparison of the measured $F_2$ with QCD
fits.}
\end{figure}

We have seen that QCD can be used to calculate the value of
the Wilson coefficients as well as the scale
dependence of the non-perturbative parton distributions. In practice,
when one compares DIS data with theoretical predictions, one needs
only to adjust the value of the parton distributions at a relatively low
initial scale $Q_0$, and then one uses the DGLAP evolution equations
in order to obtain their value at a higher $Q$. This program has now
been implemented to three loops (NNLO), and has been very successful
in explaining the inclusive DIS data. The agreement between QCD
and the DIS measurements is illustrated in figure \ref{fig:hera} (see
for instance \cite{Gayle1} for more details).

\section{Lecture II~: Parton evolution at small $x$ and gluon saturation}
In the first lecture, we introduced the parton model and the
evolution of parton distributions with the transverse resolution scale
$Q^2$ -- and the corresponding resummation of the powers of
$\alpha_s\log(Q^2)$. We now turn to the logarithms of
$1/x$. These logarithms are expected to be the dominant effect in
processes where the collision energy $\sqrt{s}$ is much larger than
the typical transverse momentum scale involved in the process, and may
lead to gluon saturation at very small $x$.

\subsection{Eikonal scattering}
Before going to the main subject of this lecture, let us make a detour
through an important result concerning the high energy limit of the
scattering amplitude of some state off an external field. Our
derivation here follows \cite{BjorkKS1}. Consider the 
generic $S$-matrix element
\begin{equation}
S_{\beta\alpha}\equiv
{\colorc\big<\beta_{\rm out}}\big|{\colorc\alpha_{\rm in}\big>}
=
{\colorc\big<\beta_{\rm in}}\big|U(+\infty,-\infty)\big|{\colorc\alpha_{\rm in}\big>}\, ,
\end{equation}
for the transition from a state $\alpha$ to a state $\beta$ where 
\begin{equation}
U(+\infty,-\infty)
=T_+\,\exp\Big[i\int d^4x\;{\cal L}_{\rm int}(\phi_{\rm in}(x))\Big]\; ,
\end{equation}
is the evolution operator from $t=-\infty$ to $t=+\infty$.
($T_+$ denotes an ordering in the light-cone time $x^+$.) The
interaction Lagrangian ${\cal L}_{\rm int}$ contains both the
self-interactions of the fields and their interactions with the
external field. Now apply a boost in the $z$ direction to all the
particles contained in the states $\alpha$ and $\beta$. Formally, this
can be done by multiplying the states by $\exp(-i\omega K^3)$, where
$\omega$ is the rapidity of the boost and $K^3$ the generator of
longitudinal boosts. Our goal is to compute the limit
$\omega\to+\infty$ of the transition amplitude,
\begin{equation}
    S_{\beta\alpha}^{(\infty)}\equiv\lim_{\omega\to +\infty}
    {\colorc\big<\beta_{\rm in}}\big|
    {\colorb e^{i\omega K^3}}U(+\infty,-\infty)
    {\colorb e^{-i\omega K^3}}\big|{\colorc\alpha_{\rm in}\big>}\; .
\end{equation}
The behavior of scattering amplitudes in this limit is easy to
understand. The time spent by the incoming particles in the region
where the external field is acting goes to zero as the inverse of the
collision energy $E$. If the coupling to the external field was purely
scalar, this would imply that the scattering amplitude itself goes to
zero as $E^{-1}$. However, in the case of a vector coupling, the
longitudinal component of the current increases as $E$, which
compensates the decrease in the interaction time, thereby leading to a
finite (non-zero and non infinite) high energy limit.

For this reason, let us assume that the coupling of the fields to the
external potential is of the form ${\colorc g}\, {\colord{\cal
A}_\mu(x)} J^\mu(x)$ where $J^\mu$ is a vector current built from the
elementary fields of the theory under consideration. In order to
simplify the discussion, we also assume that the external potential is
non-zero only in a finite range in $x^+$, $x^+\in[-L,+L]$ (this is to
avoid complications with long range interactions). The action of
${\colorc K^3}$ on states and operators is
      \begin{eqnarray}
      && 
      {\colorb e^{-i\omega K^3}}a^\dagger_{\rm in}(q){\colorb e^{i\omega K^3}}
      =a^\dagger_{\rm in}({\colorb e^\omega} q^+,{\colorb e^{-\omega}}q^-,\vec\q_\perp)\nonumber\\
      &&
      {\colorb e^{-i\omega K^3}}\big|\vec\p\cdots{}_{\rm in}\big>
      =\big|({\colorb e^\omega} p^+,\vec\p_\perp)\cdots{}_{\rm in}\big>
      \nonumber\\
      &&\vphantom{\int}
      {\colorb e^{i\omega K^3}}\phi_{\rm in}(x){\colorb e^{-i\omega K^3}}
      =\phi_{\rm in}({\colorb e^{-\omega}} x^+,{\colorb e^{\omega}}x^-,\vec\x_\perp)
      \; ,
    \end{eqnarray}
namely, it multiplies the $+$ component of momenta by $e^\omega$ and
their minus component by $e^{-\omega}$, while keeping the transverse
components unchanged. The external potential ${\colord{\cal
A}_\mu(x)}$ is unaffected by ${\colorc K^3}$, and the components of
$J^\mu(x)$ are changed as follows:
    \begin{eqnarray}
      &&
      {\colorb e^{i\omega K^3}}J^i(x){\colorb e^{-i\omega K^3}}
      =J^i({\colorb e^{-\omega}} x^+,{\colorb e^{\omega}}x^-,\vec\x_\perp)
      \nonumber\\
      &&
      {\colorb e^{i\omega K^3}}J^-(x){\colorb e^{-i\omega K^3}}
      ={\colorb e^{-\omega}}\,J^-({\colorb e^{-\omega}} x^+,{\colorb e^{\omega}}x^-,\vec\x_\perp)
      \nonumber\\
      &&
      {\colorb e^{i\omega K^3}}J^+(x){\colorb e^{-i\omega K^3}}
      ={\colorb e^\omega}\,J^+({\colorb e^{-\omega}} x^+,{\colorb e^{\omega}}x^-,\vec\x_\perp)
      \nonumber
    \end{eqnarray}
Because $K^3$ does not modify the ordering in
$x^+$, we can write
    \begin{equation}
      {\colorb e^{i\omega K^3}}U(+\infty,-\infty){\colorb e^{-i\omega K^3}}
      =
      T_+ \exp i\int d^4x
      \;{\cal L}_{\rm int}({\colorb e^{i\omega K^3}}
      \phi_{\rm in}(x){\colorb e^{-i\omega K^3}})\; .
    \end{equation}
In addition, we can split the evolution operator into three factors
  \begin{equation}
    U(+\infty,-\infty)=U(+\infty,+L)U(+L,-L)U(-L,-\infty)
  \end{equation}
so that only the factor in the middle contains the external field. In
order to deal with the first and last factor after the boost, it is
sufficient to change variables {\colord$e^{-\omega} x^+\to x^+$},
{\colord$e^{\omega} x^-\to x^-$}. This leads to
  \begin{eqnarray}
    &&
    \lim_{\omega\to+\infty}
    {\colorb e^{i\omega K^3}}
    U(+\infty,+L)
    {\colorb e^{-i\omega K^3}}
    =U_0(+\infty,0)
    \nonumber\\
    &&
    \lim_{\omega\to+\infty}
    {\colorb e^{i\omega K^3}}
    U(-L,-\infty)
    {\colorb e^{-i\omega K^3}}
    =U_0(0,-\infty)
    \; ,
  \end{eqnarray}
where $U_0$ is the same as $U$, but with the self-interactions only.
For the factor $U(L,-L)$, the change of variables {\colord
$e^{\omega}x^-\to x^-$}  gives us 
    \begin{equation}
    \lim_{\omega\to+\infty}
    {\colorb e^{i\omega K^3}}
    U(+L,-L)
    {\colorb e^{-i\omega K^3}}=T_+\,
    \exp \Big[
      i {\colorc g}
    \int d^2\vec\x_\perp {\colord\chi(\vec\x_\perp)}{\colora\rho(\vec\x_\perp)}
    \Big]\; ,
 \end{equation}
\begin{equation}
  \mbox{with}\qquad\left\{
  \begin{aligned}
    &
    {\colord\chi(\vec\x_\perp)}\equiv \int dx^+\;
    {\colord{\cal A}^-(x^+,0,\vec\x_\perp)}\; ,
    \\
    &
    {\colora\rho(\vec\x_\perp)}\equiv \int dx^-\;{\colora J^+(0,x^-,\vec\x_\perp)}\; .
  \end{aligned}
  \right.
  \end{equation}
Only the minus component of the external vector
potential matters, because this is the component that couples to the
longitudinal current $J^+$ which is enhanced by the boost. Therefore,
the high energy limit of the transition amplitude can be written as
    \begin{equation}
    S_{\beta\alpha}^{(\infty)}=
    {\colorc\big<\beta_{\rm in}\big|}
    U_0(+\infty,0)\;T_+
      \,\exp 
      \Big[i{\colorc g}
	\int\limits_{\vec\x_\perp}{\colord\chi(\vec\x_\perp)}
		   {\colora\rho(\vec\x_\perp)}
      \Big]
    U_0(0,-\infty)
    {\colorc\big|\alpha_{\rm in}\big>}\; .
    \label{eq:eikonal}
    \end{equation}
This limit is known as the {\sl eikonal limit}. It is important to
keep in mind that this formula is the exact answer for the high-energy
limit; no perturbative expansion has been made yet, and the formula
still contains the self-interactions of the fields of the theory to
all orders. A remarkable feature of eq.~(\ref{eq:eikonal}) is that it
separates the self-interactions of the fields and their interactions
with the external potential in three different factors, a property
which is strongly suggestive of the factorization between the long and
short distance physics in high energy hadronic interactions.

\begin{figure}[htbp]
\begin{center}
\resizebox*{6cm}{!}{\includegraphics{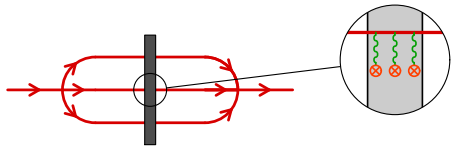}}
\end{center}
\caption{\label{fig:eikonal}Scattering off an external potential in
the high energy limit.}
\end{figure}

In order to use eq.~(\ref{eq:eikonal}) in practice, it is necessary to
make an expansion in the self-interactions of the fields, by
introducing complete sets of states between the three factors,
\begin{eqnarray}
 && 
  S_{\beta\alpha}^{(\infty)}=\smash{\sum_{{\colore\gamma},{\colore\delta}}}
  {\colorc\big<\beta_{\rm in}\big|}
  U_0(+\infty,0)
  {\colore\big|\gamma_{\rm in}\big>}
  \vphantom{\Big[}
  \nonumber\\
  &&\qquad\quad\times
  {\colore\big<\gamma_{\rm in}\big|}
    T_+\,\exp \Big[
      i{\colorc g}
      {\int\limits_{\vec\x_\perp}}{\colord\chi(\vec\x_\perp)}
      {\colora\rho(\vec\x_\perp)}
    \Big]
  {\colore\big|\delta_{\rm in}\big>
  \big<\delta_{\rm in}\big|}
  U_0(0,-\infty)
  {\colorc\big|\alpha_{\rm in}\big>}
  \; .
\end{eqnarray}
The factor $\sum_{\colore\delta} {\colore \big|\delta_{\rm
in}\big>\big<\delta_{\rm in}\big|} U(0,-\infty)
{\colorc\big|\alpha_{\rm in}\big>}$ is the {\sl Fock expansion} of the
initial state. It reflects the fact that the state $\alpha$ prepared
at $x^+=-\infty$ may have fluctuated into another state $\delta$
before it interacts with the external potential. There is also a
similar expansion for the final state. Assuming that we have performed
the Fock expansion to the desired order\footnote{The main difference
compared to the usual perturbation theory is that the integrations
over $x^+$ run only over half of the real axis, e.g. $[-\infty,0]$. In
Fourier space, this implies that the minus component
of the momentum is not conserved at the vertices, and that one gets
energy denominators instead of delta functions.}, one needs to
evaluate matrix elements such as
\begin{equation}
{\colore\big<\gamma_{\rm in}\big|}
\exp\Big[
  i{\colorc g}\int
      {\colord\chi_a(\vec\x_\perp)}{\colora\rho^a(\vec\x_\perp)}\Big]
{\colore\big|\delta_{\rm in}\big>}\; .
\label{eq:tmp1}
\end{equation}
We have reinstated color indices in this formula, since we have
applications to QCD in mind.  In order to calculate this matrix
element, the first step is to express the operator
${\colora\rho^a(\vec\x_\perp)}$ in terms of creation and annihilation
operators of the particles that can couple to the external
potential. For instance, the contribution that comes from the quarks
and antiquarks is given by
        \begin{eqnarray}
      &&
      {\colora\rho^a(\vec\x_\perp)}={\colorc t^a_{ij}}
      \smash{\int \frac{dp^+}{4\pi p^+}\frac{d^2\vec\p_\perp}{(2\pi)^2}
      \frac{d^2\vec\q_\perp}{(2\pi)^2}}
      \Big\{
      {\colorb b^\dagger_{\rm in}(p^+,\vec\p_\perp;i)
	b_{\rm in}(p^+,\vec\q_\perp;j)}
      e^{i(\vec\p_\perp-\vec\q_\perp)\cdot\vec\x_\perp}
      \nonumber\\
      &&\qquad\qquad\qquad\qquad\qquad\qquad
      -
      {\colorb d^\dagger_{\rm in}(p^+,\vec\p_\perp;i)
	d_{\rm in}(p^+,\vec\q_\perp;j)}
      e^{-i(\vec\p_\perp-\vec\q_\perp)\cdot\vec\x_\perp}
      \Big\}
      \; .
    \end{eqnarray}
(The quarks come with a positive sign and the antiquarks with a
negative sign.) The contribution of the gluons would be similar, but
the color matrix would be replaced by an element of the adjoint
representation. From this formula, we see that in eq.~(\ref{eq:tmp1}),
the states $\delta$ and $\gamma$ must have the same particle content,
because each annihilation operator in $\rho^a$ is immediately followed
by a creation operator that creates a particle of the same nature.
The $+$ component of the momenta of the particles in $\delta$ and
$\gamma$ must also be identical. The only difference between the
states $\delta$ and $\gamma$ is in the transverse momenta and in the
color of their particles. In order to recover the eikonal limit in a
more familiar form, one should go to {\sl impact parameter
representation} by performing a Fourier transformation of all the
transverse momenta in the intermediate states $\delta$ and $\gamma$,
by defining the {\sl light-cone wavefunction}
\begin{equation}
{\colord\Psi_{\delta\alpha}(\{k^+_i,\vec\x_{i\perp}\})}\equiv 
\prod_{i\in\delta}\int \frac{d^2\vec\k_{i\perp}}{(2\pi)^2}
e^{-i\vec\k_{i\perp}\cdot\vec\x_{i\perp}}
{\colore\big<\delta_{\rm in}\big|}U_0(0,\!-\infty)
{\colore\big|\alpha_{\rm in}\big>}\; .
\end{equation}
Then, from the explicit form of $\rho^a$, it is easy to check that the
only effect of the external potential is to multiply the function
${\colord\Psi_{\delta\alpha}}$ by a phase factor for each particle in
the intermediate state~:
\begin{eqnarray}
&&
{\colord\Psi_{\delta\alpha}(\{k^+_i,\vec\x_{i\perp}\})}
\;
\longrightarrow
\;
{\colord\Psi_{\delta\alpha}(\{k^+_i,\vec\x_{i\perp}\})}
\prod_{i\in\delta} {\colorb U_i(\vec\x_\perp)}
\nonumber\\
&&{\colorb U_i(\vec\x_\perp)}\equiv T_+\exp \Big[
i g_{_i}\int dx^+\;
 {\colord{\cal A}^-_a(x^+,0,\vec\x_\perp)}{\colorc t^a}
\Big]
\; .
\end{eqnarray}
In the case of non-abelian interactions, these phase factors
$U_i(\vec\x_\perp)$ are known as {\sl Wilson lines}. Wilson lines
resum multiple scatterings off the external field, as one can see by
expanding the exponential. Thus, the physical picture of high energy
scattering off some external field is that the initial state evolves
from $-\infty$ to $0$, multiply scatters during an infinitesimally
short time off the external potential, and evolves again from $0$ to
$+\infty$ to form the final state, as illustrated in figure
\ref{fig:eikonal}.
In terms of light-cone wavefunctions and of Wilson lines, the high
energy limit of the transition amplitude reads
\begin{equation}
S_{\beta\alpha}^{(\infty)}
=
\sum_{{\colore \delta}}
\int\Big[\prod_{i\in\delta}
\frac{dk_i^+}{4\pi k_i^+}d^2\x_{i\perp}
\Big]
{\colore\Psi_{\delta\beta}^\dagger(\{k_i^+,\vec\x_{i\perp}\})}
\Big[\prod_{i\in\delta}{\colord U_{i}(\vec\x_{i\perp})}\Big]
{\colore \Psi_{\delta\alpha}(\{k_i^+,\vec\x_{i\perp}\})}
\; .
\end{equation}

\subsection{BFKL equation}
Let us now derive the BFKL equation. Our derivation is inspired from
\cite{Weige1,Kovch5,Kovch3,Balit1,Muell5}.  Consider the forward
scattering off an external field of a state $\alpha$ whose simplest
Fock component is a color singlet quark-antiquark pair. Thus, the
transition amplitude can be written as\setbox1\hbox to 2cm{\hfil
\begin{feynman}{14mm}
\diagram{16}{
	-3 0.95 translate
	1 setlinejoin
	[] 0 setdash
	1 setlinecap
	0 setgray
	/Width 0.08 def
	colore
	1 0 0.6 0 beglfermion 2 copy
	/Width 0.05 def
	colorb
	0.6 180 90 begcfermion 0.6 0 beglscalar pop pop
	0.6 -90 -180 endcfermion
	0.6 -180 -90 begcscalar
	0.6 0 beglscalar
	0.7 setgray
	/Width 0.05 def
	colorb
	0.6 0 beglscalar
	0.6 0 -90 endcfermion 
	0.6 -90 0 begcscalar
	2 copy
	/Width 0.08 def
	colore
	0.6 0 beglfermion pop pop
	/Width 0.05 def
	colorb
	0.6 90 0 endcfermion
	0.6 0 90 begcscalar
	0.6 180 beglscalar
	0.3 setgray
	2.7 -1 moveto 0.2 0 rlineto 
	0 2 rlineto -0.2 0 rlineto 0 -2 rlineto fill
	0.02 setlinewidth
	0 setgray
	2.7 -1 moveto 0.2 0 rlineto 
	0 2 rlineto -0.2 0 rlineto 0 -2 rlineto stroke
}\end{feynman}\hfil}
\begin{equation}
\raise -5mm\box1 = 
{\colore\left|\Psi^{(0)}(\vec\x_\perp,\vec\y_\perp)\right|^2}
{\rm tr}\left[{\colorb U(\vec\x_\perp)
U^\dagger(\vec\y_\perp)}\right]
\; .
\end{equation}
We will not need to specify more the light-cone wavefunction of the
state under consideration. Note that the product of the two Wilson
lines is traced, because the state $\alpha$ is color singlet. A
crucial property of this transition amplitude is that it is completely
independent of the collision energy. However, as we shall see, a non
trivial energy dependence arises in this amplitude because of large
logarithms in loop corrections.

\begin{figure}[htbp]
\begin{center}
\resizebox*{10cm}{!}{\includegraphics{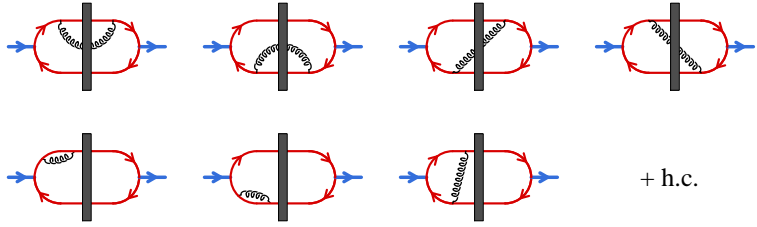}}
\end{center}
\caption{\label{fig:dip-1loop}One-loop corrections to the scattering
of a dipole off an external field. Only half of the virtual
corrections have been represented.}
\end{figure}

Consider now the 1-loop corrections to this amplitude depicted in
figure \ref{fig:dip-1loop}.
These 1-loop corrections all involve one additional gluon attached to
the quark or antiquark lines. In some of the corrections, that we
shall call {\sl real corrections}, the gluon is present in the state
that goes through the external field. In the other corrections, the
{\sl virtual corrections}, the gluon is just a fluctuation in the
wavefunction of the initial or final state.  The calculation of these
diagrams is straightforward in the impact parameter
representation. One simply needs the formula for the $q\bar{q}g$
vertex~:\setbox1\hbox to 2cm{ \hfil
\begin{feynman}{0.6cm}
\diagram{18}{
0 0 translate
1 setlinejoin
[] 0 setdash
1 setlinecap
/Width 0.05 def
colorb
0 0 0.7 0 beglfermion 2 copy
0.7 0 beglfermion pop pop
/Width 0.03 def
fgcolor
0.3 -90 halfbeglgluon1
0.37 -180 -90 amputatedbegcgluon 2 copy
0.7 0 halfbeglgluon2 pop pop
}
\end{feynman}
\hfil
}
\begin{equation}
\raise 3mm\box1\quad= \;2{\colorc g}t^a \; 
\frac{\vec{\bs \epsilon}_\lambda\cdot\vec\k_\perp}{k_\perp^2}\; ,
\end{equation}
where ${\bs\epsilon}_\lambda$ is the polarization vector of the gluon
and $\k_\perp$ its transverse momentum, and its expression in impact
parameter space,
\begin{equation}
\int \frac{d^2\vec\k_\perp}{(2\pi)^2} \; e^{i\vec\k_\perp\cdot(\vec\x_\perp-\vec\z_\perp)} \; 2{\colorc g}t^a \; 
\frac{\vec{\bs \epsilon}_\lambda\cdot\vec\k_\perp}{k_\perp^2}
=
\frac{2i{\colorc g}}{2\pi}
t^a \frac{\vec{\bs \epsilon}_\lambda\cdot(\vec\x_\perp-\vec\z_\perp)}{(\vec\x_\perp-\vec\z_\perp)^2}\; .
\end{equation}
Armed with these tools, it is easy to obtain expressions such as

\setbox1\hbox to 2cm{
\hfil
    \begin{feynman}{1cm}
      \diagram{14}{
	0 0 translate
	gsave
	-3 0 translate
	gsave
	0 0.95 translate
	1 setlinejoin
	[] 0 setdash
	1 setlinecap
	0 setgray
	/Width 0.08 def
	colore
	1 0 0.6 0 beglfermion 2 copy
	/Width 0.05 def
	colorb
	0.6 180 135 begcscalar 2 copy
	/Width 0.03 def
	fgcolor
	0.7 -110 -45 begcgluon pop pop
	/Width 0.05 def
	colorb
	0.6 135 90 begcscalar 0.6 0 beglscalar pop pop
	0.6 -90 -180 endcfermion
	0.6 -180 -90 begcscalar
	0.6 0 beglscalar
	0.7 setgray
	/Width 0.05 def
	colorb
	0.6 0 beglscalar
	0.6 0 -90 endcfermion 
	0.6 -90 0 begcscalar
	2 copy
	/Width 0.08 def
	colore
	0.6 0 beglfermion pop pop
	/Width 0.05 def
	colorb
	0.6 90 0 endcfermion
	0.6 0 90 begcscalar
	0.6 180 beglscalar
	0.3 setgray
	2.7 -1 moveto 0.2 0 rlineto 
	0 2 rlineto -0.2 0 rlineto 0 -2 rlineto fill
	0.02 setlinewidth
	0 setgray
	2.7 -1 moveto 0.2 0 rlineto 
	0 2 rlineto -0.2 0 rlineto 0 -2 rlineto stroke
	grestore
	grestore
      }
    \end{feynman}
\hfil
}
\begin{eqnarray}
\raise -4mm\box1 &=& {\colore
\left|\Psi^{(0)}(\vec\x_\perp,\vec\y_\perp)\right|^2}
{\rm tr}\left[{\colord t^a t^a}
{\colorb U(\vec\x_\perp)U^\dagger(\vec\y_\perp)}\right]
\nonumber\\
&&
\times
-2{\colorc\alpha_s}{\colora\int \frac{dk^+}{k^+}}
\int\frac{d^2\vec\z_\perp}{(2\pi)^2}
\frac{(\vec\x_\perp-\vec\z_\perp)\cdot(\vec\x_\perp-\vec\z_\perp)}
{(\vec\x_\perp-\vec\z_\perp)^2 (\vec\x_\perp-\vec\z_\perp)^2}
\; ,
\end{eqnarray}
and
\setbox1\hbox to 2cm{
\hfil
    \begin{feynman}{1cm}
      \diagram{14}{
	0 0 translate
	gsave
	-3 0 translate
	gsave
	0 0.95 translate
	1 setlinejoin
	[] 0 setdash
	1 setlinecap
	0 setgray
	/Width 0.08 def
	colore
	1 0 0.6 0 beglfermion 2 copy
	/Width 0.05 def
	colorb
	0.6 180 90 begcfermion 0.6 0 beglscalar pop pop
	0.6 -90 -180 endcfermion
	0.6 -180 -90 begcscalar
	0.6 0 beglscalar
	0.7 setgray
	/Width 0.05 def
	colorb
	0.6 0 beglscalar
	0.6 0 -90 endcfermion 
	0.6 -90 0 begcscalar
	2 copy
	/Width 0.08 def
	colore
	0.6 0 beglfermion pop pop
	/Width 0.05 def
	colorb
	0.6 90 0 endcfermion
	0.6 0 90 begcscalar
	0.6 180 beglscalar
	fgcolor
	/Width 0.03 def
	2.2 -0.6 2.5 0.6 getlparam beglgluon pop pop
	0.3 setgray
	2.7 -1 moveto 0.2 0 rlineto 
	0 2 rlineto -0.2 0 rlineto 0 -2 rlineto fill
	0.02 setlinewidth
	0 setgray
	2.7 -1 moveto 0.2 0 rlineto 
	0 2 rlineto -0.2 0 rlineto 0 -2 rlineto stroke
	grestore
	grestore
      }
    \end{feynman}
\hfil
}
\begin{eqnarray}
\raise -4mm\box1 &=& {\colore
\left|\Psi^{(0)}(\vec\x_\perp,\vec\y_\perp)\right|^2}
{\rm tr}\left[{\colord t^a} 
{\colorb U(\vec\x_\perp)U^\dagger(\vec\y_\perp)}{\colord t^a}\right]
\nonumber\\
&&\times
4{\colorc\alpha_s}
{\colora\int\frac{dk^+}{k^+}}\int\frac{d^2\vec\z_\perp}{(2\pi)^2} 
\frac{(\vec\x_\perp-\vec\z_\perp)\cdot(\vec\y_\perp-\vec\z_\perp)}
{(\vec\x_\perp-\vec\z_\perp)^2 (\vec\y_\perp-\vec\z_\perp)^2}
\; .
\end{eqnarray} 
We find that the sum of all the virtual corrections reads
\begin{eqnarray}
&&
-\frac{{\colord C_{\rm f}}{\colorc\alpha_s}}{\pi^2}
{\colora\int\frac{dk^+}{k^+}}\int d^2\vec\z_\perp\;
\frac{(\vec\x_\perp-\vec\y_\perp)^2}
{(\vec\x_\perp-\vec\z_\perp)^2 (\vec\y_\perp-\vec\z_\perp)^2}
{\colore
\left|\Psi^{(0)}(\vec\x_\perp,\vec\y_\perp)\right|^2}
{\rm tr}\left[
{\colorb U(\vec\x_\perp)U^\dagger(\vec\y_\perp)}\right]
\; ,
\nonumber\\
&&
\label{eq:bfkl-virtual}
\end{eqnarray}
where $C_{\rm f}\equiv t^a t^a=(N^2-1)/2N$ for SU(N).  In this
formula, $k^+$ is the longitudinal momentum of the gluon. As one can
see, there is a logarithmic divergence in the integration over this
variable. The lower bound should arguably be some non-perturbative
hadronic scale $\Lambda$, and the upper bound must be the longitudinal
momentum $p^+$ of the quark or antiquark that emitted the
photon. Hence we have a $\log(p^+/\Lambda)$, which is a large factor
in the limit of high-energy (strictly speaking, the high-energy limit
is ill defined because of these corrections). The calculation of the
real corrections is a bit more involved. For instance, one has
\setbox1\hbox to 2cm{ \hfil
    \begin{feynman}{1cm}
      \diagram{14}{
	0 0 translate
	gsave
	-3 0 translate
	gsave
	0 0.95 translate
	1 setlinejoin
	[] 0 setdash
	1 setlinecap
	0 setgray
	/Width 0.08 def
	colore
	1 0 0.6 0 beglfermion 2 copy
	/Width 0.05 def
	colorb
	0.6 180 90 begcfermion 0.6 0 beglscalar pop pop
	0.6 -90 -180 endcfermion
	0.6 -180 -90 begcscalar
	0.6 0 beglscalar
	0.7 setgray
	/Width 0.05 def
	colorb
	0.6 0 beglscalar
	0.6 0 -90 endcfermion 
	0.6 -90 0 begcscalar
	2 copy
	/Width 0.08 def
	colore
	0.6 0 beglfermion pop pop
	/Width 0.05 def
	colorb
	0.6 90 0 endcfermion
	0.6 0 90 begcscalar
	0.6 180 beglscalar
	fgcolor
	/Width 0.03 def
	2.2 0.6 0.6 -180 0 begcgluon pop pop
	0.3 setgray
	2.7 -1 moveto 0.2 0 rlineto 
	0 2 rlineto -0.2 0 rlineto 0 -2 rlineto fill
	0.02 setlinewidth
	0 setgray
	2.7 -1 moveto 0.2 0 rlineto 
	0 2 rlineto -0.2 0 rlineto 0 -2 rlineto stroke
	grestore
	grestore
      }
    \end{feynman}
\hfil
}
\begin{eqnarray}
\raise -4mm\box1 &=& {\colore
\left|\Psi^{(0)}(\vec\x_\perp,\vec\y_\perp)\right|^2}
{\rm tr}\left[{\colord t^a}
{\colorb U(\vec\x_\perp){\colord t^b}U^\dagger(\vec\y_\perp)}\right]
\nonumber\\
&&\!\!\!\!\!\!\!\!\!\!\!\!\!\!\!\!\!\!\!\!\!\!\!\!
\times
4{\colorc\alpha_s}{\colora\int \frac{dk^+}{k^+}}
\int\frac{d^2\vec\z_\perp}{(2\pi)^2}{\colorb {\wt U}_{ab}(\vec\z_\perp)}
\frac{(\vec\x_\perp-\vec\z_\perp)\cdot(\vec\x_\perp-\vec\z_\perp)}
{(\vec\x_\perp-\vec\z_\perp)^2 (\vec\x_\perp-\vec\z_\perp)^2}
\; ,
\end{eqnarray}
where ${\colorb {\wt U}_{ab}(\vec\z_\perp)}$ is a Wilson line in the
{\colore adjoint representation} that represents the eikonal phase
factor associated to the gluon ($\z_\perp$ is the impact parameter of
the gluon). In order to simplify the real terms, we need the following
relation between fundamental and adjoint Wilson lines,
\begin{equation}
{\colord t^a}{\colorb {\wt U}_{ab}(\vec\z_\perp)}
=
{\colorb U(\vec\z_\perp)}{\colord t^b}{\colorb U^\dagger(\vec\z_\perp)}\; ,
\end{equation}
and the Fierz identity obeyed by fundamental SU(N) matrices~:
\begin{equation}
{\colord t^b_{ij} t^b_{kl}}={\colorc\frac{1}{2}} \delta_{il} \delta_{jk}
-{\colorc \frac{1}{2N} }\delta_{ij}\delta{kl}
\; .
\end{equation}
Thanks to these identities, one can rewrite all the real corrections
in terms of the quantity ${\colorb{\bs S}(\vec\x_\perp,\vec\y_\perp)}
\equiv {\rm tr}\left[{\colorb
U(\vec\x_\perp)U^\dagger(\vec\y_\perp)}\right]/N\; .  $ Collecting all
the terms, and summing real and virtual contributions, we obtain the
following expression for the 1-loop transition amplitude
\begin{eqnarray}
&&
-\frac{{\colorc \alpha_s N^2}{\colora Y}}{2\pi^2}
{\colore\left|\Psi^{(0)}(\vec\x_\perp,\vec\y_\perp)\right|^2}
\int d^2\vec\z_\perp\;
\frac{(\vec\x_\perp-\vec\y_\perp)^2}
{(\vec\x_\perp-\vec\z_\perp)^2 (\vec\y_\perp-\vec\z_\perp)^2}
\nonumber\\
&&\qquad\qquad\qquad\qquad\qquad\times
\Big\{
{\colorb{\bs S}(\vec\x_\perp,\vec\y_\perp)}
-
{\colorb{\bs S}(\vec\x_\perp,\vec\z_\perp)}
{\colorb{\bs S}(\vec\z_\perp,\vec\y_\perp)}
\Big\}
\; ,
\label{eq:1loop}
\end{eqnarray}
where we denote $Y\equiv\ln(p^+/\Lambda)$. This correction to the
transition amplitude is not small when $\alpha_s^{-1}\lesssim Y$,
which means that $n$-loop contributions should be considered in order
to resum all the powers $(\alpha_s Y)^n$. Here, we are just going to
admit that this $n$-loop calculation amounts to exponentiating the
1-loop result. In other words, eq.~(\ref{eq:1loop}) is sufficient in
order to obtain the derivative $\partial {\bs S}/\partial Y$,
\begin{eqnarray}
&&
\frac{\partial {\colorb{\bs S}(\vec\x_\perp,\vec\y_\perp)}}{\partial{\colora Y}}
=
-\frac{{\colorc \alpha_s N_c}}{2\pi^2}
\int d^2\vec\z_\perp\;
\frac{(\vec\x_\perp-\vec\y_\perp)^2}
{(\vec\x_\perp-\vec\z_\perp)^2 (\vec\y_\perp-\vec\z_\perp)^2}
\nonumber\\
&&
\qquad\qquad\qquad\qquad\qquad\qquad\times
\Big\{
{\colorb{\bs S}(\vec\x_\perp,\vec\y_\perp)}
-
{\colorb{\bs S}(\vec\x_\perp,\vec\z_\perp)}
{\colorb{\bs S}(\vec\z_\perp,\vec\y_\perp)}
\Big\}
\; .
\label{eq:BK0}
\end{eqnarray}
It is customary to rewrite this equation in terms of $T$-matrix
elements, ${\colorb {\bs T}(\vec\x_\perp,\vec\y_\perp)} \equiv
1-{\colorb{\bs S}(\vec\x_\perp,\vec\y_\perp)}$. The BFKL
equation\cite{bfkl} describes the regime where ${\bs
T}(\x_\perp,\y_\perp)$ is small, so that we can neglect the terms that
are quadratic in ${\bs T}$. It reads~:
    \begin{eqnarray}
      &&
      \frac{\partial\,{\colorb{\bs T}(\vec\x_\perp,\vec\y_\perp)}}
      {\partial{\colora Y}}
      =
      \frac{{\colorc \alpha_s N_c}}{2\pi^2}
      \int d^2\vec\z_\perp\;
      \frac{(\vec\x_\perp-\vec\y_\perp)^2}
      {(\vec\x_\perp-\vec\z_\perp)^2 (\vec\y_\perp-\vec\z_\perp)^2}
      \nonumber\\
      &&
      \qquad\qquad\qquad\qquad\qquad\times
      \Big\{
      {\colorb{\bs T}(\vec\x_\perp,\vec\z_\perp)}
      +
      {\colorb{\bs T}(\vec\z_\perp,\vec\y_\perp)}
      -
      {\colorb{\bs T}(\vec\x_\perp,\vec\y_\perp)}
      \Big\}\; .
      \label{eq:BFKL}
    \end{eqnarray}
One can verify easily that ${\bs T}=0$ is a {\sl fixed point} of this
equation (the right hand side vanishes if one sets ${\bs T}=0$), but
that this fixed point is unstable (if one sets ${\bs T}=\epsilon>0$,
the right hand side is positive). Since there are no other fixed
points, solutions of the BFKL have an unbounded growth in the high
energy limit ($Y\to+\infty$). This behavior however is not
physical, because the unitarity of scattering amplitude implies that
${\bs T}(\x_\perp,y_\perp)$ should not become greater than unity.

\subsection{Balitsky-Kovchegov equation}
The solution to the above problem was in fact already contained in
eq.~(\ref{eq:BK0}). When written in terms of ${\bs T}$ without
assuming that ${\bs T}$ is small,
\begin{eqnarray}
      &&
      \frac{\partial\,{\colorb{\bs T}(\vec\x_\perp,\vec\y_\perp)}}
      {\partial{\colora Y}}
      =
      \frac{{\colorc \alpha_s N_c}}{2\pi^2}
      \int d^2\vec\z_\perp\;
      \frac{(\vec\x_\perp-\vec\y_\perp)^2}
      {(\vec\x_\perp-\vec\z_\perp)^2 (\vec\y_\perp-\vec\z_\perp)^2}
      \nonumber\\
      &&\!\!\!\!
      \times
      \Big\{
      {\colorb{\bs T}(\vec\x_\perp,\vec\z_\perp)}
      +
      {\colorb{\bs T}(\vec\z_\perp,\vec\y_\perp)}
      -
      {\colorb{\bs T}(\vec\x_\perp,\vec\y_\perp)}
	- {\colord{\bs T}(\vec\x_\perp,\vec\z_\perp)}
	{\colord{\bs T}(\vec\z_\perp,\vec\y_\perp)}
      \Big\}\; ,
      \label{eq:BK1}
    \end{eqnarray}
it has a non-linear term that confines ${\bs T}$ to the range
$[0,1]$. Indeed, the presence of this quadratic term makes ${\bs T}=1$
a stable fixed point of the equation. Therefore, the generic behavior
of solutions of eq.~(\ref{eq:BK1}) is that ${\bs T}$ starts at small
values at small $Y$ and asymptotically reaches the value ${\bs T}=1$
in the high energy limit. Eq.~(\ref{eq:BK1}) is known as the
{\sl Balitsky-Kovchegov equation}\cite{Kovch3,Balit1}.

The interaction of a color singlet dipole with an external color field
is a possible description of DIS, in a frame in which the virtual
photon splits into a quark-antiquark pair long before it collides with
the proton (the external color field would represent the proton
target). Although it is legitimate to treat the proton as a frozen
configuration of color field due to the brevity of the interaction, we
do not know what this field is. Moreover, since this field is created
by the partons inside the proton, that have a complicated dynamics,
this color field must be different for each collision, and should
therefore be treated as random. Therefore, in order to turn our dipole
scattering amplitude into an object that we could use to compute the
DIS cross-section at high-energy, we must average over all the
possible configurations of the external field. Let us denote by
$\big<\cdots\big>$ this average. The effect of this average on the
energy dependence of the amplitude is simply taken into account by
taking the average of eq.~(\ref{eq:BK1}). However, one sees that the
evolution equation for $\big<{\bs T}\big>$ involves in its right hand
side the average of a product of two ${\bs T}$'s, $\big<{\bs T}{\bs
T}\big>$. Therefore, we do not have a closed equation anymore. An
evolution equation for $\big<{\bs T}{\bs T}\big>$ could be obtained by
the same procedure, which would depend on yet another new object, and
so on. At the end of the day, one in fact obtains an infinite
hierarchy of nested equations, known as {\sl Balitsky's
equations}\cite{Balit1}.

It is only if one assumes that the averages of products of amplitudes
factorize into products of averages, 
\begin{equation}
\left<{\colorb{\bs T}\,{\bs T}}\right>\approx 
\left<{\colorb{\bs T}}\right>\,
\left<{\colorb{\bs T}}\right>\; ,
\end{equation}
that this hierarchy can be truncated into a closed equation which is
identical to eq.~(\ref{eq:BK1}) -- the BK equation -- with ${\bs T}$
replaced by $\big<{\bs T}\big>$. This approximation amounts to drop
certain correlations among the target fields, and is believed to be a
good approximation for a large nucleus in the limit of a large number
of colors\cite{Kovch3}.

\subsection{Gluon saturation and Color Glass Condensate}
The problem encountered with the indefinite growth of the solutions of
the BFKL can be understood in terms of the behavior of the gluon
distribution at small momentum fraction $x$. Indeed, in the regime
where the dipole scattering amplitude ${\bs T}$ is still small, it can
be calculated perturbatively,
    \begin{equation}
      {\colorb {\bs T}(\vec\x_\perp,\vec\y_\perp)}
      \propto {\colorb |\vec\x_\perp-\vec\y_\perp|^2}
      \; {\colora x}G({\colora x},{\colorb|\vec\x_\perp-\vec\y_\perp|^{-2}})
      \; ,
    \end{equation}
where ${\colora Y}\equiv \ln({\colora 1/x})$. This formula is an
example of the duality that exists in the description of scattering
processes at high energy. In the derivation of the BFKL and BK
equations, we have treated the proton target as given once for all,
and the energy dependence has been obtained by applying a boost to the
dipole projectile. But, thanks to the fact that transition amplitudes
are Lorentz invariant quantities, they can also be evaluated in a
frame where the dipole is fixed, and the boost is applied to the
proton. In this frame, the energy dependence of the scattering
amplitude comes from the $x$ dependence of the proton gluon
distribution.

Thus, an exponential behavior of ${\bs T}$ is equivalent to an
increase of the gluon distribution as a power of $1/x$~:
\begin{equation}
{\colorb {\bs T}}\sim e^{\omega {\colora Y}}
\qquad\longleftrightarrow\qquad
{\colora x}G({\colora x},Q^2)\sim \frac{1}{{\colora x}^\omega}\; .
\end{equation}
(This growth of the gluon distribution is due to gluon splittings.)
However, the gluon distribution cannot grow at this pace
indefinitely. Indeed, at some point, the occupation number of the
gluons will become large and the recombination of two gluons -- not
included in the BFKL equation -- will be favored. This phenomenon is
known as {\sl gluon saturation}\cite{sat}.
\begin{figure}[htbp]
\begin{center}
\resizebox*{4cm}{!}{\includegraphics{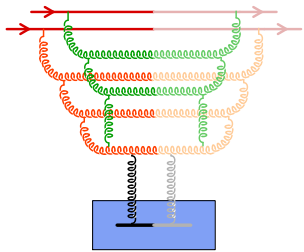}}
\end{center}
\caption{\label{fig:saturation}Gluon saturation~: merging of the
gluons ladders initiated by two valence partons. The proton target is
at the top of the picture and the probe at the bottom.}
\end{figure}
In the linear regime, described by the BFKL equation, each valence
parton from the proton initiates its own gluon ladder (see figure
\ref{fig:saturation}) that evolves independently from the others. In
the saturated regime, these gluon ladders can merge, thereby reducing
the growth of the gluon distribution. The effect of these
recombinations on the scattering amplitude is taken into account by
the non-linear term of the BK equation.

A semi quantitative criterion for gluon saturation can be
obtained\cite{sat} by comparing the surface density of gluons,
$\rho\sim xG(x,Q^2)/\pi R^2$, and the cross-section for gluon
recombination, $\sigma\sim \alpha_s/Q^2$. Saturation occurs when
$1\lesssim\rho\sigma$, i.e. when
\begin{equation}
Q^2\le Q_s^2\quad,\quad\mbox{with\ \ }
{\colorb Q_s^2}\quad\sim\quad
 \frac{{\colord \alpha_s} x G(x,{\colorb Q_s^2})}{\pi
 R_{_A}^2}\quad\sim\quad {\colorb A^{1/3}} {\colorc\frac{1}{x^{0.3}}}\; .
\label{eq:sat-cond}
\end{equation}
The quantity $Q_s$ is known as the {\sl saturation momentum}. Its
dependence on the number of nucleons $A$ (in the case of a nuclear
target) comes from the fact that $xG(x,Q^2)$ scales like the volume,
while $\pi R^2$ is an area. Its $x$ dependence is a phenomenological
parameterization inspired by from fits of HERA data.
\begin{figure}[htbp]
\begin{center}
\resizebox*{6cm}{!}{\includegraphics{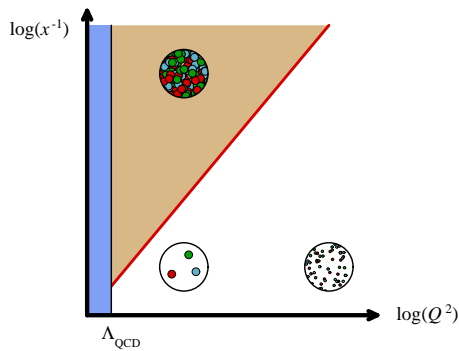}}
\end{center}
\caption{\label{fig:sat-domain}Saturation domain in the $x,Q^2$ plane.}
\end{figure}
From eq.~(\ref{eq:sat-cond}), one can divide the $x,Q^2$ in two
regions, as illustrated in figure \ref{fig:sat-domain}. The saturated
regime corresponds to the domain of low $Q$ and low $x$.

Although the BK equation describes the evolution of the dipole
scattering amplitude into the saturation regime, there is an
equivalent description of this evolution -- the {\sl Color Glass
Condensate} -- in which the central role is played by the target.  The
CGC description divides the degrees of freedom in the proton into fast
partons (large $x$) and slow partons (small $x$)\cite{mv}. The fast
partons are affected by time dilation, and do not have any significant
time evolution during the brief duration of the collision; therefore,
they are treated as static objects that carry a color source. These
color sources produce a current,
\begin{equation}
J^\mu=\delta^{\mu+}\delta(x^-)\rho(\x_\perp)\; ,
\end{equation}
written here for a projectile moving in the $+z$ direction. The
function $\rho(\x_\perp)$ describes the distribution of color charge
as a function of the impact parameter. The slow partons, on the other
hand, have a non trivial dynamics during the collision, and must be
treated as gauge fields. The only coupling between the fast and slow
partons is a coupling $A_\mu J^\mu$ between the color current of the
fast partons and the gauge fields, which allows the fast partons to
radiate slower partons by bremsstrahlung. Because the configuration of
the fast partons prior to the collision is different in every
collision, the function $\rho(\x_\perp)$ must be a stochastic
quantity, for which one can only specify a distribution
$W_{_Y}[\rho]$. Observables like cross-sections must be averaged over
all the possible configurations of $\rho$ with this distribution. In
fact, in the CGC description, this averaging procedure is
equivalent to the target average of the scattering amplitude that was
introduced in the discussion of the BK equation,
\begin{equation}
\big<\cdots\big>\equiv\int \big[D\rho\big]\;W_{_Y}[\rho]\;\cdots\; .
\end{equation}
A crucial point is that the distribution $W_{_Y}[\rho]$ depends on
$Y$, the rapidity that separates what is considered fast and
slow. Because such a separation is arbitrary, physical quantities
cannot depend on it; one can derive from this requirement a
renormalization group equation for $W_{_Y}[\rho]$ -- known as the {\sl
JIMWLK equation}\cite{jimwlk} --, of the form~:
\begin{equation}
\frac{\partial W_{_Y}[\rho]}{\partial Y}={\cal H}[\rho]\;W_{_Y}[\rho]\; .
\label{eq:jimwlk}
\end{equation}
The {\sl JIMWLK Hamiltonian} ${\cal H}[\rho]$ contains first and
second derivatives with respect to the source $\rho$,
\begin{equation}
{\cal H}[\rho]
=\int_{\x_\perp}\sigma(\x_\perp)\frac{\delta}{\delta \rho(\x_\perp)}
+\frac{1}{2}\int_{\vec{\x}_\perp,\vec{\y}_\perp}
{\colorc \chi(\vec{\x}_\perp,\vec{\y}_\perp)}
\frac{\delta^2}{\delta {\colorb \rho(\vec{\x}_\perp)}\delta\colorb \rho(\vec{\y}_\perp)}\;,
\label{eq:H}
\end{equation}
where $\sigma(\x_\perp)$ and $\chi(\x_\perp,\y_\perp)$ are known
functionals of $\rho$. In fact, the JIMWLK equation is equivalent to
the infinite hierarchy of Balitsky's equations -- of which the BK is
an approximation that neglects some correlations. In the CGC
description of scattering processes, the energy dependence of
amplitudes arises from the $Y$ dependence of the distribution
$W_{_Y}[\rho]$. For instance, the dipole scattering amplitude would be
written as
\begin{equation}
\left<{\colorb {\bs T}(\vec\x_\perp,\vec\y_\perp)}\right>
=
\int\left[D\rho\right]\;
{\colorb W_{_Y}[\rho]}\;
\left[1-\frac{1}{N_c}{\rm tr}(U(\vec\x_\perp)U^\dagger(\vec\y_\perp))\right]
\; ,
\end{equation}
where the Wilson line $U$ is evaluated in the color field generated by
the configuration $\rho$ of the color sources. This formula is very
similar -- at least in spirit -- to the standard collinear
factorization in DIS. The functional $W_{_Y}[\rho]$ can be seen as an
extension of the usual concept of parton distribution, that contains
information about parton correlations beyond the mere number of
partons, while the square bracket is the analogue of the
``perturbative cross-section''.  This formula is a Leading Logarithm
(LL) factorization formula in the sense that it resums all the powers
$(\alpha_s Y)^n$. Moreover, it also resums all the rescattering
corrections, in $(Q_s/p_\perp)^p$, a feature which is not included in
collinear factorization.

Eq.~(\ref{eq:jimwlk}) predicts the energy dependence of the
distribution of sources. However, it must be supplemented by an
initial condition at some $Y_0$. As with the DGLAP equation, the
initial condition is non-perturbative, and one must in general model
it or guess it from experimental data. In the case of large nuclei,
one often uses the {\sl McLerran-Venugopalan model}, which assumes
that $W_{_{Y_0}}[\rho]$ is a Gaussian\cite{mv,Kovch1,JeonV}~:
\begin{equation}
{\colorb W_{_{Y_0}}[\rho]}
=
\exp\left[
  -\int d^2\vec\x_\perp \frac{{\colorb \rho(\vec\x_\perp)\rho(\vec\x_\perp)}}{2{\colora \mu^2(\vec\x_\perp)}}
\right]\; .
\end{equation}
The idea behind this model is that the color charge per unit area,
$\rho(\x_\perp)$, is the sum of the color charges of the partons that
sit at approximately the same impact parameter. In a large nucleus,
this will be the sum of a large number of random charges; for $N_c=3$,
this leads to a Gaussian distribution for $\rho$ plus a small (albeit
physically very relevant) contribution from the cubic
Casimir~\cite{JeonV}. The fact that this Gaussian has only
correlations local in impact parameter is a consequence of
confinement~: color charges separated by more than the nucleon size
cannot be correlated. The MV model is generally used at a moderately
small $x$, of the order of $10^{-2}$. If the problem under
consideration requires smaller values of $x$, one should use the BK or
JIMWLK equations, with the MV distribution as the initial condition.

\subsection{Analogies with reaction-diffusion processes}
There are interesting analogies between the evolution equations that
govern the energy dependence of scattering amplitude in QCD and simple
models of {\sl reaction-diffusion processes}\cite{MunieP}. The
simplest setting in which these correspondences can be seen is to
consider the dipole scattering amplitude off a large nucleus, and to
assume translation and rotation invariance in impact parameter
space. It is useful to define its Fourier transform as
\begin{equation}
{\colord N(Y,k_\perp)}\equiv
2\pi\int d^2\vec\x_\perp
\;e^{i\vec\k_\perp\cdot\vec\x_\perp}\;
\frac{\left<{\colorb {\bs T}(0,\vec\x_\perp)}\right>_{_Y}}{x_\perp^2}\; .
\end{equation}
(Note the factor $1/x_\perp^2$ included in this definition.) It turns
out that for this object $N$, the BK equation has a very simple
non-linear term,
\begin{equation}
      \frac{\partial {\colord N({\colora Y},k_\perp)}}
	   {\partial{\colora Y}} =\frac{{\colorc \alpha_s N_c}}{\pi}
	   \Big[ {\colorc \chi(-\partial_L)} {\colord N({\colora
	   Y},k_\perp)}-{\colord N^2({\colora Y},k_\perp)} \Big]\; .
\end{equation}
In this equation, $L\equiv {\colorb\ln(k_\perp^2/k_0^2)}$ and
${\colorc \chi(\gamma)}\equiv {\colorc
2\psi(1)-\psi(\gamma)-\psi(1-\gamma)}$ with $\psi(\z)\equiv
d\ln\Gamma(z)/dz$. The function $\chi(\gamma)$ has poles at $\gamma=0$
and $\gamma=1$, and a minimum at $\gamma=1/2$. By expanding it up to
quadratic order around its minimum, and by defining new variables,
    \begin{eqnarray}
      &&{ t}\sim {\colora Y}\nonumber\\
      &&{z}\sim L +\frac{{\colorc\alpha_s N_c}}{2\pi}
      \chi^{\prime\prime}(1/2)\;{\colora Y}\; ,
    \end{eqnarray}
the BK equation simplifies into
\begin{equation}
{\partial_t}{\colord N} 
  ={\partial_z^2}{\colord N} +{\colord N}-{\colord N}^2
\; ,
\end{equation}
known as the {\sl{\colora Fisher-Kolmogorov-Petrov-Piscounov} (FKPP)
equation}. This equation has been extensively studied in the
literature, because it is the simplest realization of the so-called
reaction-diffusion processes. It describes the evolution of a number
$N$ of objects that live in one spatial dimension. The diffusion term
${\partial_z^2}{\colord N}$ describes the fact that these entities can
hop from one location to neighboring locations. The positive linear
term $+N$ means that an object can split into two, and the negative
quadratic term $-N^2$ that two objects can merge into a single one.
One can easily check that this equation has two fixed points, $N=0$
which is unstable and $N=1$ which is stable.

\begin{figure}[htbp]
\begin{center}
\resizebox*{6cm}{!}{\includegraphics{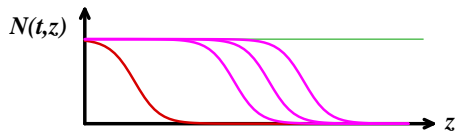}}
\end{center}
\caption{\label{fig:trav-wave}Travelling wave solutions of the FKPP
equation. Red~: initial condition. Magenta~: solution at equally
spaced  times.}
\end{figure}

An important property of this equation is that it admits {\sl
asymptotic travelling waves} as solutions. Let us assume that the
initial condition $N(t_0,z)$ goes to $1$ at $z\to -\infty$ and to $0$
at $z\to +\infty$, with an exponential tail
$N(t_0,z)\empile{\sim}\over{z\to+\infty} \exp(-\beta z)$. If the slope
of the exponential obeys $\beta> 1$, the solution at late time depends
only on a single variable,
  \begin{equation}
    {\colord N}(t,z)
    \empile{\sim}\over{t\to +\infty} 
	   {\colord N}(z-2t-\frac{3}{2}\ln(t))\; .
  \end{equation}
When $t\to+\infty$, the logarithm can be neglected in front of the
term linear in time, and one has a travelling wave moving at a
constant velocity $dz/dt=2$ without deformation (see figure
\ref{fig:trav-wave}). 
Moreover, this velocity is independent of the details of the initial
condition for a large class of initial conditions.

Going back to the dipole scattering amplitude, this result implies
the following scaling behavior at large $Y$~:
\begin{equation}
\left<{\colorb{\bs T}(0,\vec\x_\perp)}\right>_{_Y}
= {\colorb T}({\colorb Q_s(Y)}x_\perp)\; ,
\label{eq:scaling1}
\end{equation}
with a saturation scale of the form
\begin{equation}
    {\colorb Q_s^2(Y)}
    =k_0^2 
    \;{\colorb Y}^{-\delta}\;
    e^{\omega{\colorb Y}}\; .
\end{equation}
(The exponential comes from the constant in the velocity of the
travelling wave, and the power law correction comes from the
subleading logarithm.) 
\begin{figure}[htbp]
\begin{center}
\resizebox*{5.5cm}{!}{\includegraphics{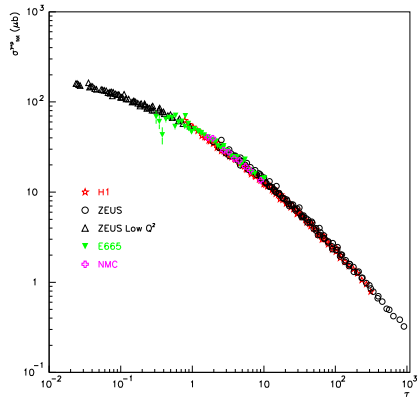}}
\end{center}
\caption{\label{fig:geom-scal}Photon-proton total cross-section
measured at HERA, displayed against $\tau\equiv Q^2/Q_s^2(Y)$.}
\end{figure}
This scaling property has an interesting phenomenological consequence
for the inclusive DIS cross-section, that one can express in terms of
the forward dipole scattering amplitude thanks to the optical
theorem~:
\begin{equation}
 {\colord \sigma_{\gamma^*p}}({\colora Y},{\colorc Q^2})= \sigma_0
      \int {\colord d^2\vec\x_\perp}
      \int_0^1 dz \left|\psi(z,{\colord x_\perp},{\colorc Q^2})\right|^2 
      \left<{\colorb{\bs T}}(0,{\colord\vec\x_\perp})\right>_{_Y}
      \; .
\end{equation}
In this formula, $\psi(z,{\colord x_\perp},{\colorc Q^2})$ is the
light-cone wave function for a photon of virtuality $Q^2$ that splits
into a quark-antiquark dipole of size $\x_\perp$, the quark carrying
the fraction $z$ of the longitudinal momentum of the photon. This
wavefunction can be calculated in QED, and its only property that we
need here is that it depends only on the combination
$[m^2+Q^2z^2(1-z)^2]\x_\perp^2$ where $m$ is the quark mass. If one
neglects the quark mass, then eq.~(\ref{eq:scaling1}) implies a simple
scaling for the $\gamma^* p$ cross-section itself~:
\begin{equation}
{\colord \sigma_{\gamma^*p}}({\colora Y},{\colorc Q^2})
=
{\colord \sigma_{\gamma^*p}}({\colorc Q^2}/Q_s^2(Y))\; .
\end{equation}
Such a {\sl geometrical scaling}\cite{geomscal} has been found in the
DIS experimental results\footnote{In addition to explaining
geometrical scaling, saturation inspired fits of DIS data are quite
successful at small $x$. See \cite{fits}.}, as shown in figure
\ref{fig:geom-scal}.  A comment is in order here; as the approach
based on collinear factorization and the DGLAP equation succeeds at
reproducing much of the inclusive DIS data, it certainly also
reproduces this scaling that is present in the data. However, this
approach does not provide an explanation for the scaling. It arises
via some fine tuning of the initial condition for the DGLAP
evolution. In contrast, in the Color Glass Condensate description of
DIS, this scaling is almost automatic.

\section{Lecture III~: Nucleus-nucleus collisions in the CGC framework}
\subsection{Introduction}
Up to now, we only considered DIS, in which a possibly saturated
proton or nucleus is probed by an elementary
object\footnote{Proton-nucleus collisions also belong to this
category. Examples of processes have been studied in\cite{pA}.}  -- a
virtual photon that has fluctuated into a quark-antiquark dipole. In
such a situation, the scattering amplitude can be written in closed
form as a product of Wilson lines, and its energy dependence can be
obtained either from Balitsky's equations or from the JIMWLK evolution
of the distribution of sources that produce the color field of the
proton. There are however interesting problems that involve two
densely occupied projectiles.
\begin{figure}[htbp]
\begin{center}
\hskip 5mm
\resizebox*{5.5cm}{!}{\includegraphics{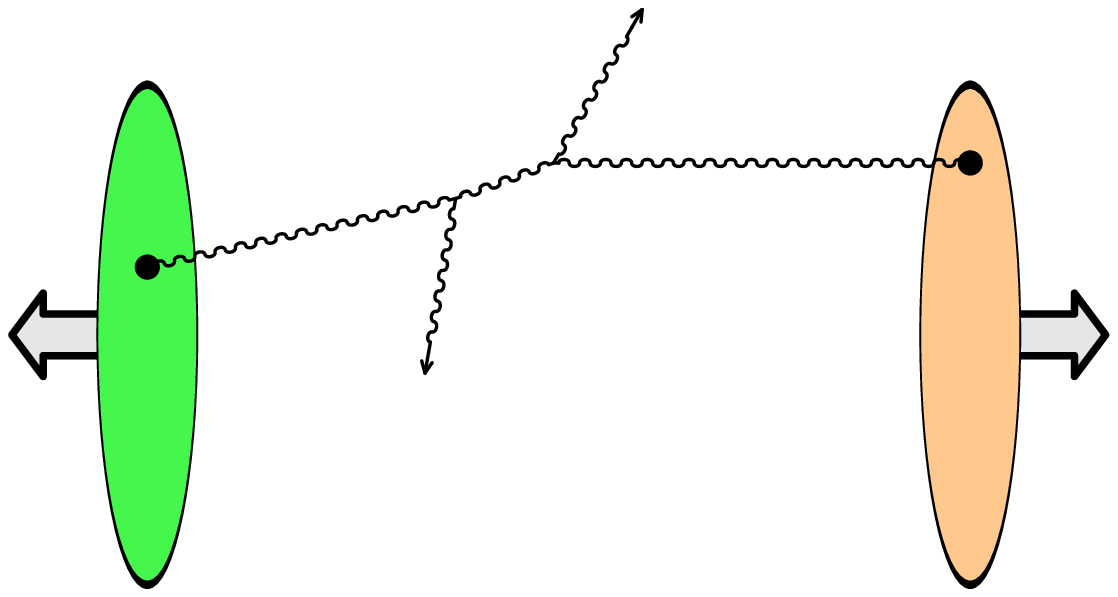}}
\hfill
\resizebox*{5.5cm}{!}{\includegraphics{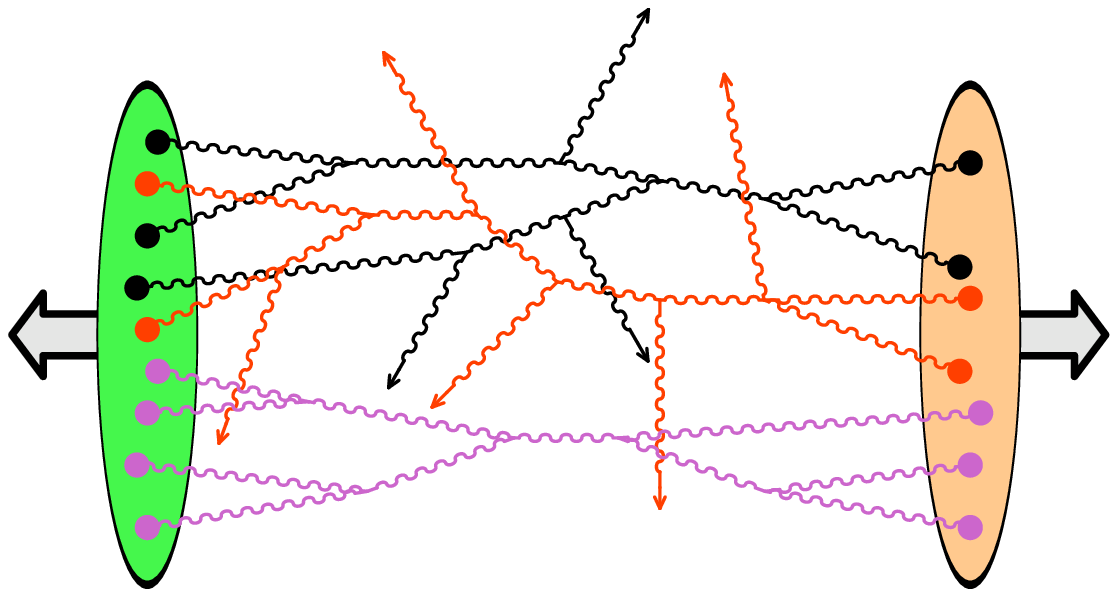}}
\hskip 5mm
\end{center}
\caption{\label{fig:AA}Typical contributions to gluon production in
hadronic collisions. The dots denote the color sources. Left: dilute
regime. Right: saturated regime.}
\end{figure}
The archetype of such a situation is a  high-energy nucleus-nucleus
collision. In these collisions, one of the main challenges is to
calculate the multiplicity of the particles (gluons at leading order)
that are produced at the impact of the two nuclei. In the Color Glass
Condensate framework, one has to couple the gauge fields to a current
that receives contributions from the color sources of the two
projectiles,
\begin{equation}
J^\mu=\delta^{\mu+}\delta(x^-)\rho_1(\x_\perp)
+\delta^{\mu-}\delta(x^+)\rho_2(\x_\perp)\; .
\label{eq:J}
\end{equation}
The fact that there are two strong sources leads to complications that
are two-fold~:
\begin{itemize}
\item there is no explicit formula that gives the multiplicity (or any
other observable) in terms of Wilson lines in the collision of two
saturated projectiles,
\item if one is interested by the particle spectrum at some rapidity
$Y$, one must evolve the two projectiles from their respective beam
rapidity to $Y$. The question of the factorization of the large
logarithms of $1/x$ is now much more complicated than in DIS.
\end{itemize}
The kind of complications one is facing in this problem is illustrated
in figure \ref{fig:AA}. In the saturated regime, reactions initiated
by more than one parton (color source in the CGC description) in each
projectile become important. Moreover, there can be a superposition of
many independent scatterings, that will appear as disconnected graphs.

\subsection{Power counting and bookkeeping}
In the saturated regime, the color density $\rho$ (represented by dots
in figure \ref{fig:AA}) is non-perturbatively large $\rho\sim
g^{-1}$. This is due to the fact that the occupation number,
proportional to $\big<\rho\rho\big>$, is of order $\alpha_s^{-1}$ in
this regime. Thus for a {\sl connected} graph, the order in $g$ is
given by
\begin{equation}
{\colord\frac{1}{g^2}}\; 
{\colord g}^{n_{\rm g}}\; 
{\colord g}^{2 n_{_L}}\; ,
\label{eq:counting}
\end{equation}
where $n_{\rm g}$ is the number of produced gluons and $n_{_L}$ the
number of loops. One can see that this formula is independent of the
number of sources $\rho$ attached to the graph. Indeed, since each
source brings a factor $g^{-1}$ and is attached at a vertex that
brings a factor $g$, each source counts as a factor $1$. If the
diagram under consideration is made of several disconnected subgraphs,
one should apply eq.~(\ref{eq:counting}) to each of them separately.

Among all the diagrams that appear in the calculation of particle
production, a special role is played by the so-called {\sl vacuum
diagrams} -- diagrams that have $n_{\rm g}=0$ external gluons. They
only connect sources of the two projectiles, and are thus
contributions to the vacuum-to-vacuum amplitude $\big<0{}_{\rm
out}\big|0{}_{\rm in}\big>$, hence their name. The order of connected
vacuum diagrams is $g^{2(n_{_L}-1)}$. An extremely useful property is
that the sum of all the vacuum diagrams (connected or not) is the
exponential of those that are connected (that we denote $iV[j]$ where
$j$ is the external current due to the color sources of the two
projectiles)
\begin{equation}
\sum \left({\colora{\rm all\ the\ vacuum}\atop{\rm diagrams}}\right)
=
\exp \left\{
\sum\; 
\Big(
{\colora{{\rm connected}\atop{\rm vacuum\ diagrams}}}
\Big)
\right\}
\equiv e^{i{\colorc V[j]}}\; .
\end{equation}
The reason why vacuum diagrams are important in our problem is that it
is possible to write all the time ordered products of fields -- that
enter in the reduction formulas for gluon production amplitudes -- as
derivatives of $\exp(iV[j])$
\begin{equation}
{\colorc\big<0_{\rm
      out}\big|}T{\colord A(x_1)}
  \cdots{\colord A(x_n)}{\colorc\big|0_{\rm in}\big>}
=
\frac{\delta}{i\delta{\colord j(x_1)}}
\cdots
\frac{\delta}{i\delta{\colord j(x_n)}}\;\;
e^{i{\colorc V[j]}}\; .
\end{equation}
Thanks to this property, one can write a very compact formula for the
probability $P_n$ of producing exactly $n$ gluons in the
collision\cite{GelisV2,GelisV3,GelisV4}, 
\begin{equation}
{\colora P_n}=\frac{1}{n!}
      \left.{\colorb {\cal D}^{\colora n}}\;e^{i{\colorc V[j_+]}}\;
        e^{-i{\colorc V^*[j_-]}}
        \right|_{j_+=j_-=j}\; ,
\label{eq:Pn}
\end{equation}
where the operator ${\cal D}$ is defined by\footnote{We are a bit
careless here with the Lorentz indices, polarization vectors, etc,
because our main goal is to highlight the general techniques for
keeping track of the diagrams that contribute to particle production
in the saturated regime.}
\begin{equation}
  \left\{
     \begin{aligned}
      &
      {\colorb{\cal D}}\equiv
      \int_{x,y}\; {\colorc G_{+-}^0(x,y)}\; 
      \square_x\square_y\;
      \frac{\delta}{\delta {\colord j_+(x)}}
      \frac{\delta}{\delta {\colord j_-(y)}}&&&\; ,
      \nonumber\\
      &
      {\colorc G_{+-}^0(x,y)}\equiv
      \int \frac{d^3\vec\p}{(2\pi)^32E_p}
      \; e^{ip\cdot(x-y)}&&&
      \; .
    \end{aligned}
    \right.
\label{eq:D}
\end{equation}
An important point to keep in mind about eq.~(\ref{eq:Pn}) is that the
external currents must be kept distinct in the amplitude and complex
conjugate amplitude until all the derivatives contained in ${\cal D}$
have been taken. Only then one is allowed to set $j_+$ and $j_-$ to
the physical value of the external current. The propagator $G_{+-}^0$,
that has only on-shell momentum modes, is the usual cut propagator
that appears in {\sl Cutkosky's cutting
rules}\cite{PeskiS1,Cutko1}. The operator ${\cal D}$ acts on cut
vacuum graphs by removing two sources (one on each side of the cut,
i.e. a $j_+$ and a $j_-$), and by connecting the points where they
were attached by the cut propagator $G_{+-}^0$.  In fact, since $P_n$
is obtained by acting $n$ times with the operator ${\cal D}$, it is
the sum of all the cut vacuum diagrams in which exactly $n$
propagators are cut. Eq.~(\ref{eq:Pn}) also makes obvious the fact
that the probabilities $P_n$ do not have a meaningful perturbative
expansion in the saturated regime, because the sum $iV[j]$ of the
connected vacuum diagrams starts at the order $g^{-2}$.

By summing eq.~(\ref{eq:Pn}) from $n=0$ to $\infty$ while keeping
$j_+$ and $j_-$ distinct, one obtains the sum of all the cut vacuum
diagrams with the current $j_+$ in the amplitude and $j_-$ in the
complex conjugate amplitude to be
\begin{eqnarray}
    \sum\left({\colora{\rm all\ the\ {\colore cut}}\atop{\rm vacuum\ diagrams}}\right)
    \;\;=\;\;
    e^{\colord{\cal D}}\; e^{i{\colorb V[j_+]}}\; e^{-i{\colorb V^*[j_-]}}
    \; .
\label{eq:AGK}
\end{eqnarray}
When we set $j_+=j_-$, this sum becomes $\sum_n P_n$, and therefore it
should be equal to 1 because of unitarity. Eq.~(\ref{eq:Pn}) is very
useful, because it allows to replace infinite sets of Feynman diagrams
by simple algebraic equations. Similarly, the fact that
eq.~(\ref{eq:AGK}) is 1 when $j_+=j_-$ corresponds to a cancellation
of an infinite set of graphs\footnote{This cancellation is closely
related to the {\sl Abramovsky-Gribov-Kancheli
cancellation}\cite{AbramGK1}.}, that would be very difficult to see at
the level of diagrams.

\subsection{Inclusive gluon spectrum}
\begin{figure}[htbp]
\centerline{
\hskip 10mm
\resizebox*{4.5cm}{!}{\includegraphics{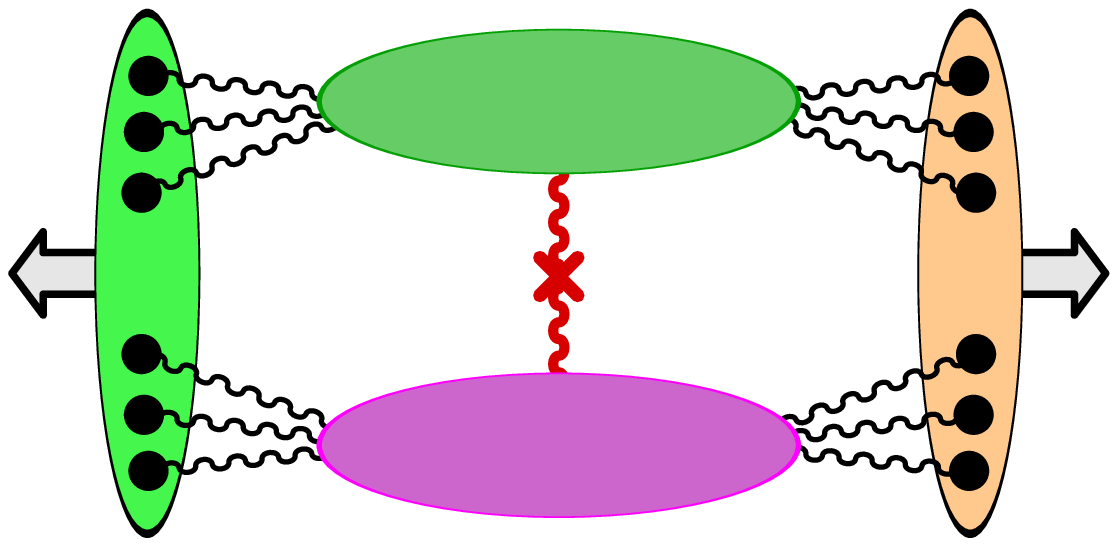}}
\hfill
\resizebox*{4.5cm}{!}{\includegraphics{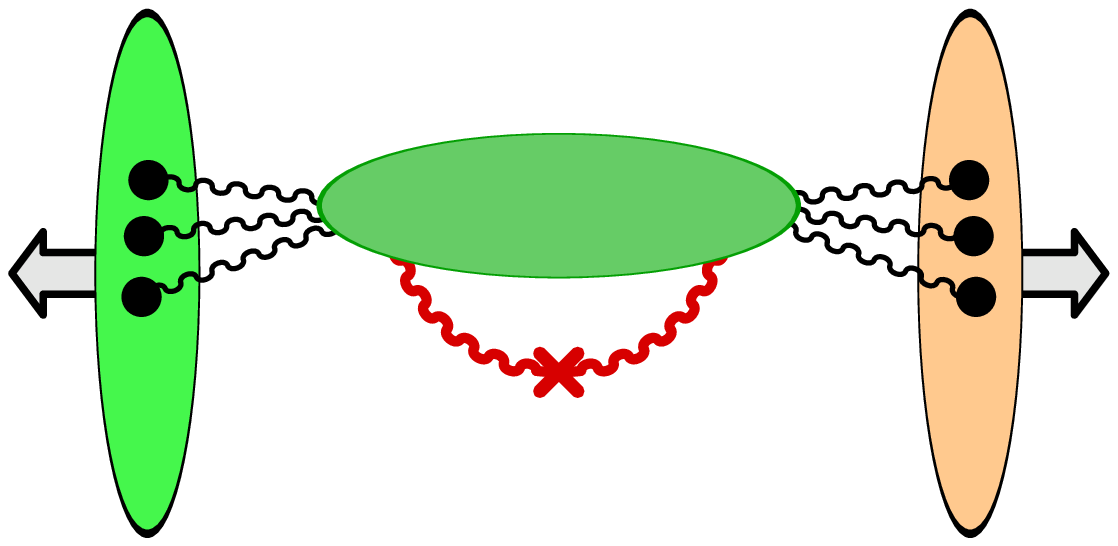}}
\hskip 10mm
}
\caption{\label{fig:nbar}The two topologies contributing to the
average gluon multiplicity $\overline{N}$. In each blob, one must sum
over all the possible ways of cutting the propagators.}
\end{figure}

Eq.~(\ref{eq:Pn}) leads to compact formulas for moments of the
distribution of produced particles. The first moment -- the average
multiplicity -- reads\cite{GelisV2}
\begin{eqnarray}
  {\colorc\overline{N}}\ =\ \sum_{n=0}^\infty n\;{\colorc P_n}
  &=&
  {\colord{\cal D}} \left\{\;e^{\colord{\cal D}}\; 
  e^{i{\colorb V[j_+]}}\; e^{-i{\colorb V^*[j_-]}}\right\}_{j_+=j_-=j}
  \; .
\end{eqnarray}
With the help of eq.~(\ref{eq:AGK}), this formula tells us that
$\overline{N}$ is given by the action of the operator ${\cal D}$ on
the sum of all the cut vacuum diagrams. In plain english, this
translates into~: take a cut vacuum diagram (connected or not), remove
a source on each side of the cut, and put a cut propagator where the
sources were attached. Depending on whether the cut vacuum diagram one
starts from is connected or not, one gets two different topologies,
displayed in figure \ref{fig:nbar}.
Each of the blobs in these diagrams can be any connected graph, and
must be cut in all the possible ways\footnote{Note that by not
performing the $d^3\p$ integration contained in the explicit cut
propagator, one obtains the {\sl inclusive gluon spectrum}
$d\overline{N}/d^3\p$ instead of the integrated multiplicity.}. Thus,
only connected graphs contribute to the multiplicity.

An important point is that, even though the perturbative expansion for
the $P_n$ is not well defined, the multiplicity (and more generally
any moment of the distribution $P_n$) can be organized in a sensible
perturbative series\footnote{The fact that this is possible for
$\overline{N}$ but not for the $P_n$'s themselves is due to the fact
that the only graphs that contribute to $\overline{N}$ are
connected. This is a consequence of the AGK cancellation.}. The
Leading Order is obtained by keeping only the leading order vacuum
graphs, i.e. those that have no loops~:\setbox1\hbox to
  4cm{\hfil
\begin{feynman}{1.5cm}
\diagram{10}{
  -2.5 2.4 translate
  1 setlinejoin
  [] 0 setdash
  1 setlinecap
  0 setgray
  /arw {
    /ll exch def
    /yy exch def
    /xx exch def
    0.9 setgray
    xx yy moveto
    0 0.2 rlineto
    ll 0 rlineto
    0 0.2 rlineto
    ll 0.3 mul -0.4 rlineto
    ll 0.3 mul neg -0.4 rlineto
    0 0.2 rlineto
    ll neg 0 rlineto
    0 0.2 rlineto fill
    0 setgray
    0.07 setlinewidth
    xx yy moveto
    0 0.2 rlineto
    ll 0 rlineto
    0 0.2 rlineto
    ll 0.3 mul -0.4 rlineto
    ll 0.3 mul neg -0.4 rlineto
    0 0.2 rlineto
    ll neg 0 rlineto
    0 0.2 rlineto
    stroke
  } def
  -1 0.35 -1 arw
  gsave
  -1 0.35 translate
  0.2 1 scale
  0.2713 0.9684 0.2986 setrgbcolor
  0 0 2.5 0 360 arc fill
  0 setgray
  0.08 setlinewidth
  0 0 2.5 0 360 arc stroke
  grestore
  6.9 0.35 +1 arw
  gsave
  6.9 0.35 translate
  0.2 1 scale
  1.0000 0.7901 0.5563 setrgbcolor
  0 0 2.5 0 360 arc fill
  0 setgray
  0.08 setlinewidth
  0 0 2.5 0 360 arc stroke
  grestore
  gsave
  2.95 2 translate
  1.6 1.6 scale
  -1 0 1.6 200 beglphoton blob pop pop
  -1 0 1.5 187 beglphoton blob pop pop
  -1 0 1.47 174 beglphoton blob pop pop
  1 0 1.6 -20 beglphoton blob pop pop
  1 0 1.5 -7 beglphoton blob pop pop
  1 0 1.47 +6 beglphoton blob pop pop
  /Width 0.06 def colorb
  0 0 1.05 -90 beglphoton 
  2 copy -90 bigcross
  1.05 -90 beglphoton
  /Width 0.03 def fgcolor
  1 1.6 div 0.3 1.6 div scale
  graya
  0 0 2.3 0 360 arc fill
  colora
  0.05 setlinewidth
  0 0 2.3 0 360 arc stroke
  grestore
  gsave
  2.95 -2 0.7 add translate
  1.6 1.6 scale
  -1 0 1.6 160 beglphoton blob pop pop
  -1 0 1.5 173 beglphoton blob pop pop
  -1 0 1.47 186 beglphoton blob pop pop
  1 0 1.6 20 beglphoton blob pop pop
  1 0 1.5 7 beglphoton blob pop pop
  1 0 1.47 -6 beglphoton blob pop pop
  1 1.6 div 0.3 1.6 div scale
  grayc
  0 0 2.3 0 360 arc fill
  colorc
  0.05 setlinewidth
  0 0 2.3 0 360 arc stroke
  grestore
  0.8 tb
  2.2 1.8 moveto (tree) show
  2.2 -1.55 moveto (tree) show
}
\end{feynman}
\hfil}
\begin{equation}
{\colorc\overline{N}}_{_{\colorb LO}}=
\sum_{{\colora\rm trees}}\;\;\sum_{{\colora\rm cuts}}\raise -9mm\box1\; .
\label{eq:N_LO}
\end{equation}
Thus $\overline{N}$ starts at the order $g^{-2}$. In
eq.~(\ref{eq:N_LO}), for each tree diagram, one must sum over all the
possible ways of cutting its lines. The simplest way of doing this is
to use Cutkosky's rules~:
\begin{itemize}
  \item assign $+$ or $-$ labels to each vertex and source of the
  graph, in all the possible ways (there are $2^n$ terms for a graphs
  with $n$ vertices and sources). A $+$ vertex has a coupling $-ig$
  and a $-$ vertex has a coupling $+ig$,
  \item the propagators depend on which type of labels they
  connect. In momentum space, they read~:
    \begin{eqnarray}
      && G_{++}^0(p)=i/(p^2+i\epsilon)\qquad(\mbox{standard Feynman
      propagator}) \nonumber\\ &&
      G_{--}^0(p)=-i/(p^2-i\epsilon)\qquad(\mbox{complex conjugate
      of } G_{++}^0(p))
      \nonumber\\
      &&
      G_{+-}^0(p)=2\pi\theta(-p^0)\delta(p^2)
      \nonumber\\
      &&
      G_{-+}^0(p)=2\pi\theta(p^0)\delta(p^2)
      \; .
    \end{eqnarray}
\end{itemize}
A quick analysis shows that, when one sets $j_+=j_-$, summing over the
$\pm$ labels at each vertex produces combinations of
propagators,
\begin{eqnarray}
&&
G_{++}^0(p)-G_{+-}^0(p)= {G_{_R}^0(p)}
\nonumber\\
&&
G_{-+}^0(p)-G_{--}^0(p)= G_{_R}^0(p)\; ,
\end{eqnarray}
where $G_{_R}^0(p)$ is the {\sl retarded propagator}\footnote{In
momentum space, $G_{_R}^0(p)=i/(p^2+i\,{\rm
sign}(p_0)\,\epsilon)$. Therefore, in coordinate space, it is
proportional to $\theta(x^0-y^0)$, hence its name.}.  Thus, for a
given tree graph, doing the sum over the cuts simply amounts to
replacing all its propagators by retarded propagators. The last step
is to perform the sum over all the trees. It is a well known result
that the sum of all the tree diagrams that end at a point $x$ is a
solution of the classical equations of motion of the field theory
under consideration. In our case, this sum is a color field ${\cal
A}^\mu(x)$ that obeys the Yang-Mills equations
\begin{equation}
\left[{\cal D}_\mu,{\cal F}^{\mu\nu}\right]=J^\nu\; ,
\end{equation}
where $J^\nu$ is the color current associated to the sources
$\rho_{1,2}$ that represent the incoming projectiles (see
eq.~(\ref{eq:J})). The boundary conditions obeyed by ${\cal A}^\mu(x)$
depend on the nature of the propagators that entered in the sum of
tree diagrams. When these propagators are all retarded, one gets a
retarded solution of the Yang-Mills equations, that vanishes in the
remote past, $\lim_{x_0\to -\infty}{\cal A}^\mu(x)=0$. The precise
formula for the gluon spectrum in terms of this solution of the
Yang-Mills equations reads
\begin{equation}
\frac{d{{\colora\overline{N}}_{_{LO}}}}{dY d^2\vec\p_\perp}=
\frac{1}{16\pi^3}\int d^4x d^4y\; {\colorc e^{ip\cdot (x-y)}}\;
\square_x\square_y\;
\sum_\lambda \epsilon^\mu_\lambda \epsilon^\nu_\lambda\;\;
{\colord {\cal A}_\mu(x)}{ {\cal A}_\nu(y)}
\; .
\label{eq:spectrum}
\end{equation}
Note that, although the integrations over $x$ and $y$ look
4-dimensional, they can be rewritten as 3-dimensional integrals
evaluated at $x_0\to+\infty$, thanks to the identity
	\begin{equation}
	  \int d^4x \; e^{ip\cdot x}\;\square_x\,
          {\colord{\cal A}_\mu(x) } = \lim_{x^0\to +\infty}
	  \int d^3\vec\x \;e^{ip\cdot x}\; 
	  \left[\partial_{0}-iE_p\right]
	  {\colord{\cal A}_\mu(x) }\; .
	\end{equation}
Solving the Yang-Mills equations is an easy problem in the case of a
single source $\rho$, but turns out to be very challenging when there
are two sources moving in opposite directions. The Schwinger gauge,
defined by the constraint ${\cal A}^\tau\equiv x^+{\cal A}^-+x^-{\cal
A}^-=0$, is quite useful because it alleviates the need to ensure that
the current $J^\nu$ is covariantly conserved\footnote{In general
gauges, one has to enforce the condition $\big[{\cal
D}_\mu,J^\mu\big]=0$ (this is a consequence of Jacobi's identity for
commutators). Because this relation involves a covariant derivative
rather than an ordinary derivative, the radiated field leads to
modifications of the current itself.}. In this gauge, ${\cal A}^+=0$
where $J^-\not=0$ and conversely, which makes this condition
trivial. Moreover, in this gauge, one can find the value of the gauge
field on a time-like surface just above the light-cone (at a proper
time $\tau=0^+$) simply by matching the singularities across the
light-cone.  These initial conditions\cite{KovneMW2} can be
written as\footnote{An interesting feature of the gauge fields at
early times after the collision -- a phase recently named ``glasma''
-- is that the chromo-electric and magnetic fields are purely
longitudinal, while they were transverse to the beam axis just before
the collision\cite{LappiM1}.}
\begin{eqnarray}
&&
{\colorb {\cal A}^i(\tau=0,\vec\x_\perp)}=
{\colord {\cal A}^i_1(\vec\x_\perp)}+{\colord {\cal A}^i_2(\vec\x_\perp)}
\nonumber\\
&&{\colorb {\cal A}^\eta(\tau=0,\vec\x_\perp)}
=
-\frac{ig}{2}\left[
{\colord {\cal A}^i_1(\vec\x_\perp)}\;,
\;{\colord {\cal A}^i_2(\vec\x_\perp)}
\right]
\nonumber\\
&&{\colorb {\cal A}^\tau}=0 \quad\mbox{(gauge\ condition)}
\; ,
\label{eq:init-cond}
\end{eqnarray}
where ${\cal A}^\eta\equiv \tau^{-2}(x^-{\cal A}^+-x^={\cal A}^-)$. In
this formula, ${\cal A}_{1}^i(\x_\perp)$ and ${\cal
A}_{2}^i(\x_\perp)$ are the gauge fields created by each
nucleus%
\setcounter{footnote}{0}\footnote{Because retarded solutions are causal, the field
below the light-cone cannot depend simultaneously on $\rho_1$ and
$\rho_2$.}  below the light-cone~:
\begin{eqnarray}
&&
{\colorb {\cal A}^i_1} = \frac{i}{g}
{\colord U_1(\vec\x_\perp)}\partial^i {\colord U^\dagger_1(\vec\x_\perp)}
\quad,\qquad {\colord U_1(\vec\x_\perp)}=T_+ \,\exp\; ig \int dx^+ T^a \frac{1}{{\bs\nabla}_\perp^2} {\colord\rho^a_1(x^+,\vec\x_\perp)}
\nonumber\\
&&
{\colorb {\cal A}^i_2} = \frac{i}{g}
{\colord U_2(\vec\x_\perp)}\partial^i {\colord U^\dagger_2(\vec\x_\perp)}
\quad,\qquad {\colord U_2(\vec\x_\perp)}=T_- \,\exp\; ig \int dx^- T^a \frac{1}{{\bs\nabla}_\perp^2} {\colord\rho^a_2(x^-,\vec\x_\perp)}\; .
\nonumber\\
&&
\end{eqnarray}
Therefore, the problem of solving the Yang-Mills equations from
$x_0=-\infty$ to $x_0=+\infty$ is reduced to solving them in the
forward light-cone from a known initial condition\footnote{Note that
at $\tau>0$, the YM equations are the vacuum ones, since all the
sources are located on the light-cone.}.

\begin{figure}[htbp]
\begin{center}
\resizebox*{6cm}{!}{\includegraphics{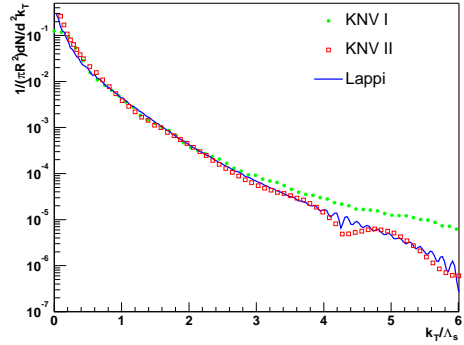}}
\end{center}
\caption{\label{fig:gluon-spectrum}The gluon spectrum at leading order
in the CGC framework.}
\end{figure}

Since our problem is invariant under boosts in the $z$ direction, one
can completely eliminate the space-time rapidity $\eta$ from the
equations of motion (and the initial conditions in
eq.~(\ref{eq:init-cond}) are also $\eta$-independent).  Thus, in the
forward light-cone, one has to solve numerically\cite{classglue}
equations of motion in time and two spatial dimensions, and then to
evaluate eq.~(\ref{eq:spectrum}). The result of this computation is
displayed in figure \ref{fig:gluon-spectrum}.
In this computation, the MV model was used as the distribution of the
sources $\rho_1$ and $\rho_2$. Therefore, the dependence of the
spectrum on the momentum rapidity $Y$ of the produced gluon cannot be
obtained in this calculation, and only the $k_\perp$ dependence is
shown. The main effect of gluon recombinations on this spectrum is
that it reduces the yield at low transverse momentum, $k_\perp\lesssim
Q_s$. Indeed, in a fixed order calculation in perturbative QCD, the
spectrum would behave as $k_\perp^{-4}$. In the CGC picture, the
singularity of the spectrum at low $k_\perp$ is only
logarithmic\footnote{If the final Fourier decomposition is performed
at a finite time $\tau$, the spectrum is completely regular when
$k_\perp\to 0$.}, and is therefore integrable.

\subsection{Inclusive quark spectrum}
\begin{figure}[htbp]
\centerline{
\hskip 10mm
\resizebox*{4.5cm}{!}{\includegraphics{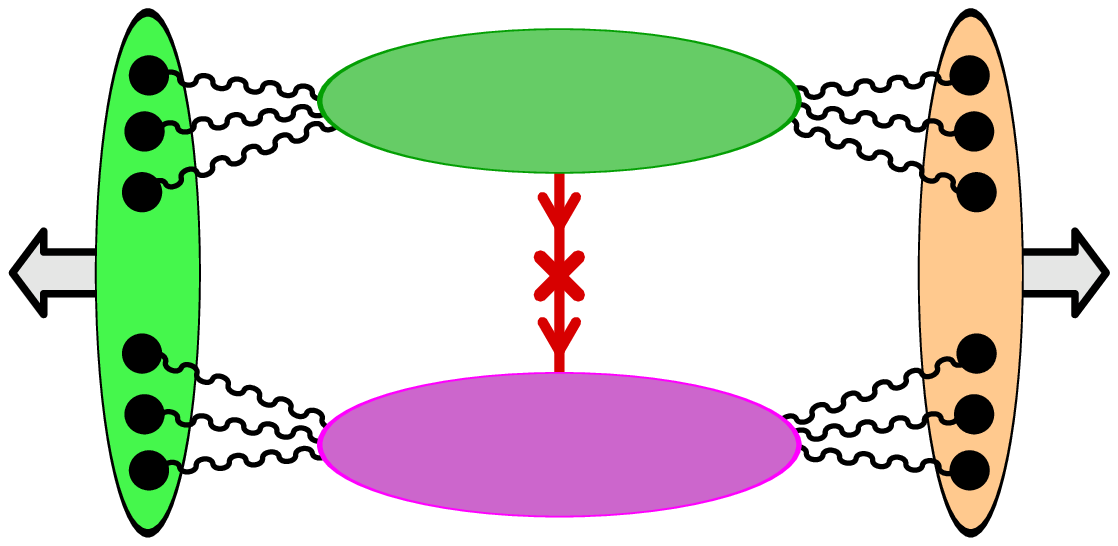}}
\hfill
\resizebox*{4.5cm}{!}{\includegraphics{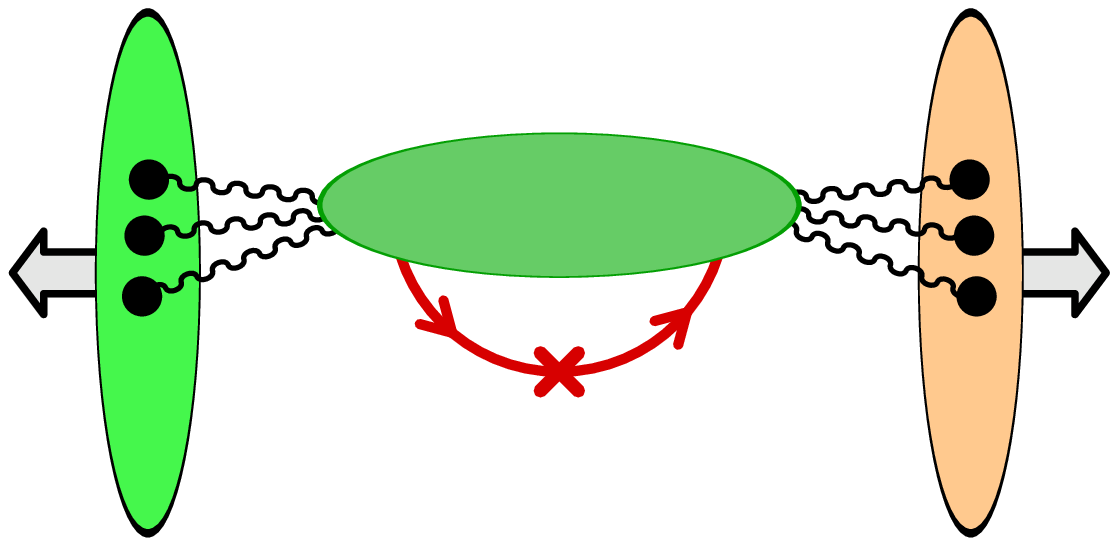}}
\hskip 10mm
}
\caption{\label{fig:nbar_q}Topologies corresponding to eq.~(\ref{eq:nbar_q}).}
\end{figure}

A similar study has also been performed for the initial production of
quarks in nucleus-nucleus collisions\cite{quarks}. The starting point
is to construct for quarks an operator ${\cal D}_q$ that plays the
same role as the operator ${\cal D}$ defined in eq.~(\ref{eq:D})~:
\begin{equation}
{\cal D}_q\equiv \int_{x,y} S^0_{+-}(x,y)\;\slpartial_x\slpartial_y
\frac{\delta}{\delta \eta_+(x)}
\frac{\delta}{\delta \overline{\eta}_-(y)}\; ,
\end{equation}
where $S^0_{+-}(x,y)$ is the free cut fermionic propagator and where $\eta$
is a Grassmanian current that couples to the spinors. In terms of this
operator, the probability of producing $n$ quarks is given by~:
\begin{equation}
P_n^{(q)}=\frac{1}{n!}{\cal D}_q^n
\;
\left.
e^{\cal D}\;
 e^{iV[j_+,\eta_+]}
\;
e^{-iV^*[j_-,\eta_-]}
\right|_{{j_+=j_-=j}\atop{\eta_+=\eta_-=0}}\; .
\label{eq:Pnq}
\end{equation} The first thing to note is that now the
connected vacuum diagrams, whose sum is $iV$, depend on both the
source $j$ and on the source $\eta$. However, the latter is set to
zero at the end of the calculation, because in the CGC one assumes
that the color sources in the wavefunction of the projectiles couple
only to the gluons. Therefore, the source $\eta$ serves only as an
intermediate bookkeeping device. Another important point in this
formula is the presence of the factor $\exp({\cal D})$. This factor
means that we are calculating an {\sl inclusive probability}, for
producing exactly $n$ quarks possibly accompanied by an arbitrary
number of gluons\footnote{Without this factor, we would be calculating
the probability of producing $n$ quarks and $0$ gluons. Note that in
principle, we should also modify our definition of the probability of
producing $n$ gluons by a factor $\exp({\cal D}_q)$. However, the
quarks are a subleading correction compared to the gluons, and this
change would not affect the gluon spectrum at leading order.}. In
practice, this fact means that one must sum over all the possible ways
of cutting the gluons lines in the diagrams that contribute to quark
production. From eq.~(\ref{eq:Pnq}), one obtains the following formula
for the average number of produced quarks
\begin{equation}
\overline{N}_q={\cal D}_q\; 
\left.
\underline{e^{{\cal D}_q}\;
e^{\cal D}\;
 e^{iV[j_+,\eta_+]}
\;
e^{-iV^*[j_-,\eta_-]}}
\right|_{{j_+=j_-=j}\atop{\eta_+=\eta_-=0}}\; .
\label{eq:nbar_q}
\end{equation}
In this formula, the underlined factors represent the sum of all
(connected or not) the cut vacuum diagrams made of quarks and gluons,
with sources $j_+,\eta_+$ on one side of the cut, and sources
$j_-,\eta_-$ on the other side. Acting on a term of this sum with
${\cal D}_q$ removes a source $\eta_+$ and a source $\eta_-$, and
connect the points where these sources were attached by a cut fermion
propagator. Diagrammatically, this corresponds to the two topologies
displayed in figure \ref{fig:nbar_q}.
Note however that the topology that appears on the left of figure
\ref{fig:nbar_q} cannot exist because it has a quark line which is not
closed onto itself (this is forbidden since we set the fermionic
sources $\eta_\pm$ to zero at the end of this calculation). Thus, we
only have the second family of diagrams, that have at least one
loop. This means that the average number of quarks is of order $g^0$,
compared to the number of gluons which is of order $g^{-2}$.

The leading contribution to the quark multiplicity is obtained by
including only tree diagrams in the blob. Thus, we have to sum all the
graphs that have one quark loop (with an explicit cut on it) and any
number of gluonic trees attached to it, and all the cuts thereof.  The
sum of all the gluonic trees and their cuts has already been
encountered in the computation of the gluon multiplicity~: it is equal
to the retarded solution ${\cal A}^\mu(x)$ of the Yang-Mills equations
that vanish in the remote past. Therefore, the quark spectrum is given
by
\begin{equation}
    \frac{d{\colora\overline{N}_{\rm q}}}{dYd^2\vec\p_\perp}
    =
    \frac{1}{16\pi^3}
    \int_{x,y}e^{ip\cdot x}\,
    {\colorc\overline{u}(\vec\p)}\,\stackrel{\rightarrow}{\slpartial}_x\,
    {\colorb S_{+-}(x,y)}\,
    \stackrel{\leftarrow}{\slpartial}_y
    \,{\colorc u(\vec\p)}\,e^{-ip\cdot y}\; ,
\label{eq:nbar_q2}
\end{equation}
where $S_{+-}$ is the cut quark propagator on which the retarded
classical field ${\cal A}^\mu$ has been resummed. This resummed
propagator can be obtained as the solution of the equation
    \begin{equation}
      {\colord S_{\epsilon\epsilon^\prime}(x,y)} = {\colorc
      S^0_{\epsilon\epsilon^\prime}(x,y)} -i{\colorc
      g}\sum_{{\colorc\eta=\pm}}
      (-1)^\eta\int d^4z \,{\colorc S^0_{\epsilon\eta}(x,z)}
      {\colorb{\cal A}_\mu(z)}\gamma^\mu
      {\colord S_{\eta\epsilon^\prime}(z,y)}\; ,
    \end{equation}
where $\epsilon,\epsilon^\prime=\pm$ (we need only the combination
$\epsilon=+,\epsilon^\prime=-$ in eq.~(\ref{eq:nbar_q2}), but the four
terms get mixed when one resums the background field).  It is possible
to decouple these equations by performing a ``rotation'' on the
$\epsilon,\epsilon^\prime$ indices\cite{BaltzgMP1},
    \begin{align}
      &{\colorc S_{\epsilon\epsilon^\prime}}&\to&&&
      {\colorc{\bs S}_{\alpha\beta}}
      \equiv\sum_{\epsilon,\epsilon^\prime=\pm}
      U_{\alpha\epsilon}U_{\beta \epsilon^\prime}
      {\colorc S_{\epsilon\epsilon^\prime}}&&
      \nonumber\\
      &(-1)^\epsilon \delta_{\epsilon\epsilon^\prime}
      &\to&&&
	  {\bs \Sigma}_{\alpha\beta}\equiv\sum_{\epsilon=\pm}U_{\alpha \epsilon}U_{\beta \epsilon}(-1)^\epsilon&&
      \; ,
      \label{eq:rotation}
    \end{align}
\begin{equation}
\mbox{with}\qquad
U=\frac{1}{\sqrt{2}}\begin{pmatrix}1 & -1 \cr 1 & 1
\cr\end{pmatrix}\; .
\end{equation}
After this rotation, the propagator matrix becomes triangular,
    \begin{equation}
      {\colorc{\bs S}_{\alpha\beta}}=
      \begin{pmatrix}
        0 & {\colorc S_{_A}} \cr
        {\colorc S_{_R}} & {\colorc S_{_D}} \cr
      \end{pmatrix}
      \quad,\qquad
      {\bs \Sigma}_{\alpha\beta}=
      \begin{pmatrix}
        0 & 1 \cr 1 & 0 \cr
      \end{pmatrix}
    \end{equation}
with $S_{_R}$ and $S_{_A}$ the resummed retarded and advanced
propagators and where ${\colorc S_{_D}^0(p)}=2\pi
\slp\delta(p^2)$. The main simplification comes from the fact that the
product of the free matrix propagator and of ${\bs\Sigma}$ is the sum
of a diagonal and a nilpotent matrix, which makes the calculation of
its $n$-th power very easy\footnote{Indeed, with very formal
notations, the resummed matrix propagator is
\begin{equation*}
{\bs S}={\bs S}^0\sum_{n=0}^\infty (-ig{\cal A})^n\Big[{\bs\Sigma}{\bs S}^0\Big]^n\; .
\end{equation*}}. In particular, one finds that the equations
that lead to the retarded (and also the advanced) propagator do not
mix with anything else,
    \begin{equation}
      {\colord S_{_{R}}(x,y)}={\colorc S^0_{_{R}}(x,y)}
      -i\,{\colorc g}\int d^4z\,{\colorc S^0_{_{R}}(x,z)}{\colorb{\cal A}_\mu(z)}\gamma^\mu{\colord S_{_{R}}(z,y)}\; ,
    \end{equation}
and that the resummed $S_{_D}$ can be expressed in terms of
$S_{_{R,A}}$ as\footnote{The $*$ symbol denotes the convolution of
2-point functions: $(A*B)(x,y)=\int_z A(x,z)B(z,y)$.}
    \begin{equation}
      {\colord S_{_D}}={\colord S_{_R}}*{\colorc S_{_R}^0{}^{-1}}*{\colorc S_{_D}^0 }*{\colorc S_{_A}^0{}^{-1}}*{\colord S_{_A}}\; .
    \end{equation}
At this point, one must invert the rotation done in
eq.~(\ref{eq:rotation}) in order to obtain $S_{+-}$ which is needed in
the formula for the quark spectrum. This gives the quark spectrum in
terms of retarded quantities,
    \begin{equation*}
      \frac{d{\colora \overline{N}_{\rm q}}}{dY d^2\vec\p_\perp}
        =\frac{1}{16\pi^3}
        \int\frac{d^3\vec\q}{(2\pi)^3 2E_q}
        \Big|
          {\cal T}_{_R}(\vec\p,\vec\q)
        \Big|^2\; ,
    \end{equation*}
where ${\cal T}_{_R}$ is the ``scattering part'' of the retarded
propagator, related to $S_{_R}$ by
\begin{equation}
S_{_R}=S_{_R}^0+S_{_R}^0 * {\cal T}_{_R} * S_{_R}^0\; .
\end{equation}
The last step is to write this object in terms of retarded solutions
of the Dirac equation in the background field ${\cal A}^\mu$. It is
easy to check that
\begin{eqnarray}
&&{\cal T}_{_R}(\vec\p,\vec\q)
=\lim_{x^0\to+\infty}
\int d^3{\vec{\x}}\; {\colora e^{i p\cdot x}\; u^\dagger(\vec\p)}
{\colorb \psi_{_\q}(x)}\nonumber\\
&&(i\slpartial_x-g\,{\colord \slcalA(x)}){\colorb\psi_{_\q}(x)}=0\; ,\;\;
{\colorb\psi_{_\q}(x^0,\vec{\x})}\empile{=}\over{x^0\to-\infty} {v(\vec{\q})}
e^{iq\cdot x}\; .
\label{eq:proj}
\end{eqnarray}
In this formula, $u(\p)$ and $v(\q)$ are the usual free spinors. Note
that the initial condition for the Dirac equation is a {\sl negative
energy spinor}, and that the projection performed at the final time is
with a positive energy spinor. In the vacuum, there would be no
overlap between these spinors. However, since in our problem the
spinor travels on top of a time-dependent background field, it
acquires positive energy modes which make ${\cal T}_{_R}$ non zero.

\begin{figure}[htbp]
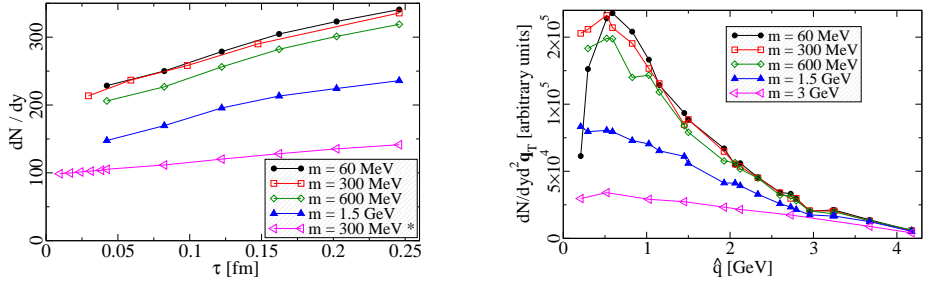

\begin{center}
\hskip 5mm
\resizebox*{5.5cm}{!}{\includegraphics{AA-taudep2gevplusbp.ps}}
\hfill
\resizebox*{5.5cm}{!}{\includegraphics{AA-pspectqs20p.ps}}
\hskip 5mm
\end{center}
\caption{\label{fig:quarks}Numerical results on quark production from
the CGC. Left: time evolution of the quark yield. Right: quark
$k_\perp$ spectra for different masses.}
\end{figure}
This formulation of quark production in the CGC framework has been
implemented numerically, also with the MV model for the average over
the configurations of the color sources $\rho_{1,2}$. Similarly to
what happened with the gluons, one can obtain analytically the value
of the spinors just above the light-cone. Hence, the numerical
resolution of the Dirac equation is only needed in the forward
light-cone. However, there is a major difference compared to the
gluons at LO~: even though the background color field does not depend
on rapidity, this is not true of the solutions of Dirac
equation\footnote{This has nothing to do with the fact that we are
considering fermions, but rather with the quark spectrum being a NLO
quantity -- that involves a loop in the background of the classical
field.}. Indeed, the momentum $\q$ in the initial condition renders
the spinors dependent on the space-time rapidity $\eta$ (the boost
invariance of the background field implies that the spinors depend
only on the difference $\eta-y_\q$ where $y_\q$ is the rapidity of the
momentum $\q$). This difference makes the computation of the quark
spectrum much more computationally intensive relative to that of the
gluon spectrum, because one has to keep the three dimensions of
space. Some of the results obtained are displayed in figure
\ref{fig:quarks}. On the left plot is shown the time dependence of the
quark yield, for different quark masses (i.e. the yield obtained by
performing the projection in eq.~(\ref{eq:proj}) at a finite time
instead of taking the limit $x_0\to+\infty$). One can see that a good
fraction of the quarks are produced at $\tau=0$, when the two nuclei
pass through each other\footnote{In the analogous QED problem of
$e^+e^-$ production in the high-energy collision of two electrical
charges, all the electrons are produced at $\tau=0$ and their number
does not change at $\tau>0$. This is because in QED, the
electro-magnetic potential in the forward light-cone is a pure gauge,
that could be made to vanish by a gauge transformation.} and that the
number slightly increases in time afterwards due to the color field
present in the forward light-cone. The right panel of figure
\ref{fig:quarks} shows the $k_\perp$ dependence of the spectrum for
various quark masses. As expected, the spectrum is harder for larger
quark masses. Note that the tail of the curves is probably affected by
important lattice artifacts due to a too coarse lattice.

\subsection{Loop corrections to the gluon spectrum}
Thus far, we  limited ourselves to the leading order contribution
for both the gluons and the quarks. However, we {\it a priori} know from
figures \ref{fig:nbar} and \ref{fig:nbar_q} what diagrams 
contribute to the gluon and quark multiplicities to all orders. There
is therefore a well defined and systematic procedure to
compute corrections to the previous results. Loop corrections to gluon production are 
very relevant for the following reasons:
\begin{itemize}
\item They contain terms that are
divergent due to unbounded integrals over longitudinal momenta, very
similar to the divergences encountered in the derivation of the BK
equation. One should verify whether these divergences can be absorbed
in the distributions $W[\rho_1]$ and $W[\rho_2]$ of the color sources
of each projectile. This {\sl factorization} is crucial for the
internal consistency of the CGC framework.
\item It has been noted recently that the boost invariant solution
${\cal A}^\mu(x)$ of the Yang-Mills equations is
unstable\footnote{This instability is very similar to the Weibel
instability that occurs in anisotropic
plasmas\cite{YMinsta}${}^{,}$\cite{weibel}.}; rapidity dependent
perturbations to this solution grow exponentially in time. Loop
corrections generate this kind of rapidity dependent
perturbations. Tracking all these terms and resumming them is very
important in order to get meaningful answers from the CGC regarding
the momentum distribution of the produced gluons, and may be relevant
in the problem of thermalization in heavy ion collisions.
\end{itemize}
Note that these two items address very different stages of
the collision process. The first relates to the incoming
wavefunctions (and as such should be independent of the subsequent
collision), while the second issue is about what happens in the final
state after the collision. Therefore, we should aim at writing the
1-loop corrections in a way that separates the initial and final state
as clearly as possible.

\begin{figure}[htbp]
\centerline{
\hskip 10mm
\resizebox*{4.5cm}{!}{\includegraphics{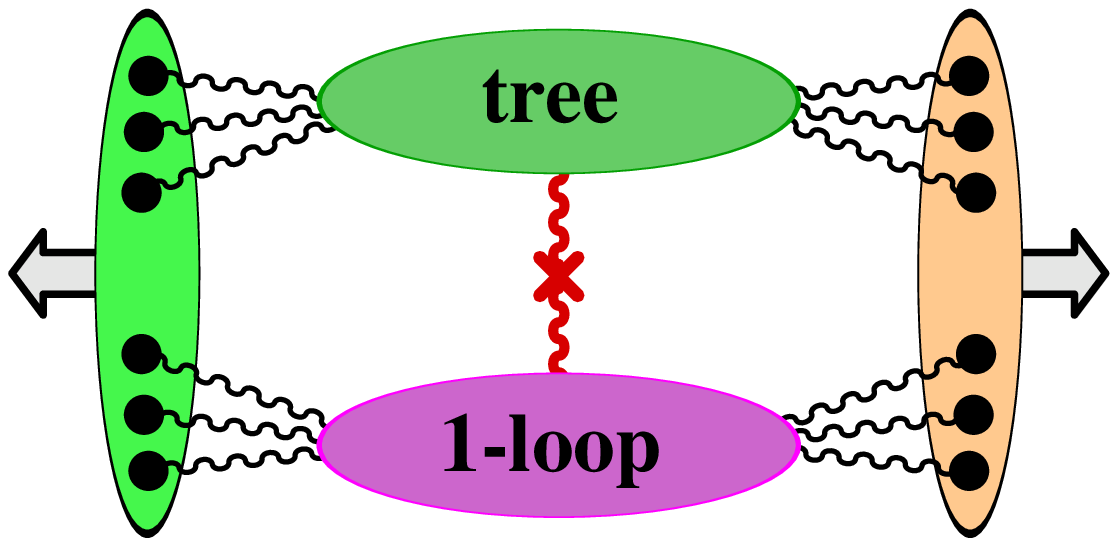}}
\hfill
\resizebox*{4.5cm}{!}{\includegraphics{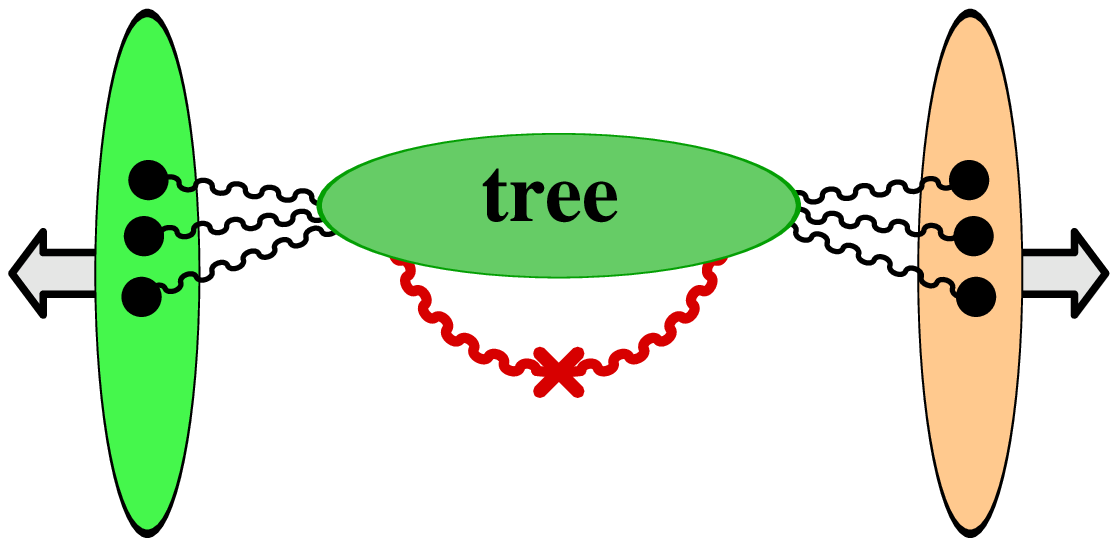}}
\hskip 10mm
}
\caption{\label{fig:nbar_1loop}1-loop diagrams contributing to the gluon spectrum.}
\end{figure}

Let us start by listing the relevant diagrams~: the 1-loop corrections
to the average multiplicity are shown in the diagrams of figure
\ref{fig:nbar_1loop}.
The topology on the left is very similar to the one already
encountered at tree level, except that one of the blobs has now a loop
correction in it. The topology on the right is new; but it is in fact
similar to what we had to evaluate in the case of the quark spectrum,
except that the fermionic cut propagator $S_{+-}$ must be replaced by
the cut propagator $G_{+-}$ of a gluon. The NLO contribution to the
gluon spectrum can be written as
\begin{eqnarray}
\smash{\frac{d\overline{N}_{_{NLO}}}{dY d^2\p_\perp}}
&=&\smash{\frac{1}{16\pi^3}
\int d^4x d^4y\; e^{ip\cdot(x-y)}\;
\square_x\square_y\sum_\lambda \epsilon^\lambda_\mu \epsilon^\lambda_\nu
\;
\Big[
{\cal A}^\mu(x)\delta{\cal A}^\nu(y)
\!+\!
\delta{\cal A}^\mu(x){\cal A}^\nu(y)}
\nonumber\\
&&\qquad\qquad\qquad\qquad\qquad\qquad\qquad\qquad\qquad\qquad
+G_{+-}^{\mu\nu}(x,y)
\Big]\; .
\label{eq:nlo0}
\end{eqnarray}
The two terms of the first line are the contribution of the diagram on
the left of figure \ref{fig:nbar_1loop} (the loop can be in either of
the two blobs), and the term on the second line comes from the diagram
on the right. The field $\delta{\cal A}$ that appears on the first
line is the 1-loop correction to ${\cal A}$; and it obeys the
linearized equation of motion for small fluctuations.

Let us now illustrate how one can separate the initial state from the
final state in the term that contains $G^{\mu\nu}_{+-}(x,y)$. First,
by analogy with the case of the quarks, we can write
\begin{eqnarray}
&&
\int d^4x d^4y e^{ip\cdot(x-y)}\;
\square_x\square_y\sum_\lambda \epsilon^\lambda_\mu \epsilon^\lambda_\nu
\;G_{+-}^{\mu\nu}(x,y)
=\sum_{\lambda,\lambda^\prime}
\int\frac{d^3\q}{(2\pi)^3 2E_\q}\left|{\cal T}^{\lambda\lambda^\prime}_{_R}(\p,\q)\right|^2\; ,
\nonumber\\
&&
{\cal T}^{\lambda\lambda^\prime}_{_R}(\p,\q)\equiv\lim_{x_0\to+\infty}
\int d^3\x\;e^{ip\cdot x}\;(\partial_x^0-iE_\p)\;
\epsilon_\mu^\lambda\, a_{\lambda^\prime\q}^\mu(x)\; ,
\label{eq:nlo1}
\end{eqnarray}
where $a_{\lambda^\prime\q}^\mu(x)$ is a small fluctuation of the
gauge field on top of ${\cal A}^\mu$, with initial condition
$\epsilon_{\lambda^\prime}^\mu e^{iq\cdot x}$ when $x_0\to-\infty$.
The equation of motion of this fluctuation is obtained by writing the
Yang-Mills equations for ${\cal A}+a$ and by linearizing it in $a$.  A
central formula in order to separate the initial and final states is
the following\footnote{To avoid encumbering the equations with indices
of various kinds, we are suppressing all the indices in this and the
following formula.}
\begin{equation}
a(x)=\int\limits_{\tau=0^+}d^3\y\;
\Big[a(0,\y)\cdot {\bs T}_{\y}\Big]\;{\cal A}(x)\; ,
\label{eq:magic}
\end{equation}
where $(0,\y)$ denotes a point located on the light-cone ($\tau=0$)
($\y$ represents any set of three coordinates that map the
light-cone.)  In this formula, the classical field ${\cal A}$ is
considered as a functional of its initial condition ${\cal A}(0,\y)$
on the light-cone. The notation $\Big[a(0,\y)\cdot {\bs T}_{\y}\Big]$
is a shorthand for
\begin{equation}
\Big[a(0,\y)\cdot {\bs T}_{\y}\Big]
\equiv
{\colorb a(0,\vec\y)}\frac{\delta}{\delta {\colorb{\cal A}(0,\vec\y)}}
+
\Big[(n\cdot\partial_y){\colorb a(0,\vec\y)}\Big]
\frac{\delta}{\delta(n\cdot\partial_y){\colorb {\cal A}(0,\vec\y)}}
\; .
\label{eq:T}
\end{equation}
(In this formula, the 4-vector $n^\mu$ is a vector
normal\footnote{$n^\mu dx_\mu=0$ if $dx_\mu$ is a small
displacement on the light-cone at the point under consideration.} to
the light-cone.) The proof of eq.~(\ref{eq:magic}) is
straightforward\footnote{Write the Green's formula that expresses
${\cal A}(x)$ in terms of the initial ${\cal A}(0,\y)$,
insert it in eq.~(\ref{eq:magic}), and check that this leads to
the Green's formula that relates $a(x)$ to its initial condition
$a(0,\y)$.}, but its diagrammatic interpretation is more interesting.
\begin{figure}[htbp]
\begin{center}
\resizebox*{10cm}{!}{\includegraphics{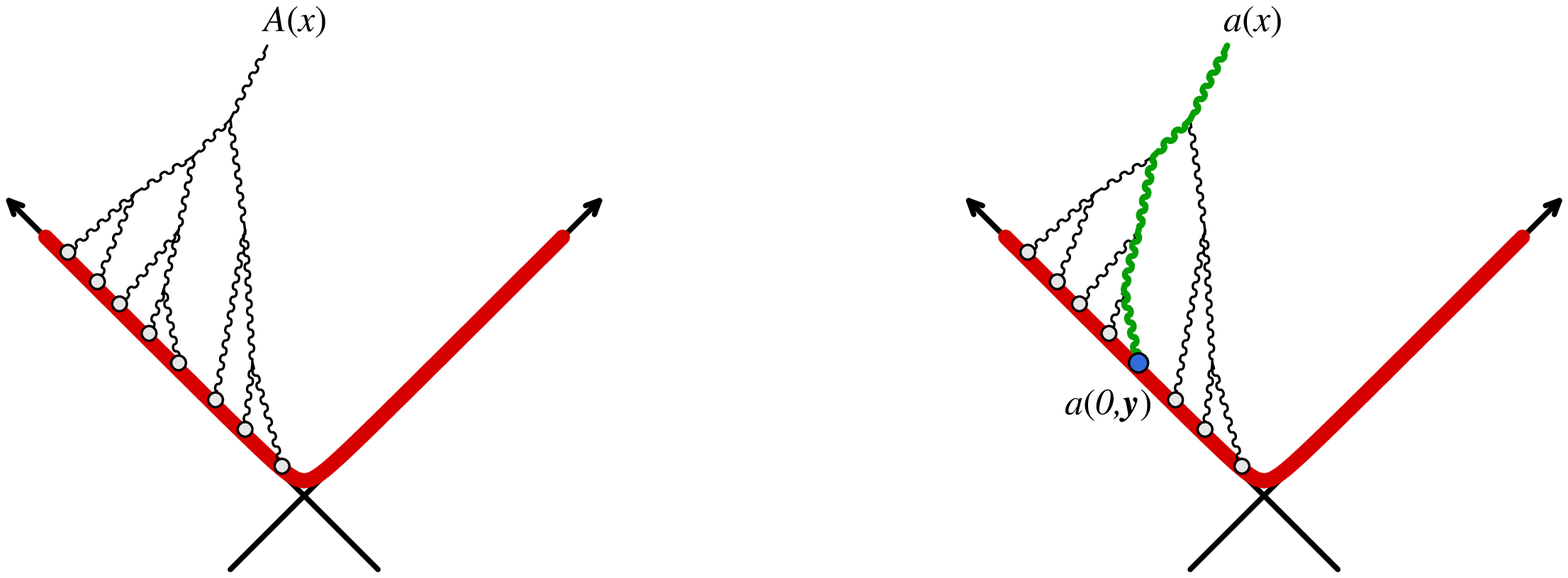}}
\end{center}
\caption{\label{fig:YM-init}Left~: diagrammatic representation of
${\cal A}$ as a function of its initial condition on the light-cone
(the open dots represent the initial ${\cal A}(0,\y)$). Right~:
propagation of a small fluctuation on top of the classical field.}
\end{figure}
Note first that ${\cal A}^\mu(x)$, seen as a functional of its initial
condition on the light-cone, can also be represented by tree diagrams,
as illustrated in the left panel of figure \ref{fig:YM-init}. (This can
be seen from the Green's formula
for ${\cal A}(x)$.) The action of the operator defined in
eq.~(\ref{eq:T}) on the classical field ${\cal A}(x)$ is to replace
one of the open dots in figure \ref{fig:YM-init} by the fluctuation
$a(0,\y)$, represented by a filled dot in the right panel of figure
\ref{fig:YM-init}. The diagram one gets after this is nothing but a
contribution to the propagation of a small fluctuation over the
classical field. Plugging eq.~(\ref{eq:magic}) in eq.~(\ref{eq:nlo1}),
this quantity becomes
\begin{eqnarray}
&&
\lim_{x_0=y_0\to+\infty}\int d^3\x d^3\y\;
e^{ip\cdot(x-y)}\;
(\partial_x^0-iE_\p)(\partial_y^0+iE\p)
\sum_\lambda \epsilon^\lambda_\mu \epsilon^\lambda_\nu
\nonumber\\
&&\times\!\!\!
\int\limits_{\tau=0^+}\!\!\!d^3\u d^3\v
\sum_{\lambda^\prime}\!\!
\int\frac{d^3\q}{(2\pi)^3 2E_\q}
\left[
\Big[a_{\lambda^\prime\q}(0,\u)\cdot {\bs T}_{\u}\Big]\;{\cal A}^\mu(x)
\right]
\left[
\Big[a^*_{\lambda^\prime\q}(0,\v)\cdot {\bs T}_{\v}\Big]\;{\cal A}^\nu(y)
\right]\; .
\nonumber\\
&&
\end{eqnarray}
The brackets are crucial in this formula, in order to limit the scope
of the derivatives contained in the operators ${\bs T}_\u$ and ${\bs
T}_\v$.  Note that, if it were not for these brackets, the first line
and the two ${\cal A}$'s of the second line would be nothing but the
LO gluon spectrum. It turns out that, after one adds the contribution
of the first line in eq.~(\ref{eq:nlo0}), the NLO correction to the
spectrum can be written as
\begin{eqnarray}
&&
\frac{d\overline{N}_{_{NLO}}}{dY d^2\p_\perp}
=
\left[\,
\int\limits_{\tau=0^+}\!\!\!
d^3\u\;
\Big[\delta{\cal A}(0,\u)\cdot {\bs T}_\u\Big]
+\!\!\!
\int\limits_{\tau=0^+}\!\!\!d^3\u d^3\v\;
\Big[\Sigma(\u,\v)\cdot{\bs T}_\u{\bs T}_\v\Big]
\right]
\frac{d\overline{N}_{_{LO}}}{dY d^2\p_\perp}\; ,
\nonumber\\
&&
\label{eq:nlo2}
\end{eqnarray}
where the LO spectrum is considered as a functional of the initial
classical field on the light-cone. In this equation, $\delta{\cal
A}(0,\u)$ is the value of $\delta{\cal A}$ on the light-cone, and the
2-point $\Sigma(\u,\v)$ is defined as
\begin{equation}
\Sigma(\u,\v)\equiv
\sum_{\lambda^\prime}
\int\frac{d^3\q}{(2\pi)^3 2E_\q}\;
a_{\lambda^\prime\q}(0,\u)a^*_{\lambda^\prime\q}(0,\v)\; .
\label{eq:sigma}
\end{equation}
Note that $\delta{\cal A}(0,\u)$ and $\Sigma(\u,\v)$ are in principle
calculable analytically. 
\begin{figure}[htbp]
\begin{center}
\resizebox*{8cm}{!}{\includegraphics{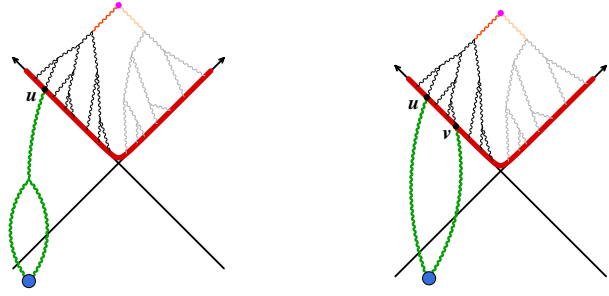}}
\end{center}
\caption{\label{fig:factor}Illustration of eq.~(\ref{eq:nlo2}). The 1-
and 2-point functions below the light-cone are respectively
$\delta{\cal A}(0,\u)$ and $\Sigma(\u,\v)$.}
\end{figure}
Eq.~(\ref{eq:nlo2}) realizes the separation we were seeking of the
initial and final states. Indeed, the operator in the square bracket
depends only on what happens below the light-cone, i.e. before the
collision. On the contrary, the LO spectrum seen as a functional of
the initial classical field ${\cal A}$ depends only on the final state
dynamics.  The other benefit of this formula is that is expresses the
NLO correction as a perturbation of the LO one; this property -- that
seems generalizable to other inclusive observables -- suggests the
universality of the initial state divergences and their
factorizability.

From eq.~(\ref{eq:nlo2}), it is easy to see what are the potential
sources of divergences. A first issue is that the coefficients
$\delta{\cal A}(0,\u)$ and $\Sigma(\u,\v)$ are infinite. For
$\Sigma(\u,\v)$ for instance, the integration over the longitudinal
component of the momentum $\q$ in eq.~(\ref{eq:sigma}) diverges. A
similar divergence occurs in the loop contained in $\delta{\cal
A}(0,\u)$. The fact that these divergences arise in the first factor
of eq.~(\ref{eq:nlo2}) indicates that they are related to the
evolution of the initial projectiles prior to the collision. These
divergences can be momentarily regularized by introducing cutoffs
$Y_0,Y^\prime_0$ in rapidity around the rapidity $Y$ at which the
spectrum is calculated. Thus, $\delta{\cal A}(0,\u)$ and
$\Sigma(\u,\v)$ become finite but depend on these unphysical cutoffs.
To be consistent, the distribution of the sources $\rho_1$ and
$\rho_2$ should be evolved from the beam rapidities to $Y_0$ and
$Y^\prime_0$ respectively. Thus, the complete formula for the LO+NLO
spectrum, including the average over the sources, should be
\begin{eqnarray}
&&
\frac{d\overline{N}_{_{LO+NLO}}}{dY d^2\p_\perp}
=
\int \big[D\rho_1\big]\big[D\rho_2\big]\;
W_{_{Y_{\rm beam}-Y_0}}[\rho_1]
W_{_{Y_{\rm beam}+Y^\prime_0}}[\rho_2]
\nonumber\\
&&\times\underbrace{
\left[1+\!\!\!
\int\limits_{\tau=0^+}\!\!\!
d^3\u\;
\Big[\delta{\cal A}(0,\u)\cdot {\bs T}_\u\Big]
+\!\!\!
\int\limits_{\tau=0^+}\!\!\!d^3\u d^3\v\;
\Big[\Sigma(\u,\v)\cdot{\bs T}_\u{\bs T}_\v\Big]
\right]_{Y^\prime_0}^{Y_0}}
\frac{d\overline{N}_{_{LO}}}{dY d^2\p_\perp}\; ,
\nonumber\\
&&\hskip 43mm {\cal O}_{Y^\prime_0}^{Y_0}[\rho_1,\rho_2]
\end{eqnarray}
where the subscript $Y^\prime_0$ and superscript $Y_0$ indicate that
the momentum integrals contained in the bracket have cutoffs in
rapidity. Recall that the LO spectrum in the right hand side is a
function of ${\cal A}$ on the light-cone, which is itself a function
of $\rho_{1,2}$. The factorizability of these divergences in the
initial state is equivalent to the independence of the previous
formula with respect to the unphysical $Y_0$ and $Y^\prime_0$. Let us
for instance change $Y_0$ into $Y_0+\Delta Y_0$. According to the
JIMWLK equation, the distribution of $\rho_1$ is modified by
\begin{equation}
W_{_{Y_{\rm beam}-Y_0}}[\rho_1]\quad\to\quad
\Big[1+\Delta Y_0{\cal H}[\rho_1]\Big]\;W_{_{Y_{\rm beam}-Y_0}}[\rho_1]\; .
\end{equation}
At the same time, the operator in the right hand side varies by
\begin{equation}
{\cal O}_{Y^\prime_0}^{Y_0}[\rho_1,\rho_2]\quad\to\quad
{\cal O}_{Y^\prime_0}^{Y_0}[\rho_1,\rho_2]
+\Delta Y_0\;\frac{\partial {\cal O}_{Y^\prime_0}^{Y_0}[\rho_1,\rho_2]}{\partial Y_0}\; .
\end{equation}
At this point, one can verify that the terms linear\footnote{Note
that, since we have only considered 1-loop corrections, this
independence can only be satisfied for small variations of the cutoff,
at linear order in these variations.} in $\Delta Y_0$ cancel provided
that
\begin{equation}
\frac{\partial {\cal O}_{Y^\prime_0}^{Y_0}[\rho_1,\rho_2]}{\partial Y_0}
=
{\cal H}^\dagger[\rho_1]\; .
\end{equation}
Similar considerations on the $Y^\prime_0$ dependence give another
condition~:
\begin{equation}
\frac{\partial {\cal O}_{Y^\prime_0}^{Y_0}[\rho_1,\rho_2]}{\partial
Y^\prime_0}
=
-
{\cal H}^\dagger[\rho_2]\; .
\end{equation}
Therefore, in order to check whether one can factorize these
divergences in the JIMWLK evolution of the incoming sources, one must
calculate the coefficients $\delta{\cal A}(0,\u)$ and $\Sigma(\u,\v)$
-- or at least their divergent part -- and remap the operator ${\cal
O}_{Y^\prime_0}^{Y_0}[\rho_1,\rho_2]$ into the JIMWLK
Hamiltonian. Although this program has not been fully implemented yet,
one can already note that the structure of ${\cal
O}_{Y^\prime_0}^{Y_0}[\rho_1,\rho_2]$ makes this outcome very plausible.

Eq.~(\ref{eq:nlo2}) also allows us to discuss the issue of the
instability of the boost invariant classical solution. These
instabilities manifest themselves in the fact that the action of ${\bs
T}_\u$ on ${\cal A}(x)$ diverges when the time $x_0$ goes to
infinity. Indeed,
\begin{equation}
{\bs T}_\u {\cal A}(x)\sim\frac{\delta {\cal A}(x)}{\delta {\cal A}(0,\y)}
\end{equation}
is a measure of how ${\cal A}(x)$ is sensitive to its initial
condition. 
\begin{figure}[htbp]
\begin{center}
\resizebox*{6cm}{!}{\rotatebox{-90}{\includegraphics{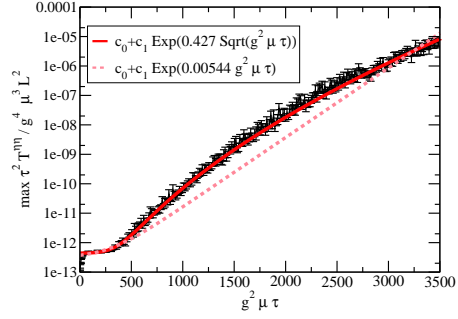}}}
\end{center}
\caption{\label{fig:insta}Time dependence of small fluctuations on top
of the boost independent classical field.}
\end{figure}
Therefore, if the solution ${\cal A}(x)$ is unstable, small
perturbations of its initial condition lead to exponentially growing
changes in the solution. From the numerical study of these
instabilities (see figure \ref{fig:insta}), one gets\cite{YMinsta}
\begin{equation}
{\bs T}_\u {\cal A}(x)\sim e^{\sqrt{\mu\tau}}\; ,
\end{equation}
where $\mu$ is of the order of the saturation momentum. This means
that, although the 1-loop corrections are suppressed by a factor
$\alpha_s$ compared to the LO, some of these corrections are enhanced
by factors that grow exponentially in time after the collision. At
first sight, one may expect a complete breakdown of the CGC
description at
\begin{equation}
\tau_{\rm max}\sim Q_s^{-1}\ln^2\left(\frac{1}{\alpha_s}\right)\; ,
\label{eq:taumax}
\end{equation}
i.e. the time at which the 1-loop corrections become as large as the
LO contribution. The only way out of this conclusion is to resum all
these enhanced corrections in the hope that the resummed series is
better behaved when $\tau\to+\infty$. Let us assume for the time being
that we have performed this resummation, and that the sum of these
enhanced terms generalize
eq.~(\ref{eq:nlo2}) to read 
\begin{equation}
\frac{d\overline{N}_{\rm resummed}}{dY d^2\p_\perp}
=
Z[{\bs T}_\u]\;\frac{d\overline{N}_{_{LO}}[{\cal A}(0,\u)]}{dY d^2\p_\perp}\; ,
\end{equation}
where $Z[{\bs T}_\u]$ is a certain functional of the operator ${\bs
T}_\u$. In the right hand side, we have emphasized the fact that the
LO spectrum is a functional of the initial classical field on the
light-cone. This formula can be written in a more intuitive way by
performing a Fourier transform of $Z[{\bs T}_\u]$, 
\begin{equation}
Z[{\bs T}_\u]\equiv
\int\big[Da(\u)\big]\;
e^{i\int_{\tau=0^+}d^3\u\;\big[a(\u)\cdot {\bs T}_\u\big]}
\;
\widetilde{Z}[a(\u)]\; .
\end{equation}
In this formula, the functional integration $[Da(\u)]$ is in fact an
integration over two fields~: the fluctuation $a(\u)$ itself and its
derivative normal to the light-cone $(n\cdot\partial_u)a(\u)$. Thanks
to the fact that ${\bs T}_\u$ is the generator of translations of the
initial conditions on the light-cone, the exponential in the previous
formula is the translation operator itself. Therefore, when this
exponential acts on a functional of the initial classical field ${\cal
A}(0,\u)$, it gives the same functional evaluated with a shifted
initial condition ${\cal A}(0,\u)+a(\u)$. Therefore,
we can write
\begin{equation}
\frac{d\overline{N}_{\rm resummed}}{dY d^2\p_\perp}
=\int\big[Da(\u)\big]\;\widetilde{Z}[a(\u)]\;
\frac{d\overline{N}_{_{LO}}[{\cal A}(0,\u)+a(\u)]}{dY d^2\p_\perp}\; .
\end{equation}
We see that the effect of the resummation is simply to add
fluctuations to the initial conditions of the classical field, with a
distribution that depends on the details of the
resummation\footnote{In a recent work by one of the authors, using a
completely different approach, the spectrum of initial fluctuations
was found to be Gaussian\cite{FukusGM1}.}. It is easy to understand why these
fluctuations are crucial in the presence of instabilities~: despite
the fact that they are suppressed by an extra power of $\alpha_s$, the
instabilities make them grow and eventually become as large as the
LO. One can also see that the resummation has the effect of lifting
the time limitation of eq.~(\ref{eq:taumax}). Indeed, after the
resummation, the fluctuation $a(\u)$ has entered in the initial
condition for the full Yang-Mills equation, whose non-linearities
prevent the solution from blowing up.  A very important question is
whether these instabilities fasten the local thermalization of the
system formed in heavy ion collisions.

\subsection{Summary and outlook}
If the initial state factorization works as expected, and after the
resummation of the leading contributions of the instability, the
formula for the gluon spectrum should read
\begin{eqnarray}
\frac{d{\colorc\overline{N}}}{dYd^2\vec\p_\perp}
&=&
\int \big[D\rho_1]\;[D\rho_2 \big]
\;\;
{\colord W_{_{Y_{\rm beam}-Y}}[\rho_1]}\;{\colord W_{_{Y_{\rm beam}+Y}}[\rho_2]}
\nonumber\\
&&\qquad\times
\int\big[Da\big]\;{\colorb\widetilde{Z}[a]}\;
\frac{d{\colorc\overline{N}_{_{LO}}}[{\cal A}(0,\u){\colorb +a(\u)}]}
{dYd^2\vec\p_\perp}
\; .
\label{eq:final}
\end{eqnarray}
This formula resums the most singular terms at each order in
$\alpha_s$. Because of their relation to the physics of the initial
and final state respectively, the distributions $W[\rho]$ generalize
parton distributions, while ${\colorb\widetilde{Z}[a]}$ plays a role
similar to that of a fragmentation function\footnote{Naturally, this
function has nothing to do with a gluon fragmenting into a
hadron. Instead, it is related to how classical fields become
gluons.}.

Note that, even after the resummations performed in the initial and
final states of eq.~(\ref{eq:final}), this formula still suffers from
the usual problem of collinear gluon splitting in the final
state. This is not a serious concern in heavy ion collisions though,
because collinear singularities occur only when one takes the
$\tau\to+\infty$ limit, and we do expect to have to switch to another
description (like hydrodynamics) long before this becomes a
problem. In fact, the initial condition for hydrodynamics should be
specified in terms of the energy-momentum tensor, which is infrared
and collinear safe because it measures only the flow of energy and
momentum.

A more important problem, that has still not received a satisfactory
answer, is to understand how the initial particle spectrum -- or the
local energy momentum-tensor -- become isotropic. This requires formulating a kinetic 
theory of the glasma which describes how particles emerge from the decaying classical field and their subsequent 
interactions both with the classical field and with other particles. Recently, such a kinetic equation has been derived 
for a scalar field theory coupled to strong sources~\cite{GelisJV1}. Extending this work to QCD and exploring its consequences-
in particular, the approach of the particle+field system towards equilibration remains a challenging problem.

\section*{Acknowledgements}
FG would like to thank the organizers -- and in particular
D. P. Menezes -- of the Xth Hadron Physics Workshop held in
Florianopolis, Brazil, for their invitation to give these lectures and
for the nice and stimulating atmosphere of the meeting, as well as the
hospitality of M.B. Gay-Ducati at the UFRGS, and of E.S. Fraga and
T. Kodama at the UFRJ. FG also acknowledges the financial support of
CAPES-COFECUB under project \#443-04. TL and RV are supported by DOE Contract No. DE-AC02-98CH10886.


\end{document}